\documentclass[a4paper, 12pt]{article}

\usepackage{tikz}
\usepackage{latexsym,amsmath,amsfonts,amssymb}
\usepackage{mathrsfs}
\usepackage[american]{babel}
\usepackage{bbm}
\usepackage[nosort]{cite}    
\usepackage[pdfencoding=auto]{hyperref}
\hypersetup{colorlinks, citecolor=[rgb]{.7,0,0}, linkcolor=[rgb]{0,0,0.7}, urlcolor=[rgb]{0,0,0.5}}

\renewcommand{\baselinestretch}{1.2}
\newcommand{\wt}{\widetilde}
\newcommand{\wh}{\widehat}
\newcommand{\wb}{\overline}
\newcommand{\matht}[1]{\texorpdfstring{\ensuremath{\boldsymbol{#1}}}{#1}}
\newcommand{\ts}{\textstyle}
\newcommand{\ds}{\displaystyle}
\newcommand{\str}{\rule{0pt}{0.6em}}

\newcommand{\eg}{\textit{e.g.}}

\newcommand{\ie}{\textit{i.e.}}

\numberwithin{equation}{section}

\newcommand{\Dslash}{D\hspace{-7.6pt}\raisebox{0.9pt}{\slash}\,}
\newcommand{\nn}{\nonumber}
\newcommand{\mat}[1]{\begin{pmatrix} #1 \end{pmatrix}}
\newcommand{\smat}[1]{\big( \begin{smallmatrix} #1 \end{smallmatrix} \big)}
\newcommand{\be}{\begin{equation}} \newcommand{\ee}{\end{equation}}
\newcommand{\bea}{\begin{equation} \begin{aligned}} \newcommand{\eea}{\end{aligned} \end{equation}}
\newcommand{\ba}{\begin{array}} \newcommand{\ea}{\end{array}}

\newcommand{\cC}{\mathcal{C}}
\newcommand{\cD}{\mathcal{D}}

\newcommand{\cF}{\mathcal{F}}
\newcommand{\cG}{\mathcal{G}}

\newcommand{\cI}{\mathcal{I}}

\newcommand{\cK}{\mathcal{K}}
\newcommand{\cL}{\mathcal{L}}

\newcommand{\cN}{\mathcal{N}}
\newcommand{\cO}{\mathcal{O}}

\newcommand{\cQ}{\mathcal{Q}}

\newcommand{\cV}{\mathcal{V}}

\newcommand{\cY}{\mathcal{Y}}
\newcommand{\cZ}{\mathcal{Z}}

\newcommand{\bC}{\mathbb{C}}

\newcommand{\bP}{\mathbb{P}}

\newcommand{\bR}{\mathbb{R}}

\newcommand{\bZ}{\mathbb{Z}}
\newcommand{\ff}{\mathfrak{f}}
\newcommand{\fg}{\mathfrak{g}}
\newcommand{\fh}{\mathfrak{h}}

\newcommand{\fm}{\mathfrak{m}}

\newcommand{\fn}{\mathfrak{n}}

\newcommand{\fr}{\mathfrak{r}}

\newcommand{\sT}{{\sf{T}}}
\newcommand{\sW}{\mathsf{W}}

\newcommand{\unit}{\mathbbm{1}}

\newcommand{\rSU}{\mathrm{SU}}
\newcommand{\rU}{\mathrm{U}}
\newcommand{\rSL}{\mathrm{SL}}

\newcommand{\fsu}{\mathfrak{su}}

\DeclareMathOperator{\Tr}{Tr}

\DeclareMathOperator{\rk}{rk}
\DeclareMathOperator{\re}{\mathbb{R}e}
\DeclareMathOperator{\im}{\mathbb{I}m}

\DeclareMathOperator{\Li}{Li}

\makeatletter
\def\blfootnote{\gdef\@thefnmark{}\@footnotetext}
\makeatother

\begin{document}

\thispagestyle{empty}
\begin{flushright}
SISSA  19/2022/FISI
\end{flushright}
\vspace{13mm}  
\begin{center}
{\huge  A quantum mechanics for magnetic horizons} 
\\[13mm]
{\large Francesco Benini$^{1,2,3}$, Saman Soltani$^{1,2}$, and Ziruo Zhang$^{1,2}$}

\bigskip
{\it
$^1$ SISSA, Via Bonomea 265, 34136 Trieste, Italy \\[.2em]
$^2$ INFN, Sezione di Trieste, Via Valerio 2, 34127 Trieste, Italy \\[.2em]
$^3$ ICTP, Strada Costiera 11, 34151 Trieste, Italy \\[.2em]
}


\vspace{2cm}

{\bf Abstract}\\[8mm]
\end{center}

\noindent
We construct an $\cN=2$ supersymmetric gauged quantum mechanics, by starting from the 3d Chern-Simons-matter theory holographically dual to massive Type IIA string theory on AdS$_4 \times S^6$, and Kaluza-Klein reducing on $S^2$ with a background that is dual to the asymptotics of static dyonic BPS black holes in AdS$_4$. The background involves a choice of gauge fluxes, that we fix via a saddle-point analysis of the 3d topologically twisted index at large $N$. The ground-state degeneracy of the effective quantum mechanics reproduces the entropy of BPS black holes, and we expect its low-lying spectrum to contain information about near-extremal horizons. Interestingly, the model has a large number of statistically-distributed couplings, reminiscent of SYK models.

\newpage
\pagenumbering{arabic}
\setcounter{page}{1}
\setcounter{footnote}{0}
\renewcommand{\thefootnote}{\arabic{footnote}}

{\renewcommand{\baselinestretch}{.87} \parskip=0pt
\setcounter{tocdepth}{2}
\tableofcontents}


\section{Introduction}
\label{sec: intro}

In the context of counting the quantum microstates of black holes \cite{Strominger:1996sh}, a lot of work has been done over the years for what concerns the supersymmetric (or BPS) sector, both in flat space and in anti-de-Sitter (AdS) space.
Much less is known about non-supersymmetric black holes. With the development of our understanding of 2d JT gravity \cite{Teitelboim:1983ux, Jackiw:1984je, Maldacena:2016upp, Saad:2019lba} and the SYK model \cite{Sachdev:1992fk, Kitaevtalk:2015, Maldacena:2016hyu}, though, progress has been possible for near-BPS and near-extremal black holes. In particular, in a series of papers \cite{Iliesiu:2020qvm, Heydeman:2020hhw, Boruch:2022tno, Iliesiu:2022onk} the authors were able to derive the contribution to the behavior of the density of states of those black holes above extremality, coming from the dynamics of gravitational zero-modes in the near-horizon region. The analysis revealed the presence of a gap above extremality for BPS black holes, and a strong suppression of the density of states for extremal black holes in the non-supersymmetric case. For black holes in AdS, where the overall entropy of BPS black holes can be determined from the dual field theory at large $N$ (see, \eg, \cite{Benini:2015eyy}), it would be desirable to reproduce the results above about near-extremal black holes from a field theory computation. In the case of AdS$_3$, indeed, it has been possible to extract the density of near-extremal states from a beautiful and general analysis of CFT$_2$'s \cite{Ghosh:2019rcj}, but no similar computation is available in higher dimensions.

In this paper we make a step in that direction, by constructing a supersymmetric gauged quantum mechanics (QM) that we expect to capture information about near-extremal black hole horizons. We work in a very specific setup: massive Type IIA string theory on $S^6$, which is dual to a 3d $\cN=2$ $\rSU(N)_k$ Chern-Simons-matter theory \cite{Guarino:2015jca}.%
\footnote{The theory has three adjoint chiral multiplets and a superpotential. It is essentially the 4d $\cN=4$ $\rSU(N)$ super-Yang-Mills theory reduced to 3d and deformed by an $\cN=2$ Chern-Simons term. The Chern-Simons level $k$ is proportional to the quantized Romans mass $F_{0}$ in massive type IIA string theory.}
The supergravity admits asymptotically-AdS$_4$ static magnetic (or topologically twisted) BPS black holes \cite{Cacciatori:2009iz, Guarino:2017eag, Guarino:2017pkw}, that we aim to describe. The quantum mechanics is then obtained by reducing the dual 3d field theory on $S^2$, with a specific background that corresponds to the black hole asymptotics.%
\footnote{The background is dual to the black-hole chemical potentials, or charges, depending on the ensemble.}

More specifically, the entropy of static%
\footnote{To be precise, here we work in the grand-canonical ensemble at zero chemical potential for the angular momentum quantum number. This means that the BPS states of rotating magnetically-charged black holes contribute as well. However, at large $N$, the index is dominated by the states of static (\ie, with vanishing angular momentum) black holes. It could be interesting to study the refinement of the TT index by a chemical potential for angular momentum \cite{Benini:2015noa}.}
magnetically-charged BPS black holes in AdS$_4$ is captured by the topologically twisted (TT) index \cite{Benini:2015noa, Closset:2016arn} of the dual 3d boundary theory \cite{Benini:2015eyy, Hosseini:2016tor, Benini:2016hjo, Benini:2016rke, Liu:2017vll, Jeon:2017aif, Azzurli:2017kxo}, see in particular \cite{Benini:2017oxt, Hosseini:2017fjo, Liu:2018bac} for the specific example in massive Type IIA studied here. In the Lagrangian formulation, the TT index is the Euclidean partition function of the theory on $S^2 \times S^1$, in the presence of a supersymmetric background that holographically reflects the asymptotics of the BPS black hole. The background can be thought of as a topological twist on $S^2$ that preserves two supercharges, or equivalently as an external magnetic flux for the R-symmetry. One observes that the TT index takes the form of the Witten index of a quantum mechanics, obtained by reducing the 3d theory on $S^2$ with the twisted background. This fact is not a coincidence: the TT index is robust under continuous deformations, in particular under the flow to low energies, where one only remains with the light 1d degrees of freedom contributing to the Witten index. Up to exponentially small corrections at large $N$, the index is the grand canonical partition function for the BPS ground states of that quantum mechanics. In other words, the ground states of that quantum mechanics are the microstates of a BPS black hole with given charges, and one expects the excited states to describe near-extremal black holes. The goal of this paper is to construct such a quantum mechanics.

The procedure we outlined has a technical complication: the formula for the TT index --- schematically in (\ref{index original expression}) --- has an infinite sum over gauge fluxes on $S^2$. For each term in the sum, one obtains a different quantum mechanics upon reduction. Thus it appears that, even at finite $N$, one has to deal with a quantum mechanical model with an infinite number of sectors, over which we do not have good control.%
\footnote{This is partially due to the fact that the reduction is in the grand canonical ensemble for the electric charges (though it is micro-canonical for the magnetic charges), with fixed chemical potentials. Therefore, the states of all BPS and near-BPS black holes are mixed up together.}
Nevertheless, in the large $N$ limit we expect one sector to dominate the entropy%
\footnote{We are grateful to Juan M. Maldacena for suggesting this possibility to us years ago.}
and thus to contribute the majority of the states. We determine such a sector by performing a saddle-point evaluation of the index in the sum over fluxes. This gives us an $\cN=2$ supersymmetric gauged quantum mechanics with a finite number of fields (at finite $N$).

The resulting $\cN=2$ QM, that we exhibit in Section~\ref{sec: QM}, has some interesting features. It has $\rU(1)^N$ gauge group, and a number of fields that scales as $N^\frac{7}3$. It has an $\rSU(2)$ global symmetry, dual to the isometry of the $S^2$ black-hole horizon. More importantly, it has a large number of couplings among the fields, expressed in terms of Clebsch-Gordan coefficients (arising in the reduction from the overlap of Landau-level wave-functions on $S^2$). Therefore, although the quantum mechanics is specific and well defined, we expect that at large $N$ its couplings could be approximated by random variables following a suitable statistical distribution. This makes us hopeful that the IR dynamics might have some traits in common with supersymmetric SYK models \cite{Fu:2016vas, Heydeman:2022lse}. The idea of obtaining a supersymmetric QM with fixed, but statistically distributed, couplings that could describe near-extremal horizons already appeared in \cite{Anninos:2016szt} in the context of asymptotically-flat black holes in string theory.

In the large $N$ saddle-point evaluation of the TT index, we noticed that there is actually a series of saddle points --- one of which dominates the large $N$ expansion. These saddle points are labelled by shifts of the chemical potentials by $2\pi$, and likely correspond to a series of complex supergravity solutions with the very same boundary conditions, as in \cite{Dijkgraaf:2000fq, Aharony:2021zkr}.

The paper is organized as follows. In Section~\ref{sec: saddle point} we re-examine the large $N$ limit of the TT index by performing a saddle-point approximation both in the integration variables as well as in the sum over fluxes. This analysis already appeared recently in \cite{Hosseini:2022vho}. Section~\ref{sec: KK reduction}, which is the most technical one, is devoted to the dimensional reduction of the 3d theory on $S^2$ in the presence of gauge magnetic fluxes. This reduction involves a judicious choice of gauge fixing. In Section~\ref{sec: QM} we exhibit the effective $\cN=2$ gauged quantum mechanics; the hurried reader who is only interested in the final result can directly jump there. Finally, in Section~\ref{sec: stability} we comment on which type of classical and quantum corrections to our analysis one might expect. Many of the technical details are collected in appendices.

\section{Saddle-point approach to the TT index}
\label{sec: saddle point}

We begin by re-examining the evaluation of the TT index of 3d $\cN=2$ gauge theories at large $N$. The localization formula for the index found in \cite{Benini:2015noa} involves a sum over gauge fluxes $\fm$ on $S^2$, as well as a contour integral in the space of complexified gauge connections $u$ on $S^1$. At large $N$, we apply a saddle-point approximation both to the integral over $u$ as well as to the sum over fluxes, treated as a continuous variable $\fm$. The idea to compute a supersymmetric index in this way was put forward, for instance, in \cite{Hosseini:2018uzp, Choi:2019zpz} (see also \cite{Jain:2021sdp, Hosseini:2021mnn, Hosseini:2022vho}).%
\footnote{In particular, the evaluation of the (refined) TT index of the specific model studied here, through a saddle-point approximation of the sum over fluxes, has recently already appeared in \cite{Hosseini:2022vho}.}
The upshot is to identify a specific gauge flux sector that dominates the index and, via holography, the BPS black hole entropy. In Section~\ref{sec: KK reduction} we will use that flux sector to perform a reduction of the 3d theory on $S^2$ down to a quantum mechanics.

The analysis in this and the following sections is performed in a specific (and simple) model, presented in Section~\ref{sec: the model}. This choice is made for the sake of concreteness, but other theories (for instance ABJM \cite{Aharony:2008ug}) could be studied in a similar way.

\subsection{The basic idea}
\label{sec: basic idea}

We are interested in the topologically twisted index \cite{Benini:2015noa} of the theory, because this quantity is known to reproduce the entropy of a class of BPS AdS$_4$ dyonic black holes \cite{Benini:2017oxt, Hosseini:2017fjo, Liu:2018bac}. The localization formula for the index can be written schematically as
\be
\label{index original expression}
\cI_{S^2\times S^1} = \frac{1}{\lvert \sW \rvert} \, \sum_{\fm\in\Gamma_{\fh}} \, \oint_{\mathcal{C}} \, \prod_{i=1}^{N} \, \frac{du^i}{2\pi} \; e^{\fm V'(u) \,+\, \Omega(u)} \;.
\ee
Here $\lvert \sW \rvert$ is the order of the Weyl group, $\Gamma_\fh$ is the co-root lattice, $N$ is the rank of the gauge group, and $\cC$ is an appropriate integration contour for the complexified Cartan-subalgebra-valued holonomies $\{u^i\}\in \fh_\bC / 2\pi \Gamma_\fh$. Let us outline three different approaches to this expression at large $N$.

\begin{enumerate}
\item The approach developed in \cite{Benini:2015noa} was to resum over $\fm$, schematically
\be
\cI_{S^2\times S^1} = \frac{1}{\lvert \sW \rvert} \oint_{\mathcal{C}} \prod_{i=1}^{N}\frac{du^i}{2\pi} \, \frac{e^{\Omega(u)}}{1-e^{V'(u)}} \;,
\ee
then determine the positions $\bar u$ of the poles by solving the ``Bethe Ansatz Equations" (BAEs)
\be
\label{schematic BAEs}
e^{V'(\bar u)} = 1 \;,
\ee
and finally take the residues
\be
\cI_{S^2\times S^1}^{\text{BAE}} = \frac{1}{\lvert\sW\rvert} \sum_{\bar{u}\in \text{BAE}} \, \frac{e^{\Omega(\bar u)}}{i^N \, V''(\bar u)} \;.
\ee

\item Alternatively, we can evaluate both the sum over $\fm$ and the integral over $u$ in (\ref{index original expression}) in the saddle-point approximation, treating $\fm$ as a continuous variable. The simultaneous saddle-point equations for $\fm$ and $u$ are, schematically:
\be
\label{basic system}
\left\{ \begin{aligned} 0 &= V'(\bar u) \\ 0 &= \bar\fm V''(\bar u) + \Omega'(\bar u) \;. \end{aligned} \right.
\ee
Taking into account that $V'(u)$ in (\ref{index original expression}) is defined up to integer shifts by $2\pi i$, the first set of equations is exactly the set of BAEs (\ref{schematic BAEs}), while the second set of equations uniquely fixes $\bar\fm$ in terms of $\bar u$. The Jacobian at the saddle point is
\be
J^{\text{3d}}(\fm,u)=\det \mat{ 0 & V''(u) \\ V''(u) & \fm V'''(u) + \Omega''(u) } = - \bigl( V''(u) \bigr)^2 \;.
\ee
Therefore, in the saddle-point approximation:
\be
\cI^\text{saddle}_{S^2 \times S^1} \,\simeq\, \frac1{\lvert \sW \rvert} \sum_{\bar u \in \text{saddles}} \frac{e^{\Omega(\bar u)}}{ \sqrt{J^\text{3d}} } = \frac1{\lvert \sW \rvert} \sum_{\bar u \in \text{BAEs}} \frac{e^{\Omega(\bar u)}}{ i^N \, V''(\bar u) } \;.
\ee
This method gives exactly the same answer as the previous method.

\item A more rough approximation is to fix $\fm$ in (\ref{index original expression}) to the value determined by the equations (\ref{basic system}), 
\be
\label{reduced integral no sum}
\cI^\text{fix $\bar\fm$}_{S^2 \times S^1} \,\simeq\, \cI_{S^1} \,\equiv\, \frac1{\lvert \sW \rvert} \, \oint_\cC \, \prod_{i=1}^N \frac{du^i}{2\pi} \; e^{\bar \fm V'(u) \,+\, \Omega(u)} \;,
\ee
and then solve the integral in $u$ in the saddle-point approximation. The saddle-point equations are $\bar \fm V''(u) + \Omega'(u) = 0$, therefore all solutions $\bar u$ of (\ref{basic system}) are also saddle points of (\ref{reduced integral no sum}). Assuming that there are no other solutions, we find
\be
\cI_{S^1} \,\simeq\, \frac1{\lvert \sW \rvert} \sum_{\bar u \in \text{BAEs}} \frac{e^{\Omega(\bar u)}}{\sqrt{J^{\text{1d}}}} \;.
\ee
The Jacobian in this case is $J^{\text{1d}}=\bar \fm V'''(\bar u) + \Omega''(\bar u) = V'' \bigl( \frac{\Omega'}{V''} \bigr)'(\bar u)$ and is different from before, however as long as the Jacobian is subleading with respect to the exponential contribution, this approach captures the leading behavior.
\end{enumerate}
In our setup we will find a series of saddle points $(\bar u, \bar\fm)$, and the expression $\cI_{S^1}$ in (\ref{reduced integral no sum}) evaluated on the dominant one will turn out to be the Witten index of an effective quantum mechanics that we will construct. In order to do so, we will first have to find the saddle-point flux $\bar\fm$, and then reduce the 3d theory on $S^2$ in the presence of such a flux.

\subsection{The model}
\label{sec: the model}

We consider the AdS/CFT pair discovered in \cite{Guarino:2015jca}, that was used in \cite{Benini:2017oxt, Hosseini:2017fjo, Liu:2018bac} to study certain magnetic black holes in massive type IIA on AdS$_4 \times S^6$ \cite{Cacciatori:2009iz, Guarino:2017eag, Guarino:2017pkw}. The field theory is a 3d $\cN=2$ Chern-Simons-matter theory with gauge group $\rSU(N)_k$, coupled to three chiral multiplets $\Phi_{a=1,2,3}$ in the adjoint representation. We can simplify the computation by considering a $\rU(N)_k$ gauge theory, with no sources for the new topological symmetry. No field is charged under $\rU(1)\subset \rU(N)$ and thus the only effect of this is to introduce a decoupled sector, whose Hilbert space on a Riemann surface $\Sigma_\fg$ consists of $k^\fg$ states, which is a single one in the case of $S^2$. The theory has a superpotential
\be
\label{3d superpot}
W = \lambda_\text{3d} \Tr \Phi_1 \, [\Phi_2, \Phi_3] \;.
\ee
The global symmetry is $\rSU(3)\times \rU(1)_R$. We parameterize its Cartan subalgebra with three R-charges $R_a$, characterized by the charge assignment $R_a(\Phi_b)\equiv(R_a)_b= 2\delta_{ab}$. We choose the Cartan generators of the flavor symmetry to be $q_{1,2}=(R_{1,2}-R_3)/2$. In this basis, all fields have integer charges. Notice that $e^{i\pi R_a} = (-1)^F$ for $a=1,2,3$.

To study AdS$_4$ BPS dyonic black holes, we place the theory on%
\footnote{One could also study the theory on a Riemann surface $\Sigma_\fg$ \cite{Benini:2016hjo, Closset:2016arn}, but here we will focus on the sphere.}
$S^2\times\bR$ using a topological twist on $S^2$, so that one complex supercharge is preserved \cite{Maldacena:2000mw}. This is precisely the background of the topologically twisted index in \cite{Benini:2015noa}. In other words, there is a background gauge field $A_R$ corresponding to an R-symmetry that is equal and opposite to the spin connection when acting on the top component of the supersymmetry parameter $\epsilon$:
\be
\label{R-symmetry flux}
\frac{1}{2\pi}\int_{S^2} dA_R = -1\,.
\ee
The R-symmetry used for the twist must have integer charge assignments, and a generic such R-charge can be written as $q_R=R_3-\fn_1 q_1-\fn_2 q_2$ for $\fn_{1,2}\in\bZ$. Note that $\sum_a (q_R)_a=2$ and the superpotential correctly has R-charge 2. Under these inequivalent twists, the scalar component of $\Phi_a$ experiences a flux $\fn_a = (q_R)_a \int_{S^2} \frac{dA_R}{2\pi} = - (R_3)_a + \fn_1 (q_1)_a + \fn_2 (q_2)_a$.
This formula provides a definition of $\fn_3\equiv -2-\fn_1-\fn_2$. Thus, twisting by a generic R-symmetry with integer charge assignments is the same as twisting with respect to $R_3$ and simultaneously turning on background gauge fields $A_{1,2}$ coupled to the flavor charges $q_{1,2}$ with
\be
\label{flavor flux}
\frac{1}{2\pi}\int_{S^2} dA_{1,2} = \fn_{1,2} \;.
\ee

The theory that we are considering has a UV Lagrangian consisting of various building blocks which are individually supersymmetric off-shell. The vector multiplet $V$ (in Wess-Zumino gauge) contains the adjoint-valued fields $(\sigma,\lambda,\wb\lambda,A_\mu,D)$, where $\sigma$ is a dynamical real scalar field and $D$ a real auxiliary field. We consider a supersymmetrized Chern-Simons Lagrangian for it, but we also add the super-Yang-Mills Lagrangian as a regulator. The chiral multiplets $\Phi_a$ contain the adjoint-valued fields $(\Phi_a,\Psi_a,F_a)$, for which we consider the kinetic Lagrangian and the superpotential term. These Lagrangians, in Lorentzian signature and Wess-Zumino gauge, are:
\begin{align}
\label{initial 3d Lagrangians}
\cL_\text{YM} &= \frac{1}{2e_\text{3d}^2} \Tr \biggl[ -\frac{1}{2}F_{\mu\nu}F^{\mu\nu} - D_\mu\sigma D^\mu\sigma + D^2-i \overline\lambda \bigl( \Dslash - \sigma \bigr) \lambda \biggr] \;, \\
\cL_\text{CS} &= \frac{k}{4\pi} \Tr \biggl[ - \epsilon^{\mu\nu\rho} \Bigl( A_\mu\partial_\nu A_\rho-\frac{2i}{3}A_\mu A_\nu A_\rho \Bigr) -i \overline\lambda \lambda - 2D \sigma \biggr] \;, \nn \\
\cL_\text{chiral} &= - D_\mu\Phi^\dagger_a D^\mu\Phi_a - \Phi^\dagger_a \bigl( \sigma^2 + D \bigr) \Phi_a + F^\dagger_a F_a
- i \overline\Psi_a \bigl( \Dslash + \sigma \bigr) \Psi_a + i \overline\Psi_a \lambda \Phi_a + i \Phi^\dagger_a \overline\lambda \Psi_a \;, \nn \\
\cL_\text{W} &= \frac{\partial W}{\partial\Phi_a}F_a + \frac{1}{2} \, \frac{\partial^2 W}{\partial \Phi_a\partial \Phi_b} \, \overline{\Psi_b^c} \, \Psi_a + \text{c.c.}\;, \nn
\end{align}
where we used the convention $\Psi^c\equiv i \sigma_1 \Psi^*$ for the conjugated spinor. The superpotential must be a gauge-invariant holomorphic function of R-charge $2$. The supersymmetry variations preserved by these Lagrangians are in Appendix~\ref{app: 3d SUSY variations}.

In order to obtain a microscopic description of the black hole entropy, one counts the ground states of this theory. It is convenient to work in the grand canonical ensemble, in which one introduces a set of chemical potentials $\Delta_a$, $a=1,2$ for each flavor Cartan generator. As for the fluxes, it is useful to introduce a third chemical potential $\Delta_3$ such that
\be
\Delta_1+\Delta_2+\Delta_3 \,\in\, 2\pi\mathbb{Z} \;,
\ee
where all chemical potentials are only defined modulo $2\pi$. This constraint \cite{Benini:2016rke} is required in order for $q_a\Delta_a$ to commute with the supersymmetry generators. Computing the thermal partition function is hard because the theory is strongly coupled in the IR, therefore one can start from a quantity protected by supersymmetry: the topologically twisted index
\be
\label{TTI}
\cI_\text{3d}(\fn,\Delta) = \Tr\, (-1)^F \, e^{-\beta H} \, e^{iq_a\Delta_a} \;,
\ee
where $F$ is the Fermion number, $H$ the Hamiltonian on the sphere $S^2$ in the presence of the magnetic fluxes (\ref{R-symmetry flux})-(\ref{flavor flux}), and the trace is over the Hilbert space of states. This quantity only gets contributions from the ground states of the theory. It was argued in \cite{Benini:2015eyy}, exploiting the $\fsu{(1,1|1)}$ superconformal symmetry algebra expected to emerge from the $\text{AdS}_2\times S^2$ near-horizon region in gravity, that the BPS states of a pure single-center black hole have constant statistics $(-1)^F$ in each charge sector, meaning that the index gets non-interfering contributions (at least at leading order in $N$) and can account for the black hole entropy.%
\footnote{This expectation was confirmed for rotating black holes in AdS$_5$ in \cite{Boruch:2022tno}.}

The TT index (\ref{TTI}) can be computed exactly using supersymmetric localization techniques \cite{Benini:2015noa, Closset:2016arn}, and for the model considered here one obtains \cite{Benini:2017oxt, Hosseini:2017fjo}:
\begin{multline}
\label{index full expression from papers}
\cI_\text{3d}(\fn,\Delta) = \frac{(-1)^N}{N!} \prod_{a=1}^3 \frac{y_a^{N^2(\fn_a+1)/2}}{(1-y_a)^{N(\fn_a+1)}} \, \sum_{\fm\in\Gamma_{\fh}} \, \oint_\text{JK} \, \prod_{i=1}^N \frac{dz_i}{2\pi iz_i} \, z_i^{k\fm_i} \times {} \\
{} \times \prod_{i\neq j}^N \biggl( 1-\frac{z_i}{z_j} \biggr) \prod_{a=1}^3 \prod_{i\neq j}^N \biggl( \frac{z_i-y_az_j}{z_j-y_az_i} \biggr)^{\fm_i} \biggl( 1-y_a\frac{z_i}{z_j} \biggr)^{-\fn_a-1} \;.
\end{multline}
Here $z_i\equiv e^{iu_i}$ and $y_a\equiv e^{i\Delta_a}$. This expression can be conveniently compiled into the same form as \eqref{index original expression}:
\be
\label{index detailed expression}
\cI_\text{3d}(\fn,\Delta)= \frac{1}{N!} \, \sum_{\fm\in\Gamma_{\fh}} \, \oint_{\text{JK}} \Biggl(\, \prod_{i=1}^{N}\frac{du_i}{2\pi} \Biggr) \; e^{\sum_i \fm_i V_i'(u,\Delta) \,+\, \Omega(u,\fn,\Delta)} \;.
\ee
The two functions appearing in the exponent are
\begin{align}
\label{V prime with m and Omega}
\sum_{i=1}^N\fm_i V'_i(u,\Delta)
&= \sum_{i=1}^N \fm_i \Biggl\{ ik u_i + \sum_{j=1}^N \sum_{a=1}^3 \Bigl[ \Li_1\Bigl( e^{i ( u_{ji} - \Delta_a )} \Bigr) - \Li_1 \Bigl( e^{i ( u_{ji}+\Delta_a )} \Bigr) \Bigr] + i \pi \bigl( N - 2n_i \bigr) \Biggr\} \;, \nn \\
\Omega(u,\fn,\Delta) &= \sum_{a=1}^3(1 +\fn_a)\sum_{i,j}^N \Li_1 \Bigl(e^{i(u_{ij}+\Delta_a)} \Bigr)- \sum_{i\neq j}^N\Li_1\bigl(e^{iu_{ij}}\bigr) \\
&\quad + i\frac{N^2}{2}\sum_{a=1}^3(1+\fn_a)\Delta_a+ \pi i(2M+N) \;,  \nn
\end{align}
where $u_{ji}=u_j - u_i$ whilst $n_i$ and $M$ are integer ambiguities. The JK integration contour is the so-called Jeffrey-Kirwan residue \cite{JeffreyKirwan}. We used the polylogarithm function
\be
\Li_1(z) = - \log(1-z) \;,
\ee
while more properties are in Appendix~\ref{subsec: polylogs}.

\subsection[The large \texorpdfstring{$N$}{N} limit]{The large \matht{N} limit}

To obtain the saddle-point equations, we first formulate \eqref{index detailed expression} in a large $N$ continuum description as in \cite{Herzog:2010hf}, and subsequently take functional derivatives. The Weyl symmetry permuting the discrete Cartan subalgebra index $i$ can be used to order the holonomies $u_i$ such that $\im u_i$ increases with $i$. The discrete index $i$ is then substituted with a continuous variable $t\in[t_-,t_+]$, after which $u$ and the flux $\fm$ become functions of $t$. The reparameterization symmetry in $t$ is fixed by identifying, up to normalization, $t$ with $\im u(t)$:
\be
\label{holonomy ansatz}
u(t) = N^\alpha \, \bigl( it+v(t) \bigr) \;.
\ee
This introduces the density
\be
\rho(t)\equiv \frac{1}{N} \, \frac{di}{dt} \;,
\ee
in terms of which any sum will be replaced by an integral: $\sum_i \to N\int\! dt \,\rho(t)$. The density $\rho$ must be real, positive, and integrate to $1$ in the defining range. The $N^\alpha$ scaling is introduced in such a way that $u(t)$ is an $N$-independent continuous function. This ansatz is also motivated by the fact that dual black holes have an entropy scaling with a power law in $N$.

We perform the large $N$ computation in Appendix~\ref{app: large N}. In (\ref{final mV' app}) and (\ref{final Omega app}) we find:
\bea
\label{mV' and Omega large N}
\int\! dt\, \fm \, V' &= ikN \!\int\! dt\, \rho \, \fm \, u + iN^{2-2\alpha} \, G(\Delta) \!\int\! dt\, \frac{\dot\fm \, \rho^2}{(1-i\dot{v})^2} + \cO\bigl( \fm N^{2-3\alpha} \bigr) \;, \\
\Omega &= - N^{2-\alpha} \, f_+(\fn,\Delta) \int\! dt \, \frac{\rho^2}{ 1-i\dot{v} } + \cO\bigl( N^{2-2\alpha} \bigr) \;,
\eea
where a dot means $\frac{d}{dt}$ and we introduced the functions
\be
\label{def G and f+}
G(\Delta)= \sum_{a=1}^3g_+(\Delta_a) \;,\qquad\qquad
f_+(\fn,\Delta) = -\sum_{a=1}^3 (1+\fn_a) \Bigl( g'_+(\Delta_a) - g'_+(0) \Bigr) \;,
\ee
and
\be
\label{def g+}
g_+(\Delta) = \frac{1}{6}\Delta^3 - \frac{\pi}{2}\Delta^2 + \frac{\pi^2}{3} \Delta \;.
\ee
The entire exponent in the integrand of \eqref{index detailed expression} is the functional:
\begin{multline}
\label{extremization functional}
\cV = i k N^{1+\alpha} \int\! dt\, \rho \, \fm \, (it+v) +iN^{2-2\alpha} \, G(\Delta) \int\! dt\, \frac{\dot\fm \, \rho^2}{(1-i\dot{v})^2} + {} \\
{} - N^{2-\alpha} \, f_+(\fn,\Delta) \int\! dt\, \frac{\rho^2}{1-i\dot{v}} + N^{2-\alpha} \, \mu \biggl( \int\! dt\, \rho-1 \biggr) \;,
\end{multline}
where we added a Lagrange multiplier $\mu$ to enforce the normalization of $\rho$. In order for the terms in $\cV$ to compete and give us a (non-trivial) saddle-point, we need to set $\alpha=\frac{1}{3}$ and $\fm(t)=N^\frac{1}{3}\wh{\fm}(t)$, where $\wh{\fm}(t)$ is an $N$-independent function.

To find the saddle-point configurations at large $N$, we extremize $\cV$ with respect to $\rho$, $v$, $\wh\fm$ and $\mu$. After some massaging, the saddle-point equations are:
\begin{align}
\label{extremization 1}
0 &= \frac{d}{dt}\biggl[ 2G \, \frac{\wh\fm \, \rho}{1-i\dot v} - \mu \, (it+v) \biggr] + 2i f_+ \, \rho \;, \\
\label{extremization 2}
0 &= \rho \, \wh\fm - \frac{2iG}k \, \frac{d}{dt}\biggl[ \frac{\dot{\wh\fm} \, \rho^2}{(1-i\dot v)^3}\biggr] + \frac{f_+}{G} \, \rho \, (it+v) \;, \\
\label{extremization 3}
0 &= \frac{d}{dt}\biggl[ k \, (it+v)^2 - 4iG \, \frac{\rho}{1-i \dot v} \biggr] \;,
\end{align}
together with $\int\! dt\, \rho = 1$. One can check that the functional $\cV$ is invariant under reparametri\-zations of $t$ that preserve the scaling ansatz \eqref{holonomy ansatz} for the holonomies. Such reparametrizations act as:
\bea
\label{reparametrization}
t &= t(t') \;,& v(t) &= i \bigl[t' - t(t') \bigr] + v'(t') \;, \\
\rho(t) &= \biggl( \frac{dt(t')}{dt'} \biggr)^{-1} \! \rho'(t') \;, \qquad& \wh\fm(t) &= \wh\fm'(t') \;.
\eea
Notice in particular that $v'$ becomes complex after the transformation.

As we review in Appendix~\ref{sec: sol saddle point}, the equations (\ref{extremization 1})--(\ref{extremization 3}) can be solved, yielding:
\be
\label{u m and rho final}
u(t) = \biggl( \frac{3N G}{k} \biggr)^\frac{1}{3} t \;,\quad \fm(t) = \biggl( \frac{N }{9kG^2} \biggr)^\frac{1}{3} \! f_+ \, t \;,\quad \rho(t) = \frac{3}{4} \bigl( 1-t^2 \bigr) \;,\quad t \in [-1,1] \;.
\ee
This solution is obtained after making use of the reparametrization symmetry, so in particular $v(t)$ is complex. The value of the functional $\cV$ at the saddle point for $\rho$, $v$ and $\fm$ --- which reproduces the logarithm of the index at leading order --- is
\be
\label{cV final large N}
\cV = - \frac{i N^\frac53}5 \biggl( \frac{9k}{G(\Delta)} \biggr)^{\frac13} f_+(\fn, \Delta) \;.
\ee
If $\sum_a\Delta_a=2\pi$, the two functions $G$ and $f_+$ take the particularly simple form
\be
\label{case I G and f}
G(\Delta) = \frac{1}{2} \, \Delta_1\Delta_2\Delta_3 \;,\qquad\qquad
f_+(\fn,\Delta) = -\frac{1}{2} \, \Delta_1\Delta_2\Delta_3\sum_{a=1}^3\frac{\fn_a}{\Delta_a} \;.
\ee
In this case, the saddle-point value of the logarithm of the index is
\be
\cV = \frac{i N^{\frac53}}5 \, \biggl( \frac{9k}{4} \biggr)^\frac13 \bigl( \Delta_1\Delta_2\Delta_3 \bigr)^\frac23 \, \sum_{a=1}^3 \frac{\fn_a}{\Delta_a} \;.
\ee
When the $\Delta_a$'s are real this expression matches the result of \cite{Benini:2017oxt, Hosseini:2017fjo},%
\footnote{In principle, it is not obvious whether the saddle point (\ref{u m and rho final}) contributes to the integral (\ref{index detailed expression}) along the JK contour. This is however confirmed by the fact that the result matches the one in \cite{Benini:2017oxt, Hosseini:2017fjo}, where the integral was computed as a careful sum of those residues inside the contour.}
which reproduces the black hole entropy upon performing a Legendre transform.

As mentioned above, the chemical potentials $\Delta_a$ are defined modulo $2\pi$. The expression for $\cV$ in (\ref{cV final large N}), however, is not periodic under $\Delta_a \to \Delta_a + 2\pi$. This means that we have actually found an infinite number of saddle points, parametrized by the shifts.%
\footnote{In general, only a subset of the complex saddle points contribute to the contour integral: which ones do (depending on the contour) should be determined with steepest descent.}
This suggests that --- as in AdS$_3$ \cite{Dijkgraaf:2000fq} and AdS$_5$ \cite{Aharony:2021zkr} --- there might be an infinite number of complex BPS black-hole-like supergravity solutions dual to the semiclassical expansion of the TT index. This issue deserves more study. In the following we will assume that we have identified the dominant saddle point, and we will work with it.

\section{KK reduction on a flux background}
\label{sec: KK reduction}

The next step is to perform a Kaluza-Klein (KK) reduction of the 3d $\cN=2$ gauge theory on the sphere $S^2$, in the presence of the flux background $\fm$ (\ref{u m and rho final}) determined as the saddle point of the TT index. By keeping only the light modes, we will obtain a 1d quantum mechanical model which we expect to contain information about the horizon degrees of freedom of the dyonic AdS$_4$ black holes we are interested in. This section is rather technical, and the reader only interested in the final result can directly jump to Section~\ref{sec: QM}.

Here we will first show how the full twisted theory can be seen as a gauged $\cN=2$ quantum mechanics. Afterwards, we will introduce the background of the reduction and review the standard procedure to fix the 3d gauge group down to the 1d gauge group. We will then explain why complications arise when computing the KK spectrum of the vector multiplet, and how they can be resolved by a further modification of the gauge-fixing Lagrangian.
Lastly, we will exhibit the KK spectra of the vector and chiral multiplets.

\subsection{Decomposing 3d multiplets into 1d multiplets}
\label{sec: reduction of multiplets}

After the topological twist, the theory exactly fits into the framework of a gauged $\cN=2$ quantum mechanics, and we perform various changes of variables in this section to make it explicit. A similar discussion can be found in \cite{Bullimore:2019qnt}. We give a brief review of 1d \mbox{$\cN=2$} supersymmetry in Appendix~\ref{app: 1d susy}, adapted from \cite{Hori:2014tda}, but in \ref{app: susy lagrangians} and \ref{app: twisted ym and cs} we also present new supersymmetric Lagrangians peculiar to our 3d theory.

We shall write the supersymmetry transformations in terms of anticommuting generators $Q$ and $\overline{Q}$, with the understanding that generators should be multiplied by a complex anti-commuting parameter to produce a generic supersymmetry transformation. With $\epsilon=(1,0)^\sT$, $Q$ is obtained from $\wt Q_\text{3d}$ while $\overline{Q}$ is obtained from $Q_\text{3d}$ in \eqref{3d susy chiral m} and \eqref{3d susy vector m}. Note that $Q$ and $\overline{Q}$ are related by Hermitian conjugation, that is $\overline{(QX)}= (-1)^F \, \overline{Q} \, \overline{X}\,$. The supersymmetry algebra is 
\be
\label{3d twisted susy algebra}
Q^2 = \overline{Q}^2 = 0 \;,\qquad\qquad \{Q, \overline{Q}\} = i \bigl[ \partial_t - \delta_\text{gauge}(A_t+\sigma) \bigr] \;,
\ee
where $\delta_\text{gauge}(\alpha)$ is a gauge transformation with parameter $\alpha$.
We will use frame fields $e^1_\mu$, $e^{\bar1}_\mu$ on $S^2$, which we introduce in Appendix~\ref{app: harmonics}, and write differential forms on $S^2$ with flat indices $1,\bar1$. From now on, $\overline{X}$ will denote the Hermitian conjugate of $X$ (since Dirac conjugates are no longer present anyway). After this rewriting, the supersymmetry variations and supersymmetric Lagrangians are as described below.

\paragraph{Vector multiplet.}
In Wess-Zumino gauge, the 3d vector multiplet consists of the gauge field $A_\mu$, a real scalar $\sigma$, a real auxiliary scalar $D$, and a Dirac spinor $\lambda$. The bosonic components are R-neutral while $\lambda$ has R-charge $-1$. We decompose $\lambda$ in components as
\be
\lambda = \biggl( \begin{matrix} -\overline{\Lambda}_t \\ \Lambda_{\bar1} \end{matrix} \biggr) \;,
\ee
and redefine $D$ with a shift
\be\label{D redef}
D'=D-2iF_{1\bar 1} \;.
\ee
Now, $\Lambda_{\bar1}$ has R-charge $-1$ whereas $\Lambda_t$ has R-charge $+1$. These field redefinitions have trivial Jacobian. Under the supercharges preserved by the twist, the supersymmetry variations of the vector multiplet split into 2 sets of variations. The first set (Hermitian conjugate relations being implied) is:
\bea
\label{susy vector m twist D redef}
Q A_t &= - Q\sigma = -\frac{i}{2} \, \overline\Lambda_t \;,\qquad& Q \Lambda_t &= - D_t\sigma-iD \;, \\  
QD &= -\frac{1}{2}\, (D_t-i\sigma) \, \overline\Lambda_t \;,\qquad& \overline{Q} \Lambda_t &= 0 \;.
\eea
These coincide with the supersymmetry variations \eqref{1d vect susy vars} of a 1d $\rU(N)$ vector multiplet in Wess-Zumino gauge. Note that here the fields and gauge transformations are also functions on $S^2$. The second set is:
\be
\label{1d chiral transf in 3d chiral}
Q A_{\bar1} = \frac{1}{2} \Lambda_{\bar 1} \;,\qquad \overline{Q} A_{\bar1} = 0 \;,\qquad Q\Lambda_{\bar1} = 0 \;,\qquad \overline{Q} \Lambda_{\bar1}= 2i \bigl( \partial_t A_{\bar1} - D_{\bar1} (A_t+\sigma) \bigr) \;.
\ee
These coincide with the supersymmetry variations \eqref{1d chiral susy vars WZ} of a chiral multiplet $\bigl(A_{\bar1}, \frac{1}{2}\Lambda_{\bar1} \bigr)$ in Wess-Zumino gauge, provided that the corresponding superfields
\be
\Xi_{\bar1,h} = A_{\bar1} + \frac{\theta}{2} \Lambda_{\bar1} - \frac{i}{2} \, \theta\bar\theta \, \partial_t A_{\bar1} \;,\qquad \qquad
\Xi_{1,\bar h} \,\equiv\, \overline{\Xi_{\bar1,h}} = A_1 - \frac{\bar\theta}{2} \, \overline{\Lambda}_{1} + \frac{i}{2} \, \theta\bar\theta \, \partial_t A_1
\ee
satisfying $\overline{D} \, \Xi_{\bar1,h} = D \, \Xi_{1,\bar h}=0$, transform as connections under super-gauge transformations:
\be
\label{gauge chiral super gt h}
\Xi_{\bar1,h} \,\rightarrow\, h \, \bigl( \Xi_{\bar1,h} + i\partial_{\bar1} \bigr) \, h^{-1} \;,\qquad \Xi_{1,\bar h} \,\rightarrow\, \overline{h}^{\,-1}\bigl( \Xi_{1,\bar h} + i\partial_1 \bigr) \, \overline{h} \;,
\ee
with $h = e^\chi$ and $\overline{D} \chi = 0$. We indicated as $\wb\Lambda_1$ the complex conjugate to $\Lambda_{\bar1}$.

The Yang-Mills Lagrangian is composed of two pieces, independently supersymmetric:
\begin{align}
2e_\text{3d}^2 \, \cL_\text{YM}
&= \;\, \Tr \biggl[ 4 \bigl\lvert F_{t\bar1} \bigr\rvert^2 + 4iDF_{1\bar1} - 4 \bigl\lvert D_{\bar1} \sigma \bigr\rvert^2 + i \overline\Lambda_1(D_t + i\sigma) \Lambda_{\bar1} + 2 \Lambda_t D_1 \Lambda_{\bar1} - 2 \overline\Lambda_1 D_{\bar1} \overline\Lambda_t \biggr] \nn \\
&\;\; + \Tr \biggl[ (D_t\sigma)^2 + D^2 + i\overline\Lambda_t (D_t-i\sigma) \Lambda_t \biggr] \;.
\label{3d YM in 1d susy parts}
\end{align}
Note that $2e_\text{3d}^2 \, \cL_\text{YM} = Q \overline{Q} \Tr \bigl[ -4iA_1\partial_tA_{\bar1} + 4i (A_t-\sigma) F_{1\bar1} \bigr] + Q \overline{Q} \Tr \bigl[ -\overline{\Lambda}_t \Lambda_t \bigr]$, so both terms are exact. The first piece is the appropriate kinetic term for a chiral transforming as a connection and its superspace expression is in \eqref{YM chiral part k transf}. The second piece is the standard 1d gauge kinetic term (\ref{1d vec kin term}). Likewise, the Chern-Simons Lagrangian splits into two pieces which are separately supersymmetric:
\be
\label{3d CS in 1d susy parts}
\frac{4\pi}{k} \, \cL_\text{CS}
= \Tr \Bigl[ 4iA_1\partial_tA_{\bar1} - 4i(A_t+\sigma)F_{1\bar1} + \overline{\Lambda}_1 \, \Lambda_{\bar1} \Bigr] \;+\; \Tr \Bigl[ \overline{\Lambda}_t \Lambda_t - 2 D \sigma \Bigr] \;.
\ee
The superspace expression of the first piece is given in \eqref{CS superspace}, whereas the second piece matches \eqref{1d vect mass term}.

\paragraph{Chiral multiplet.}
A 3d chiral multiplet consists of a complex scalar $\phi$ and a Dirac spinor $\Psi$. We split $\Psi$ into components as
\be
\Psi = -i \biggl( \begin{matrix} \psi \\ \eta \end{matrix} \biggr) \;.
\ee
Their R-charges are $R(\psi)=R(\eta)=R(\phi)-1$. Under the supercharges preserved by the twist, the supersymmetry variations of the 3d chiral multiplet can also be organized into two sets. The first set (Hermitian conjugate relations are again implicit) is:
\be
\label{SUSY transf of twisted chiral 1}
Q \phi = \psi \;,\qquad \overline{Q} \phi = 0 \;,\qquad Q \psi = 0 \;,\qquad \overline{Q} \psi = i (D_t-i\sigma) \phi \;.
\ee
They coincide with the supersymmetry variations \eqref{1d chiral susy vars WZ} of a 1d chiral multiplet $(\phi,\psi)$ in Wess-Zumino gauge, with corresponding superfield $\Phi_h = \phi + \theta\psi - \frac{i}{2} \, \theta\bar\theta \, \partial_t\phi$.
The second is:
\be
\label{SUSY transf of twisted chiral 2}
Q \eta = - f \;,\quad\; \overline{Q} \eta = - 2D_{\bar1} \phi \;,\quad\; Q f = 0 \;,\quad\; \overline{Q} f = - i (D_t-i\sigma) \eta - 2D_{\bar 1}\psi+i\Lambda_{\bar 1}\phi \;.
\ee
They match the variations \eqref{1d fermi susy vars WZ} of a 1d Fermi multiplet $(\eta,f)$ in Wess-Zumino gauge, whose corresponding superfield
\be
\cY_h = \eta - \theta f + 2 \bar\theta D_{\bar1} \phi + \theta\bar\theta \, \Bigl( -\frac{i}{2} \, \partial_t\eta-2D_{\bar1} \psi + i \Lambda_{\bar1} \phi \Bigr)
\ee
satisfies
\be
\label{3d E term}
\overline{D} \, \cY_h = E\bigl( \Phi_h, \Xi_{\bar1,h} \bigr) = -2 \bigl( \partial_{\bar1} - i \Xi_{\bar1,h} \bigr) \Phi_h \;.
\ee
Here $\partial_{\bar 1}$ contains the background $\rU(1)_R$ connection. In the language of 1d supersymmetry, there is an E-term superpotential for $\cY_h$. After the shift \eqref{D redef}, the kinetic term of a 3d chiral multiplet also splits into two separately supersymmetric pieces, \ie, the kinetic terms of the 1d chiral \eqref{1d chiral kin term} and of the 1d Fermi \eqref{1d Fermi kin term}:
\begin{align}
\label{chiral_lagrangian_twisted}
\cL_\text{chiral}
&= \;\, \Bigl[ \lvert D_t\phi \rvert^2 - \lvert \sigma\phi \rvert^2 - \overline{\phi} D\phi + i\overline\psi (D_t+i\sigma)\psi - i\overline{\psi}\,\overline{\Lambda}_t \phi + i\overline{\phi}\Lambda_t \psi \Bigr] \\
&\;\; + \Bigl[ i\overline\eta (D_t-i\sigma)\eta + \overline f f - |2D_{\bar1} \phi|^2 - 2\overline\psi D_1\eta + 2\overline\eta D_{\bar 1}\psi - i\overline{\eta}\Lambda_{\bar 1}\phi + i\overline{\phi} \,\overline{\Lambda}_1\eta \Bigr] \;. \nn
\end{align}
Note that $\cL_\text{chiral} = Q\overline{Q} \bigl( -i\overline{\phi} (D_t+i\sigma) \phi \bigr) + Q\overline{Q} \bigl( -\overline\eta \eta \bigr)$, so both terms are exact.

The superpotential terms can be written as $\cL_\text{W} = -Q \bigl( \eta_a \frac{\partial W}{\partial\phi_a} \bigr) + \overline{Q}\bigl( \overline\eta_a \frac{\partial \overline{W}}{\partial \overline\phi_a} \bigr)$, which in the language of 1d supersymmetry are J-terms for the Fermi multiplets $\eta_a$ with $J_a=-\frac{\partial W}{\partial\phi_a}$. Supersymmetry of the first term under $Q$, and of the second term under $\overline{Q}$, are obvious. When $\overline{Q}$ acts on the first term we get, up to a total time derivative,
\be
Q \overline{Q} \biggl( \eta_a \, \frac{\partial W}{\partial\phi_a} \biggr) = -2Q \biggl( D_{\bar 1}\phi_a \, \frac{\partial W}{\partial\phi_a} \biggr) = - 2 Q(\partial_{\bar 1}W) = -2\partial_{\bar 1}QW \;,
\ee
which is another total derivative. Thus the superpotential terms are $\bigl( Q+\overline{Q} \bigr)$-exact. The supersymmetric Chern-Simons Lagrangian is the only piece that is not exact under any supercharge.

\subsection{Reduction background}
\label{sec: reduction background}

As mentioned at the beginning of this section, we want to reduce the theory in the presence of background fluxes for the global symmetries. In particular, we turn on a (negative) unit flux for the R-symmetry $q_R$. Since it is a background for a non-dynamical field, it can be off-shell without any consequences. The presence of this background, under which the chiral multiplets are differently charged, generically breaks the $\rSU(3)$ flavor symmetry down to its diagonal subgroup $\rU(1)^2_F$. We also single out a configuration of fluxes for the dynamical gauge fields:
\be
F_{1\bar1} = \frac{i \, \fm}{4R^2} \;,\qquad \text{where $\fm$ is a constant in the Cartan subalgebra.}
\ee
The choice of $\fm$ will eventually be the one dictated by the saddle-point approximation to the topologically twisted index, discussed in Section~\ref{sec: saddle point}.
Since $F_{1\bar1}$ couples to the auxiliary field $D$ in \eqref{3d YM in 1d susy parts} like a FI parameter, the D-term equation for supersymmetric vacua is:
\be
\frac{2i}{e^2_\text{3d}} \, F_{1\bar1} + \sum_a \, [\overline{\phi}_a, \phi_a]-\frac{k}{2\pi} \sigma = 0 \;.
\ee
The background should satisfy the D-term equation in order to be supersymmetric, and it is simplest to turn on a background for $\sigma$ to cancel the background flux. This falls into the class of ``topological" vacua discussed in \cite{Intriligator:2013lca}. Moreover, since $A_t+\sigma$ appears in the algebra \eqref{3d twisted susy algebra}, we also find it appropriate to turn on a background for $A_t$, opposite to that of $\sigma$, so that the background of $A_t+\sigma$ is zero. This ensures that BPS states have zero energy even before projecting onto gauge singlets. Thus, the background we use for the reduction is:
\be
\label{bg vevs}
F_{1\bar1} = \frac{i \fm}{4R^2} \;,\qquad \sigma = -\frac{\fm}{2m_kR^2} \;,\qquad A_t = \frac{\fm}{2m_kR^2} \;,\qquad \text{where} \quad m_k\equiv\frac{k \, e^2_\text{3d} }{2\pi} \;.
\ee
One can check that all the equations of motion are satisfied on this background, except for that of $A_t+\sigma$, unless $\fm=0$. Consequently, when expanding the action, there will be a Lagrangian term linear in $A_t+\sigma$, that is
\be
\label{lin term A_t + sigma}
\Tr \biggl( \frac{k\fm}{4\pi R^2} \, (A_t+\sigma) \biggr) \;.
\ee
In other words, background fluxes produce a background electric charge in the presence of Chern-Simons terms. As we will discuss later, the presence of this linear term is crucial and it is the main source of complications when computing the vector multiplet spectrum.

We parametrize the Lie algebra $\fsu(N)$ by $N\times N$ matrices $E_{ij}$ ($i,j=1,\ldots, N$) which have a single nonzero entry $1$ in row $i$ and column $j$: $(E_{ij})_{kl}=\delta_{ik}\delta_{jl}$. Elements with $i=j$ are a basis for the Cartan subalgebra, while those with $i\neq j$ correspond to roots with root vector $(\alpha_{ij})_k=\delta_{ki}-\delta_{kj}$. The commutation relations in this basis are
\be
[E_{ij},E_{kl}]=\delta_{jk}E_{il}-\delta_{il}E_{kj} \;.
\ee
Note also that $\overline{E_{ij}} = E_{ji}$ and
\be
\Tr E_{ij} E_{kl} = \delta_{jk} \delta_{il} \;,\qquad \Tr E_{ij}[E_{kl}, E_{mn}]=\delta_{jk}\delta_{lm}\delta_{ni}-\delta_{il}\delta_{jm}\delta_{kn} \;.
\ee
We write the expansion of adjoint fields in this basis as $X=X^{ij}E_{ij}$. Note that $\wb X^{\,ij} = \wb{ X^{ji}}$.
The Cartan components will sometimes be written as $X^i\equiv X^{ii}$ for simplicity.

In the presence of global and gauge fluxes, the Lie algebra components of various fields in the vector multiplet and chiral multiplets are $\rU(1)_\text{spin}$ sections with different monopole charges $q$ (see Appendix~\ref{app: harmonics} for details). A field $\chi_q(t,\theta,\varphi)$ with monopole charge $q$ can then be expanded in a complete set of monopole harmonics $Y_{q,l,m}(\theta,\varphi)$, and the time-dependent expansion coefficients $\chi_{q,l,m}(t)$ are the 1d fields after the reduction:
\be
\chi_q(t,\theta,\varphi) = \sum_{l \geq |q|} \; \sum_{|m| \leq l} \; \chi_{q,l,m}(t) \; Y_{q,l,m}(\theta,\varphi) \;.
\ee
Defining the quantities
\be
\label{def monopole q}
q_{ij} \,\equiv\, \frac{\fm_i-\fm_j}{2} \;,\qquad\qquad q_{ij}^a \,\equiv\, \frac{\fm_i-\fm_j+\fn_a}{2} \;,
\ee
the monopole charges of the fields and their charges under the global symmetries of the theory are summarized in Table~\ref{tab: monopole charges}.

\begin{table}[t]
\centering
\begin{tabular}{|c|c|c|c|c|c|c|c|}
    \hline \rule[-0.6em]{0pt}{1.9em}
VM  & $\sigma^{ij},\, A_t^{ij},\, D^{ij}$ & $\Lambda_t^{ij}$ & $A_{\bar 1}^{ij}$ & $A_1^{ij}$ & $\Lambda_{\bar 1}^{ij}$ & $\wb\Lambda_1^{\,ij}$ \\
    \hline\rule[-0.6em]{0pt}{0pt}%
$q$ & $q_{ij}$ & $q_{ij}$& $q_{ij}+1$ & $q_{ij}-1$ & $q_{ij}+1$ & $q_{ij} -1$ \\
    \hline
$q_R$ & $0$ & $1$ & $0$ & $0$ & $-1$ & $1$ \\
    \hline
$q_1$ & $0$ & $0$ & $0$ & $0$ & $0$ & $0$ \\
    \hline
$q_2$ & $0$ & $0$ & $0$ & $0$ & $0$ & $0$ \\
    \hline
\end{tabular} \\[1em]
\begin{tabular}{|c|c|c|c|c|c|}
    \hline \rule[-0.6em]{0pt}{1.7em}
CM  & $\phi_a^{ij}$&$\psi_a^{ij}$ & $\eta_a^{ij}$&$f_a^{ij}$ \\
    \hline\rule[-0.6em]{0pt}{0pt}%
$q$ & $q_{ij}^a$ & $q_{ij}^a$ & $q_{ij}^a+1$ & $q_{ij}^a+1$ \\
    \hline
$q_R$ & $-\fn_a$ & $-\fn_a-1$ & $-\fn_a-1$ & $-\fn_a-2$ \\
    \hline
$q_1$ & $\delta_{1a}-\delta_{3a}$ & $\delta_{1a}-\delta_{3a}$ & $\delta_{1a}-\delta_{3a}$ & $\delta_{1a}-\delta_{3a}$ \\
    \hline
$q_2$ & $\delta_{2a}-\delta_{3a}$ & $\delta_{2a}-\delta_{3a}$ & $\delta_{2a}-\delta_{3a}$ & $\delta_{2a}-\delta_{3a}$ \\
    \hline
\end{tabular}
\caption{Monopole and global charges of all fields. The R-charge is $q_R$, while $q_{1,2}$ are flavor charges. Above: modes from 3d vector multiplets. The modes are defined for pairs $i,j$ such that $q_{ij}>0$. Below: modes from 3d chiral multiplets, defined for any $ij$. In both cases, the modes are in $\rSU(2)$ representations with $l \geq |q|$ and $l=q$ mod 1. 
\label{tab: monopole charges}}
\end{table}

We assume that $\fm_i\neq\fm_j$, $\forall \; i\neq j$, since this is true for the saddle-point flux, and thus $q_{ij} \neq 0$ for $i\neq j$. Given a Hermitian adjoint field $X = X^{ij} E_{ij} = \wb X$ in a vector multiplet (\ie, $A_t$, $\sigma$, $D$), its components satisfy $X^{ji} = \overline{X^{ij}}$. We parameterize the off-diagonal components in terms of complex fields $X^{ij}$ with $ij$ such that $q_{ij}>0$. For complex adjoint fields $Y = Y^{ij} E_{ij}$ in vector multiplets (\ie, $A_{\bar1}$, $A_1$, $\Lambda_{\bar1}$, $\wb\Lambda_1$), we initially parameterize the off-diagonal components in terms of complex fields $Y^{ij}$, $\wb{Y}^{\,ij}$ with $ij$ such that $q_{ij}>0$. For complex adjoint fields in chiral multiplets, instead, we simply use all components $Y^{ij}$.

The flux breaks the gauge group $\rU(N)$ to its maximal torus $\rU(1)^N$, and the 1d gauge group will consequently be $\rU(1)^N$. Indeed, the generators of 1d gauge transformations have to be constant on $S^2$, however the components $\epsilon^{ij}$ of the gauge-transformation parameter have monopole charges $q_{ij}$, and since $l\geq |q_{ij}|$, only those in the Cartan subalgebra have an $l=0$ mode which is constant on $S^2$.

\subsection{Partial gauge fixing}

In order to reduce to a gauged quantum mechanics, we need to fix the 3d gauge group to the unbroken 1d gauge group, consisting of time-dependent transformations that are constant on $S^2$. A systematic procedure to achieve that is presented in Appendix~\ref{app: partial gauge fixing} and we refer the reader to \cite{Ferrari:2013aza} for more details. We choose the Coulomb gauge with gauge-fixing function
\be\label{Coulomb gf function}
G_\text{gf} = \frac{2}{\sqrt\xi} \, \Bigl( D^B_1A_{\bar1} + D^B_{\bar1} A_1\Bigr) \;.
\ee
One can check that it leaves the 1d gauge group unfixed. The covariant derivatives above only contain the spin connection and monopole background. In general, for any $G_\text{gf}$, the gauge-fixing procedure adds the following terms to the Lagrangian:
\be
\label{FP gf}
\frac{1}{e^2_\text{3d}} \Tr \biggl[ \frac{b^2}{2} + b \, \Bigl( G_\text{gf} - \{\wt c,c\} \Bigr) + i \, \wt c \, \delta_\text{gauge}(c) \, G_\text{gf} + \frac12 \, \{\wt c,c\}^2 \biggr] \;.
\ee
Here $c$ and $\wt c$ are independent Grassmann scalars, while $b$ is a bosonic auxiliary field. Importantly, all of them are valued in the part of the gauge algebra that is broken by $G_\text{gf}$, and do not contain modes in the residual gauge algebra. In the following, a subscript $\fr$ will indicate a restriction to the residual gauge algebra, and a subscript $\ff$ a restriction to the complement containing fixed (or broken) gauge generators.%
\footnote{In the Coulomb gauge \eqref{Coulomb gf function}, $\fr$ contains diagonal transformations with $l=0$, while $\ff$ contains diagonal transformations with $l>0$ as well as all off-diagonal transformations.}
We define a BRST supercharge $s$ as:
\be
\label{brst restricted cartan}
sX = \delta_\text{gauge}(c) \, X \;,\quad sc=\frac{i}{2}\{c,c\}_{\ff} \;,\quad s\wt c = ib \;,\quad sb = \delta_\text{gauge}(R) \, \widetilde{c} \;,\quad R \equiv -\frac{1}{2} \{c,c\}_\fr \;.
\ee
One can check that
\be
s^2 = i \, \delta_\text{gauge}(R) \;,\qquad\qquad sR = 0 \;.
\ee
This allows us to define an $s$-cohomology on invariants of the residual gauge group. The terms produced by gauge fixing can then be written in a BRST-exact form:
\be
\label{def Psi_gf}
\text{(\ref{FP gf})} = \frac{1}{e^2_\text{3d}} \, s \Tr \wt c \, \biggl( -i \, G_\text{gf} - \frac{i}2 \, b + \frac{i}2 \, \{\wt c,c\} \biggr) \,\equiv\, s\Psi_\text{gf} \;.
\ee
We defined $\Psi_\text{gf}$ as the function in parentheses.
We note that there is still complete freedom in specifying the inner product in the ghost sector, \ie, the Hermiticity properties of $c$ and $\wt c$. In order for the theory to be unitary and have a consistent Hamiltonian formulation \cite{Kugo:1977yx}, one needs that $c$ and $\wt c$ are Hermitian, so that $s$ is a real supercharge and \eqref{FP gf} is real. With this choice, \eqref{FP gf} is invariant under a ghost-number symmetry valued in $\bR^*$, which acts as:
\be
c \,\mapsto\, e^{\alpha} \, c \;,\qquad \wt c \,\mapsto\, e^{-\alpha} \, \wt c \;,\qquad s \,\mapsto\, e^\alpha \, s \;,
\ee
with $\alpha \in \bR$. We say that $c$ has ghost number $n_g=1$ and $\wt c$ has $n_g=-1$. Physical observables are identified with the $s$-cohomology at $n_g=0$, since external states must be gauge invariant and cannot contain ghosts. Since $c$, $\wt c$, and $b$ are Hermitian, they are neutral under $\rU(1)_R$, and \eqref{FP gf} is invariant under $\rU(1)_R$, since $G_{\text{gf}}$ is R-neutral.

\subsection{Supersymmetrized gauge fixing}
\label{sec: SUSY gauge fixing}

As anticipated, the linear term \eqref{lin term A_t + sigma} causes complications in the computation of the KK spectrum of the vector multiplet, and the following discussion aims to explain why. The standard Faddeev-Popov gauge-fixing procedure we just reviewed generically breaks the supersymmetries that were defined on the original action because of the presence of the BRST-exact term $s\Psi_\text{gf}$, which might not be supersymmetric.
Considering a supercharge $Q$, and assuming that it does not act on the fields in the gauge-fixing complex, the transformation of $s \Psi_\text{gf}$ is $-s Q \Psi_\text{gf}$. When computing $s$-closed (\ie, gauge-invariant) quantities, this is harmless because the potentially violating term is $s$-exact, and it does not affect the result. For example, supersymmetric Ward identities can be derived for any observable in the theory, since their correlators do not depend on $s$-exact terms.

However, the spectrum of the Chern-Simons-matter theory around a monopole background is not gauge invariant, because the quadratic action is not invariant under linearized BRST transformations.%
\footnote{Although the BRST transformations are non-linear in the fields, to have a gauge-invariant spectrum, it would be enough that the quadratic action be invariant under the \emph{linearized} transformations.}
 This can be seen from the presence of the linear term \eqref{lin term A_t + sigma}. Its BRST variation is
$\frac1{4\pi R^2} \Tr\bigl( i k \fm  \, [c,A_t+\sigma] \bigr)$,
and it must cancel with the linearized BRST variation of the quadratic action, which is then nonzero. Consequently, there is no guarantee that the spectrum will be supersymmetric, because it is computed from a quadratic action that is not $s$-closed, and therefore $s$-exact terms violating supersymmetry cannot be neglected. 

A way to resolve this issue takes inspiration from \cite{Pestun:2007rz}. In addition to adding $s\Psi_\text{gf}$ to gauge fix our path integral, we can further add $\cQ\Psi_\text{gf}$. The real supercharge $\cQ$ acts as $\cQ=Q+\overline{Q}$ on physical fields, and we choose its action on the gauge-fixing complex such that $\delta\equiv(s+\cQ)$ closes on symmetries and unfixed gauge transformations. We will show that the further addition of $\cQ\Psi_\text{gf}$ does not change the expectation value of any (possibly non-supersymmetric) operator $\cO$ with ghost number $n_g\leq 0$. In particular, physical observables with $n_g=0$ are not affected. At this point, we have added $\delta\Psi_\text{gf}$ to the original action. The real supercharge $\delta$ is explicitly preserved because our choice that $\delta^2$ contains symmetries and unfixed gauge transformations implies $\delta^2\Psi_\text{gf}=0$. With this procedure, the number of preserved supercharges has not changed; while the gauge-fixed action with $s\Psi_\text{gf}$ is invariant under $s$, the gauge-fixed action with $\delta\Psi_\text{gf}$ is invariant under $\delta$. Its usefulness for computing the spectrum lies in the fact that $A_t+\sigma$ can be redefined by shifting with a quadratic combination of ghosts such that $\delta(A_t'+\sigma')=0$, making the linear term \eqref{lin term A_t + sigma} $\delta$-closed. By extension, the quadratic action which is modified by the shift is also $\delta$-closed, and its spectrum is supersymmetric.

In order for $\delta \Psi_\text{gf} = (s + \cQ) \Psi_\text{gf}$ to be invariant under $\delta$, $\delta^2$ should only contain residual gauge transformations and possibly other symmetries of $\Psi_\text{gf}$. This condition constrains how $\cQ$ can act on fields in the gauge-fixing complex. The supersymmetry transformations of the physical fields $X$ under $\cQ$ are given in \eqref{susy vector m twist D redef}-\eqref{1d chiral transf in 3d chiral} and (\ref{SUSY transf of twisted chiral 1})-(\ref{SUSY transf of twisted chiral 2}). Without specifying how $\cQ$ acts on the fields $Y$ in the gauge-fixing complex, we find:
\bea
\cQ^2 X &= \{Q,\overline{Q}\}X = i \bigl[ \partial_t - \delta_\text{gauge}(A_t+\sigma) \bigr] \, X \;,\qquad& \{\cQ, s\} X = \delta_\text{gauge}\bigl( \cQ c \bigr) \, X \;, \\
\delta^2 X&= i \Bigl[ \partial_t - \delta_\text{gauge}\bigl( A_t + \sigma + i \cQ c - R \bigr) \Bigr] \, X \;.
\eea
If we want $\delta$ to close on time translations and residual gauge transformations, the only possibility is to set $\cQ c = i(A_t+\sigma)_\ff$. Hence, physical fields satisfy the algebra:
\be
\label{delta squared algebra}
\delta^2 X = i \Bigl[ \partial_t - \delta_\text{gauge} \bigl( A_{t,\fr} + \sigma_\fr - R \bigr) \Bigr] \, X \;.
\ee
Having fixed $\cQ c$, we find that $c$ also satisfies \eqref{delta squared algebra} and specifically
\be\label{cQ s on Y}
\cQ^2 c = 0 \;,\qquad\qquad \{\cQ, s\}c = i \bigl[ \partial_t - \delta_\text{gauge}(A_{t,\fr} + \sigma_\fr) \bigr] \, c \;,
\ee
which imply \eqref{delta squared algebra}. For uniformity, we demand that \eqref{cQ s on Y} is satisfied on all fields $Y$ in the gauge-fixing complex.
Setting $\cQ \,\wt c = 0$ for simplicity, we find that this fixes $\cQ b$ and, altogether, $\cQ$ acts on the fields in the gauge-fixing complex as:
\be
\label{cQ on gf complex}
\cQ \, c = i (A_t+\sigma)_\ff \;,\qquad\qquad \cQ \, \wt c = 0 \;,\qquad\qquad \cQ \, b = \bigl[ \partial_t - \delta_\text{gauge}(A_{t,\fr} + \sigma_\fr) \bigr] \, \wt c \;.
\ee

Given $\Psi_\text{gf}$ that we defined in (\ref{def Psi_gf}), we can now determine
\be
\label{Q Psi_gf}
\cQ \, \Psi_\text{gf} = \frac{1}{e^2_\text{3d}} \Tr \biggl[ i \, \wt c \, \cQ G_\text{gf} + \frac{i}{2} \, \wt c \bigl( D_t - i\sigma \bigr) \wt c \, \biggr] \;,
\ee
where $\sigma$ acts in the adjoint representation (namely, $\sigma \wt c$ stands for $[\sigma, \wt c \,]$ in matrix notation). Hence, collecting the contributions from (\ref{FP gf}) and (\ref{Q Psi_gf}), the supersymmetrized gauge-fixing procedure requires us to add the following terms to the original Lagrangian:
\be
\delta\Psi_\text{gf} = \frac{1}{e^2_\text{3d}} \Tr \biggl[ \frac{b^2}{2} + b \, \Bigl( G_\text{gf}-\{\wt c,c\} \Bigr) + i \, \wt c \, \Bigl( \delta_\text{gauge}(c) + \cQ \Bigr) \, G_\text{gf}
+ \frac12 \{\wt c,c\}^2 +\frac{i}2 \, \wt c \, \bigl( D_t-i\sigma \bigr) \, \wt c \, \biggr] \,.
\ee
With the choice that $c$ and $\wt c$ are Hermitian, $\delta\Psi_\text{gf}$ is real.

It is important to note (following \cite{Pestun:2007rz}) that adding $\cQ\Psi_\text{gf}$ to $s\Psi_\text{gf}$ does not change the expectation values of operators with $n_g\leq 0$, even if they are \emph{not} invariant under $\cQ$. In particular, it does not change physical observables. This can be shown explicitly for the thermal partition function. We first integrate in an adjoint-valued auxiliary field $a$ to rewrite the quartic ghost interactions, after which the gauge-fixing action becomes:
\be
\delta\Psi_\text{gf} = \frac{1}{e^2_\text{3d}} \Tr \biggl[ \frac{b^2-a^2}{2} + b \, G_\text{gf} +\wt c \, \bigl[ a+b,c \bigr] + i\wt c \Bigl( \delta_\text{gauge}(c) + \cQ \Bigr) G_\text{gf} + \frac{i}2 \, \wt c \bigl( D_t-i\sigma \bigr) \wt c \biggr] .
\ee
Note that $a$ has both gauge-fixed and residual components. Since the full action is quadratic in the Grassmann fields $\{ F_\text{phys},c,\wt c \, \}$, where $F_\text{phys}$ is the set of physical fermions, we can formally perform the path integral over them, obtaining:
\be
\det\begin{pmatrix}
S_0|_{F,F} & 0 & \cQ \Psi_\text{gf}|_{F,\wt c}\\
0 & 0 & s\Psi_\text{gf}|_{c,\wt c}\\
\cQ \Psi_\text{gf}|_{\wt c,F} &  s\Psi_\text{gf}|_{\wt c,c} & \cQ \Psi_\text{gf}|_{\wt c,\wt c}
\end{pmatrix}
\,\sim\, \det\Bigl( s\Psi_\text{gf}|_{c,\wt c} \Bigr) \, \det \Bigl( S_0|_{F,F} \Bigr) \;.
\ee
All entries of the matrix on the LHS are (possibly differential) operators involving the bosons. This proves that the thermal partition function does not depend on the term $\cQ \Psi_\text{gf}$.

More generally, we prove that the expectation value of any operator $\cO$ with ghost number $n_g\leq 0$ is unchanged by the addition of $\cQ\Psi_\text{gf}$ to the Lagrangian. The key property is that $\cQ \Psi_\text{gh}$ is the sum of two terms, of ghost number $-1$ and $-2$, respectively.
Let $\langle\cdot\rangle_s$ be the path integral with $s\Psi_\text{gf}$ as gauge fixing, and let $\langle\cdot\rangle_\delta$ be the path integral with $\delta\Psi_\text{gf}$ as gauge fixing. We have
\be
\label{d eq s on res O}
\langle\cO\rangle_\delta = \bigl\langle\cO \,e^{i\cQ \Psi_\text{gf}} \bigr\rangle_s = \langle\cO\rangle_s+\sum_{n=1}^\infty\frac{(i)^n}{n!} \, \bigl\langle \cO \,(\cQ\Psi_\text{gf})^n \bigr\rangle_s = \langle \cO\rangle_s \;.
\ee
The last equality holds because ghost number is a symmetry of $\langle\cdot\rangle_s$, implying null expectation value for any correlator that has $n_g\neq 0$. Since $\cO \, (\cQ\Psi_\text{gf})^n$ has $n_g<0$, one concludes that $\langle\cO \, (\cQ\Psi_\text{gf})^n\rangle_s=0$ for every $n$. For the restricted set of operators $\cO$ with $n_g \leq 0$, one can constrain $\langle\cdot\rangle_\delta$ using the symmetries of $\langle\cdot\rangle_s$. In particular, although both supersymmetry and $\rU(1)_R$ are not symmetries of $\langle\cdot\rangle_\delta$ because $\cQ \Psi_\text{gf}$ breaks them, their Ward identities can still be used to constrain the correlators $\langle\cO\rangle_\delta$. This result will play a crucial role in Section \ref{sec: stability}.

We can now show how the linear Lagrangian term containing $A_t+\sigma$ can be made \mbox{$\delta$-invariant} using a field redefinition. This is crucial in order to have a reliable and supersymmetric spectrum. The linear term \eqref{lin term A_t + sigma} only contains modes $(A_t + \sigma)_\fr$ which are constant on $S^2$, due to the integral over $S^2$. Since $A_{t,\fr} + \sigma_\fr-R$ appears in \eqref{delta squared algebra} as a central charge, $\delta(A_{t,\fr}+\sigma_\fr-R)=0$. Therefore, by redefining
\be
\label{A_t+sigma shift}
A'_{t,\fr} + \sigma'_\fr = A_{t,\fr} + \sigma_\fr + \frac12 \{c,c\}_\fr \;,
\ee
the linear term \eqref{lin term A_t + sigma} becomes (dropping the $'$ on $A'_{t,\fr} + \sigma'_\fr$):
\be
\frac{k}{4\pi R^2} \Tr\Bigl( \fm \, (A_t + \sigma) \Bigr) + \frac{m_k}{4R^2 e^2_\text{3d}} \Tr \Bigl( c \, [\fm,c] \Bigr) \;,
\ee
where $\fm$ is diagonal and $m_k$ was defined in (\ref{bg vevs}). The first term is invariant under $\delta$, therefore after adding the second term to the quadratic action, the latter becomes invariant under $\delta$ as well, and the spectrum has to be supersymmetric (\ie, $\delta$-symmetric). Notice that the newly shifted field $A_{t,\fr} + \sigma_\fr$ is still Hermitian because $c$ is Hermitian.

\subsection{Vector multiplet spectrum}

We are now ready to compute the spectrum of the (gauge-fixed) vector multiplet action. We start by considering the off-diagonal components. The Yang-Mills, Chern-Simons, and gauge-fixing terms are expanded to quadratic order in fluctuations around \eqref{bg vevs}. After integrating out the auxiliary fields $D$ and $b$, the independent components consist of $4$ complex bosons $\bigl( A_1^{ij}, A_t^{ij},\sigma^{ij}, A_{\bar1}^{ij} \bigr)$ and $6$ complex fermions $\bigl(\, \overline{\Lambda}_1^{\,ij}, \Lambda_t^{ij}, \overline{\Lambda}_t^{\,ij}, c^{ij}, \wt c^{\;ij}, \Lambda_{\bar1}^{ij} \bigr)$ for every $i\neq j$ such that $q_{ij}>0$.%
\footnote{We have chosen to write $A_1^{ij} = \overline{A_{\bar1}^{ji}}$, $\overline{\Lambda}^{\,ij}_1 = \overline{\Lambda^{ji}_{\bar1}}$ and $\overline{\Lambda}_t^{\,ij}=\overline{\Lambda_t^{ji}}$.}
All components are then rescaled by a factor of $e_\text{3d}/R$. Moreover $A_1^{ij}$, $A_{\bar1}^{ij}$ get an extra factor of $1/\sqrt{2}$, while $\Lambda_{\bar1}^{ij}$, $\overline{\Lambda}_{1}^{\,ij}$, $\Lambda_t^{ij}$, $\overline{\Lambda}_t^{\,ij}$ get an extra factor of $\sqrt{2}$. This is to ensure that the standard 1d kinetic terms are canonically normalized. After expanding in monopole harmonics according to Table \ref{tab: monopole charges} and integrating over $S^2$, the quadratic action for off-diagonal components in momentum space becomes:
\be
\int\! \frac{dp}{2\pi} \, \sum_{i,j \,|\, q_{ij}>0} \;\; \sum_{l,\, |m|\leq l} \biggl( \, \overline{B^{ij}_{l,m}(p)} \, M_B \, B^{ij}_{l,m}(p) + \overline{F^{ij}_{l,m}(p)} \, M_F \, F^{ij}_{l,m}(p) \biggr)
\ee
where the vectors of bosonic and fermionic fields are, respectively,
\bea
B^{ij}_{l,m} &= \bigl( A_{1,l,m}^{ij} \,,\; A_{t,l,m}^{ij} \,,\; \sigma_{l,m}^{ij} \,,\; A_{\bar1,l,m}^{ij} \,\bigr)^\sT \;,\\
F^{ij}_{l,m} &= \bigl( \, \overline{\Lambda}_{1,l,m}^{\,ij} \,,\; \Lambda_{t,l,m}^{ij} \,,\; \overline{\Lambda}_{t,l,m}^{\,ij} \,,\; c_{l,m}^{ij} \,,\; \wt c_{l,m}^{\;ij} \,,\; \Lambda_{\bar1,l,m}^{ij} \,\bigr)^\sT \;.
\eea
The operators acting on the bosonic and fermionic fields are:
\be
\resizebox{\linewidth}{!}{%
$M_B = \begin{pmatrix}
p(p+m_{k}+2\sigma_0)-\dfrac{\xi+1}{\xi}\dfrac{s_-^2}{2R^2} & -\dfrac{is_-(p+m_{k}+\sigma_0)}{\sqrt{2}R} & -\dfrac{i\sigma_0s_-}{\sqrt{2}R} & \dfrac{1-\xi}{\xi} \, \dfrac{s_+s_-}{2R^2} \\[0.9em]
\dfrac{is_-(p+m_{k}+\sigma_0)}{\sqrt{2}R} & \dfrac{s_0^2}{R^2}+\sigma_0^2 & \sigma_0(p+\sigma_0) & -\dfrac{is_+(p-m_{k}+\sigma_0)}{\sqrt{2}R}
\\[0.9em]
\dfrac{i\sigma_0s_-}{\sqrt{2}R} & \sigma_0(p+\sigma_0) & (p+\sigma_0)^2-m_{k}^2-\dfrac{s_0^2}{R^2} & -\dfrac{i\sigma_0s_+}{\sqrt{2}R} \\[0.9em]
\dfrac{1-\xi}{\xi} \, \dfrac{s_+s_-}{2R^2} & \dfrac{is_+(p-m_{k}+\sigma_0)}{\sqrt{2}R} & \dfrac{i\sigma_0s_+}{\sqrt{2}R} & p(p-m_{k}+2\sigma_0)-\dfrac{\xi+1}{\xi}\dfrac{s_+^2}{2R^2}
\end{pmatrix}$}\rule[-1.86cm]{0pt}{0pt}
\ee
with
\be
\sigma_0 = -\frac{q_{ij}}{m_{k}R^2} \,,\quad s_0 = \sqrt{l(l+1)-q_{ij}^2}\,,\quad s_\pm = \sqrt{l(l+1)-q_{ij}(q_{ij}\pm 1)}=\sqrt{s_0^2\mp q_{ij}}
\ee
(notice that $\sigma_0$, $s_0$, and $s_\pm$ depend on $ij$) and
\be
\label{gauge fermions quad}
\resizebox{0.78\linewidth}{!}{%
$\ds M_F = \begin{pmatrix}
-p-m_{k}-2\sigma_0 & -\dfrac{s_-}{R} & 0 & 0 & -\dfrac{is_-}{\sqrt{2\xi}R} & 0 \\
-\dfrac{s_-}{R} & -p+m_{k} & 0 & 0 & 0 & 0 \\
0 & 0 & -p-m_{k} & 0 & 0 & -\dfrac{s_+}{R} \\
0 & 0 & 0 & \dfrac{m_{k}q_{ij}}{R^2} & \dfrac{is_0^2}{\sqrt{\xi}R^2} & 0 \\
\dfrac{is_-}{\sqrt{2\xi}R} & 0 & 0 & -\dfrac{is_0^2}{\sqrt{\xi}R^2} & - p & -\dfrac{is_+}{\sqrt{2\xi}R} \\
0 & 0 & -\dfrac{s_+}{R} & 0 & \dfrac{is_+}{\sqrt{2\xi}R} & -p+m_{k}-2\sigma_0
\end{pmatrix}$} \,.
\ee
For $l\geq q_{ij}+1$, all modes exist and are massive. Moreover, the masses of the modes%
\footnote{The counting of modes works as follows. A complex field with 2-derivative kinetic term gives two modes, with only 1-derivative kinetic term gives one mode, whereas with no kinetic term gives no modes.}
from bosons and fermions are paired thanks to the $\delta$-invariance of the action, and the ratio of fermionic to bosonic determinants is $1$. For $l=q_{ij}$, the modes of $A_{\bar1}^{ij}$ and $\Lambda_{\bar1}^{ij}$ do not exist (see Table~\ref{tab: monopole charges}), so the rightmost column and the bottom row of the matrices $M_B$, $M_F$ should be removed. In this case, there is a massless fermionic mode while the other massive modes are paired between bosons and fermions. The ratio of determinants is $- p$. For $l=q_{ij}-1$ (this case takes place only if $q_{ij}\geq 1$), modes only exist in $A_1^{ij}$ and $\overline{\Lambda}_1^{\, ij}$. The bosonic field $A_1^{ij}$ has a massless pole, and a massive pole that cancels with that of $\overline{\Lambda}_1^{\,ij}$.

The effective degrees of freedom at energies much smaller than $m_{k}$ and $\frac{1}{R}$ are the massless fermionic modes with $l=q_{ij}$ and the massless modes in $A_1^{ij}$ with $l= q_{ij}-1$ (if $q_{ij} \geq 1$). The identity of the massless fermionic modes is not immediately clear due to the off-diagonal entries in \eqref{gauge fermions quad}. We can first rescale the fields $c^{ij}_{l,m}\rightarrow R \, c^{ij}_{l,m}$, so that they have the same mass dimension as the other fermions. Defining the dimensionless ratio $\alpha = 1 / (m_k R)$ for convenience, the fermionic kinetic operator above becomes:
{\small\be
\label{gauge fermions quad l=q}
M_F \big|_{l=q_{ij}} = 
\begin{pmatrix}
-p-(1-2q_{ij}\alpha^2) \, m_{k} & -\sqrt{2q_{ij}} \, \alpha \, m_{k} & 0 & 0 & -i\sqrt{\frac{q_{ij}}{\xi}} \, \alpha \, m_{k} \\
-\sqrt{2q_{ij}} \, \alpha \, m_{k} & -p+m_{k} & 0 & 0 & 0 \\
0 & 0 & -p-m_{k} & 0 & 0 \\
0 & 0 & 0 & q_{ij} \, m_{k} & i\frac{q_{ij}}{\sqrt{\xi}} \, \alpha \, m_{k} \\
i\sqrt{\frac{q_{ij}}{\xi}} \, \alpha \, m_{k} & 0 & 0 & -i\frac{q_{ij}}{\sqrt{\xi}} \, \alpha \, m_{k} & - p
\end{pmatrix} .
\ee}%
By introducing a kinetic term $i \varepsilon \, \overline{c^{ij}} \, \partial_tc^{ij}$ by hand for the fermion $c^{ij}$, the problem of finding mass eigenstates is reduced to the usual problem of diagonalizing a mass matrix. Taking $\varepsilon\to0$ at the end of the computation, we obtain the desired $\rSL{(5,\mathbb{C})}$ transformation that diagonalizes \eqref{gauge fermions quad l=q}:
\be
\label{sl5C transformation}
\resizebox{0.9\linewidth}{!}{%
$S=\begin{pmatrix}
-\dfrac{A_-}{\sqrt{8q_{ij}^2\alpha^4\xi+A_-^2+B_-^2}} & -\dfrac{A_+}{\sqrt{8q_{ij}^2\alpha^4\xi+A_+^2+B_+^2}} & 0 & 0 & \dfrac{\alpha}{\sqrt{\xi+q_{ij}\alpha^2+2q_{ij}^2\alpha^4}} \\
\dfrac{B_-}{\sqrt{8q_{ij}^2\alpha^4\xi+A_-^2+B_-^2}} & \dfrac{B_+}{\sqrt{8q_{ij}^2\alpha^4\xi+A_+^2+B_+^2}} & 0 & 0 & \dfrac{\sqrt{2}\alpha^2}{\sqrt{\xi+q_{ij}\alpha^2+2q_{ij}^2\alpha^4}} \\
0 & 0 & 1 & 0 & 0 \\
-\dfrac{2\sqrt{2\xi}q_{ij}\alpha^3}{\sqrt{8q_{ij}^2\alpha^4\xi+A_-^2+B_-^2}} & -\dfrac{2\sqrt{2\xi}q_{ij}\alpha^3}{\sqrt{8q_{ij}^2\alpha^4\xi+A_+^2+B_+^2}} & 0 & -i\sqrt{\frac{\xi}{q_{ij}}} & \dfrac{\sqrt{\xi}\alpha}{\sqrt{\xi+q_{ij}\alpha^2+2q_{ij}^2\alpha^4}} \\
-\dfrac{i2\sqrt{2}q_{ij}\alpha^2}{\sqrt{8q_{ij}^2\alpha^4\xi+A_-^2+B_-^2}} & -\dfrac{i2\sqrt{2}q_{ij}\alpha^2}{\sqrt{8q_{ij}^2\alpha^4\xi+A_+^2+B_+^2}} & 0 & 0 & \dfrac{i}{\sqrt{\xi+q_{ij}\alpha^2+2q_{ij}^2\alpha^4}}
\end{pmatrix}\,,$}
\ee
where we have defined
\bea
A_{\pm}&=\sqrt{2q_{ij}} \alpha \left(q_{ij}\alpha^2 \left(1+2\xi\right)\pm\sqrt{q_{ij}^2\alpha^4 \left(1+2 \xi\right)^2+4 \xi(q_{ij}\alpha^2 +\xi)}\right)\\
B_{\pm}&=2\xi+q_{ij}\alpha^2 \left(1+2\xi\right)\pm\sqrt{q_{ij}^2\alpha^4 \left(1+2 \xi\right)^2+4 \xi(q_{ij}\alpha^2 +\xi)}\,.
\eea
The resulting fermionic kinetic operator is
\be
\label{gauge fermions quad l=q diag}
S^\dag \, M_F \big|_{l=q_{ij}} \,S = \begin{pmatrix}-p-\lambda_+m_{k} & 0 & 0 & 0 & 0 \\
0 & -p-\lambda_-m_{k} & 0 & 0 & 0 \\
0 & 0 & -p-m_{k} & 0 & 0 \\
0 & 0 & 0 & m_{k} & 0 \\
0 & 0 & 0 & 0 & - p\end{pmatrix}
\ee
with
\be
\lambda_\pm=\frac{q_{ij} \, \alpha^2 \left(1-2 \xi\right) \pm \sqrt{q_{ij}^2 \, \alpha^4 (1+2 \xi )^2+4 \xi \bigl( q_{ij} \, \alpha^2 +\xi \bigr)} }{2 \xi} \;.
\ee
Each row of the matrix $S$ expresses an original fermion in terms of the mass eigenstates. The linear combinations are generically complicated, but they simplify in the physical regime of interest. Since we want to reduce a Chern-Simons-matter theory on $S^2$, and the Yang-Mills term was only introduced to make propagating gauge degrees of freedom massive, we are motivated to take $m_k\gg \frac{1}{R}$ or $\alpha\to 0$. In this limit, the massless fermion at $l=q_{ij}$ is $-i\sqrt{\xi} \, \widetilde c$ (last row of $S$), and $\lambda_\pm\to\pm 1$.

The spectrum of the diagonal components can be analyzed in the same way and we will be brief. One finds that every mode is massive for $l>0$. After integrating out the $l=0$ mode of the auxiliary fields $D^i$, the quadratic Lagrangian (including the linear terms) for the remaining diagonal $l=0$ modes is:
\be
\sum_i \biggl\{ k \fm_i \, \bigl( A^i_{t,0,0} + \sigma^i_{0,0} \bigr) + 
\frac{4\pi R^2}{e^2_\text{3d}} \biggl[ \frac12 \bigl( \partial_t\sigma^i_{0,0} \bigr)^2 - \frac12 m_k^2 \bigl( \sigma^i_{0,0} \bigr)^2 + \frac12 \, \overline{\Lambda}^{\,i}_{t,0,0} \bigl( i\partial_t+m_k \bigr) \Lambda^i_{t,0,0} \biggr] \biggr\} \,.
\ee
We observe that $\sigma^i_{0,0}$ and $\Lambda^i_{t,0,0}$ have mass $m_k$ and should be integrated out at low energies $p\ll m_k$. Only the combination $\bigl( A_{t,0,0}^i+\sigma^i_{0,0} \bigr)$ remains, which is a $1$d gauge field for the gauge group $\rU(1)^N$.%
\footnote{In other words, in the language of Appendix~\ref{app: 1d susy}, we find that the superfield $V^-$ is massive, while $\Omega$ stays light and enforces gauge invariance.}

To summarize, we write the quadratic Lagrangian for the modes from the vector multiplet that contain massless poles, including fermionic partners which are necessary for supersymmetry. After having rescaled $A_{\bar 1}$ and $\Lambda_{\bar 1}$ by $m_k^{-1/2}$ we have:
\begin{multline}
\label{EFT vector multiplet provisional}
k \sum_i \fm_i \, (A_t^i+\sigma^i) + \sum_{i\neq j} \biggl\{ \Theta(q_{ij}-1) \! \sum_{|m|\leq q_{ij}-1}\bigg[\overline{A_{\bar1,q_{ij}-1,m}^{ji}} \, i\partial_t \, A^{ji}_{\bar1,q_{ij}-1,m} + \overline{\Lambda^{ji}_{\bar1,q_{ij}-1,m}} \, \Lambda^{ji}_{\bar1,q_{ij}-1,m} \\
+ \frac{1}{m_{k}} \biggl( \Bigl\lvert \partial_tA^{ji}_{\bar1,q_{ij}-1,m} \Bigr\rvert^2 + \overline{\Lambda^{ji}_{\bar1,q_{ij}-1,m}} \, i\partial_t \, \Lambda^{ji}_{\bar1,q_{ij}-1,m} \biggr) \biggr] + \Theta(q_{ij}) \! \sum_{|m|\leq q_{ij}} \Bigl( \, \overline{ \wt c^{\;ij}_{q_{ij},m}} \, i\partial_t \, \wt c^{\;ij}_{q_{ij},m} \Bigr) \biggr\}
\end{multline}
where $\Theta(n) = 1$ for $n\geq 0$ and it vanishes otherwise. Here we have changed notation, and used the fields $\bigl( A^{ji}_{\bar1}, \Lambda^{ji}_{\bar1} \bigr)$ in place of $A^{ij}_1$, $\wb\Lambda^{\,ij}_1$ because the former live in a chiral multiplet, see (\ref{1d chiral transf in 3d chiral}), while the latter in an anti-chiral multiplet. Besides, notice that there are matching degrees of freedom in $A_{\bar1}^{ji}$ and $\Lambda_{\bar1}^{ji}$ with mass $m_k$, which should not be included in the effective theory at energies $p\ll m_k$. These modes are encoded in the term proportional to $1/m_k$ and can be integrated out by neglecting that kinetic term. The workings are explained in \cite{Dunne:1989hv}. The quadratic Lagrangian for the massless modes is then:%
\footnote{Using the assumption that $q_{ij} \neq 0$ for $i\neq j$, we have substituted $\Theta(q_{ij}) \to \Theta(q_{ij} - \tfrac12)$ in (\ref{EFT vector multiplet provisional}), and consequently we have substituted $\sum_{i\neq j} \to \sum_{ij}$.}
\begin{multline}
k \sum_i \fm_i \, (A_t^i+\sigma^i) + \sum_{ij} \biggl\{ \Theta(q_{ij}-1) \! \sum_{|m|\leq q_{ij}-1} \biggl( \overline{A_{\bar1,q_{ij}-1,m}^{ji}} \, i\partial_t \, A^{ji}_{\bar1,q_{ij}-1,m} + {} \\
{} + \overline{\Lambda^{ji}_{\bar1,q_{ij}-1,m}} \, \Lambda^{ji}_{\bar1,q_{ij}-1,m} \biggr) + \Theta(q_{ij}-\tfrac12) \sum_{|m|\leq q_{ij}} \overline{ \wt c^{\;ij}_{q_{ij},m}}\, i\partial_t \, \wt c^{\;ij}_{q_{ij},m} \biggr\} \;.
\end{multline}
The bosons $A_{\bar1}^{ji}$ and the fermions $\wt c^{\;ij}$ have a 1-derivative action, while the fermions $\Lambda_{\bar1}^{ji}$ are auxiliary.

\subsection{Matter spectrum}

To find the spectrum of modes coming from the $3$d chiral multiplets, we expand the chiral multiplet Lagrangian \eqref{chiral_lagrangian_twisted} to quadratic order in fluctuations around \eqref{bg vevs}. All fields in the chiral multiplet are rescaled by $\frac{1}{R}$. After expanding in monopole harmonics according to Table~\ref{tab: monopole charges} and integrating over $S^2$, the quadratic action in momentum space is:
\begin{multline}
\int\!\frac{dp}{2\pi} \, \sum_a \sum_{i,j} \, \sum_{l, \, |m|\leq l} \Biggl\{ \biggl[ p(p+2\sigma_{0}) - \frac{s_{+,a}^2}{R^2} \biggr] \bigl\lvert \phi^{ij}_{a,l,m}(p) \bigr\rvert^2 + \bigl\lvert f^{ij}_{a,l,m}(p) \bigr\rvert^2 + {} \\
{} + \Bigl(\, \overline{\psi^{ij}_{a,l,m}(p)} \;,\; \overline{\eta^{ij}_{a,l,m}(p)} \,\Bigr)
\begin{pmatrix} -p-2\sigma_0 & \frac{s_{+,a}}{R} \\ \frac{s_{+,a}}{R} & -p \end{pmatrix}
\begin{pmatrix} \psi^{ij}_{a,l,m}(p) \\ \eta^{ij}_{a,l,m}(p) \end{pmatrix} \Biggr\}
\end{multline}
where
\be
\sigma_0 = - q_{ij} \alpha^2 m_k \,\equiv\, - \frac{m_\sigma}2 \;,\qquad\qquad s_{\pm,a}\equiv\sqrt{l(l+1)-q^a_{ij}(q^a_{ij}\pm 1)} \;.
\ee
For $l\geq |q_{ij}^a|+1$, all modes exist (see Table~\ref{tab: monopole charges}) and are massive. Moreover, the masses of bosons and fermions are paired and the ratio of determinants is 1. The modes with $l=|q_{ij}^a|$ exist in all fields if $q_{ij}^a\leq-\frac{1}{2}$, whereas they only exist in $\phi_a^{ij}$ and $\psi_a^{ij}$ if $q_{ij}^a\geq 0$. In the former case, all modes are massive. In the latter case, the field $\phi_a^{ij}$ has a massless pole, and a massive pole that cancels with that of $\psi_a^{ij}$. Provided that $q^a_{ij}\leq-1$, there exist modes with $l=|q_{ij}^a|-1 = - q^{ij}_a - 1$ in $\eta_a^{ij}$ and $f_a^{ij}$, such that $\eta_a^{ij}$ is massless while $f_a^{ij}$ is auxiliary.

To summarize, the quadratic Lagrangian for modes which contain massless poles, and that of their supersymmetry partners is
\begin{align}
& \sum_{ij,\, a} \biggl\{ \Theta(q^a_{ij}) \sum_{|m|\leq q^a_{ij}} \biggl[ m_\sigma^{ij} \Bigl( \, \overline{\phi^{ij}_{a,q^a_{ij},m}} \, i\partial_t \, \phi^{ij}_{a,q^a_{ij},m} + \overline{\psi^{ij}_{a,q^a_{ij},m}} \, \psi^{ij}_{a,q^a_{ij},m} \Bigr) + \Bigl\lvert \partial_t\phi^{ij}_{a,q^a_{ij},m} \Bigr\rvert^2 + {} \\
&\;\; + \overline{\psi^{ij}_{a,q^a_{ij},m}} \, i\partial_t \, \psi^{ij}_{a,q^a_{ij},m} \biggr] + \Theta(-q^a_{ij}-1) \!\sum_{|m|\leq -q^a_{ij}-1} \! \Bigl( \, \overline{\eta^{ij}_{a,-q^a_{ij}-1,m}} \, i\partial_t \, \eta^{ij}_{a,-q^a_{ij}-1,m} + \bigl\lvert f^{ij}_{a,-q^a_{ij}-1,m} \bigr\rvert^2 \Bigr) \biggr\}\,, \nn
\end{align}
where the $i,j$ dependence of $m_\sigma$ was made explicit. At low energies $p\ll m_\sigma^{ij}$, the quadratic kinetic term of $\phi^{ij}_{a,q^a_{ij},m}$ and the kinetic term of $\psi^{ij}_{a,q^a_{ij},m}$ can again be neglected. Note that $q^a_{ij}\geq 0$ does not exclude the possibility that $i=j$, in which case $m_\sigma^{ij}=0$. We might also have $m_\sigma^{ij}\to 0$ as $\alpha\to0$.%
\footnote{Indeed $m_\sigma \sim \alpha^2 m_k \sim \alpha/R$, therefore its scaling is not fixed by the choices we already made.}
In either case, all of $\phi^{ij}_{a,q^a_{ij},m}$ and $\psi^{ij}_{a,q^a_{ij},m}$ would be classically massless. However, quantum effects would still generically generate supersymmetric mass terms like
\be
m_{\sigma \text{(q)}}^{ij} \, \Bigl( \, \overline{\phi^{ij}_{a,q^a_{ij},m}} \, i\partial_t \, \phi^{ij}_{a,q^a_{ij},m} + \overline{\psi^{ij}_{a,q^a_{ij},m}} \, \psi^{ij}_{a,q^a_{ij},m} \Bigr) \;,
\ee
whose superspace expression is \eqref{1d chiral unconv kin term}. At scales $p\ll m_{\sigma \text{(q)}}^{ij}$, the quadratic kinetic term of $\phi^{ij}_{a,q^a_{ij},m}$ and the kinetic term of $\psi^{ij}_{a,q^a_{ij},m}$ would still be negligible. Therefore, rescaling $\phi^{ij}_{a,q^a_{ij},m}$ and $\psi^{ij}_{a,q^a_{ij},m}$ by $1/(m_\sigma^{ij})^{1/2}$ (including quantum corrections), the resulting quadratic effective Lagrangian is:
\begin{align}
\label{chiral quad Leff}
&\sum_{ij,\, a} \biggl[ \Theta(q^a_{ij}) \sum_{|m|\leq q^a_{ij}} \Bigl( \, \overline{\phi^{ij}_{a,q^a_{ij},m}} \, i\partial_t \, \phi^{ij}_{a,q^a_{ij},m} + \overline{\psi^{ij}_{a,q^a_{ij},m}} \, \psi^{ij}_{a,q^a_{ij},m} \Bigr) + {} \\
&\qquad + \Theta(-q^a_{ij}-1) \!\sum_{|m|\leq -q^a_{ij}-1}\! \Bigl( \, \overline{\eta^{ij}_{a,-q^a_{ij}-1,m}} \, i\partial_t \, \eta^{ij}_{a,-q^a_{ij}-1,m} + \bigl\lvert f^{ij}_{a,-q^a_{ij}-1,m} \bigr\rvert^2 \Bigr) \biggr] \;. \nn
\end{align}

\section{The effective Quantum Mechanics}
\label{sec: QM}

In this section we present the proposed low-energy quantum mechanical model, which is the result of setting to zero all massive modes in the gauge-fixed 3d Lagrangian while only keeping the light modes.

The gauge group is $\rU(1)^N$ and the vector multiplet only contains the gauge fields $A_t^i+\sigma^i$, with $i=1,\ldots, N$.%
\footnote{ In Wess-Zumino gauge, the only non-vanishing component of the superfield $V$ (or equivalently of $\Omega$) is $A_t+\sigma$. See Appendix~\ref{app: WZ gauge}.}
Their role is to impose Gauss's law. Because of the presence of a Wilson line of charges $k\fm_i$, coming from the 3d Chern-Simons term, Gauss's law projects onto a sector of non-vanishing gauge charges.

The matter content consists of various chiral and Fermi multiplets $X^{ij}$ with charges $+1$ under $\rU(1)_i \subset \rU(1)^N$ and $-1$ under $\rU(1)_j$. They interact with the gauge fields via the covariant derivative
\be
D_t^+X^{ij} = \Bigl( \partial_t - i \bigl( A_t^i+\sigma^i-A_t^j-\sigma^j \bigr) \Bigr) X^{ij} \;.
\ee
The matter content depends on the fluxes $\fm_i$ --- determined in (\ref{u m and rho final}) --- and $\fn_a$ through the combinations $q_{ij}$ and $q_{ij}^a$ defined in \eqref{def monopole q}. For every pair of indices $ij$, from the 3d vector multiplet we get the following matter multiplets. If $q_{ij}\leq -1$, there are 1d chiral multiplets $\Xi^{ij}_{\bar1,m} = \bigl( A_{\bar1,m}^{ij}, \Lambda^{ij}_{\bar1,m} \bigr)$ in the $\rSU(2)$ representation of highest weight $l=-q_{ij}-1$. Otherwise, if $q_{ij} \geq \frac12$, there are 1d Fermi multiplets $C^{ij}_m = \bigl( \wt c^{\;ij}_m, g^{ij}_m \bigr)$ with $l=q_{ij}$. Here we introduce the auxiliary fields $g^{ij}_m$, even though they are not present in the 3d theory, in order to make off-shell supersymmetry manifest. From the 3d chiral multiplet with flavor index $a$, we get 1d chiral multiplets $\Phi^{ij}_{a,m} = \bigl( \phi^{ij}_{a,m},\psi^{ij}_{a,m} \bigr)$ with $l=q^a_{ij}$ if $q^a_{ij}\geq 0$, and otherwise 1d Fermi multiplets $\cY^{ij}_{a,m} = \bigl( \eta^{ij}_{a,m}, f^{ij}_{a,m} \bigr)$ with $l=-q^a_{ij}-1$ if $q^a_{ij}\leq -1$. We summarize this content in Table~\ref{tab: 1d fields}, where we also list the representations and charges of each multiplet under the global symmetries $\rSU(2)$, $\rU(1)^2_F$ and $\rU(1)_R$. 

\begin{table}[t]
\centering
\begin{tabular}{|c|c|c|c|c|}
\hline \rule[-0.5em]{0pt}{1.8em}
 & $A^{ij}_{\bar1,m}$ & $\wt c^{\;ij}_m$ & $\phi^{ij}_{a,m}$ & $\eta^{ij}_{a,m}$ \\
 & {\small chiral} & {\small Fermi} & {\small chiral} & {\small Fermi} \\
\hline\rule[-0.6em]{0pt}{0pt}%
existence: & $q_{ij}\leq -1$ & $q_{ij} \geq \tfrac12$ & $q^a_{ij}\geq 0$ & $q^a_{ij}\leq -1$\\
    \hline
$l$ & $|q_{ij}|-1$ & $q_{ij}$ & $q^a_{ij}$ & $|q^a_{ij}|-1$ \\[0.2em]
    \hline
$R_3$ & $0$ & $0$ & $2\delta_{3a}$ & $2\delta_{3a}-1$\\
    \hline
    $q_1$ & $0$ & $0$ & $\delta_{1a}-\delta_{3a}$ & $\delta_{1a}-\delta_{3a}$\\
    \hline
$q_2$ & $0$ & $0$ & $\delta_{2a}-\delta_{3a}$ & $\delta_{2a}-\delta_{3a}$\\
  \hline
\end{tabular}
\caption{Matter multiplets (we indicate the bottom components) for indices $ij$ and their representations under the global symmetries. We label the $\rSU(2)$ representation by the highest weight $l\in\bZ/2$. The charges of the lowest components in each multiplet are indicated, while their superpartners have \mbox{R-charges} $R_3$ which are shifted by $-1$. 
\label{tab: 1d fields}}
\end{table}

In addition to gauge interactions, other interactions are specified by $E$ and $J$ superpotentials. We have as many $E$ and $J$ functions as there are Fermi multiplets. For a given Fermi multiplet $\eta$, $E$ is in the same gauge and flavor representation as $\eta$, and its R-charge is $R(\eta)+1$. On the contrary, $J$ is in the conjugate gauge and flavor representation with respect to $\eta$, and its R-charge is $-R(\eta)+1$. We find that the $E$ and $J$ functions are zero for the Fermi multiplets $\wt c^{\;ij}_m$. For the Fermi multiplets $\eta^{ij}_{a,m}$, the $E$ and $J$ superpotentials are:
\begin{align}
\label{E-term}
E^{ij}_{a,m} &= i\sum_k \biggl[ \;\;\; \Theta(q^a_{kj}) \sum_{|m'| \leq q^a_{kj}} e_\text{1d}^{kj} \; {\scriptstyle \sqrt{2q^a_{kj}+1}} \;\, C\bigl( \begin{smallmatrix}
|q_{ik}|-1 & q^a_{kj} & |q_{ij}^a|-1 \\
m-m' & m' & m
\end{smallmatrix}\bigr) \, A^{ik}_{\bar1,m-m'} \, \phi^{kj}_{a,m'} \hspace{1.5cm} \\
&\hspace{1.5cm} - \Theta(q^a_{ik}) \sum_{|m'|\leq q^a_{ik}} e_\text{1d}^{ik} \; {\scriptstyle \sqrt{2q^a_{ik}+1}} \;\, C\bigl(\begin{smallmatrix}
|q_{kj}|-1 & q^a_{ik} & |q_{ij}^a|-1 \\
m-m' & m' & m
\end{smallmatrix}\bigr) \,\, \phi^{ik}_{a,m'} \, A^{kj}_{\bar1,m-m'} \;\; \biggr] \;, \nn
\end{align}
\begin{align}
\label{J-term}
J^{ji}_{a,-m} &= -\sum_{b,c,k} \epsilon_{abc} \, \Theta(q^b_{jk}) \, \Theta(q^c_{ki}) \times {} \\
&\qquad \times \!\!\!\! \sum_{\substack{|m'|\leq q^b_{jk} \\ |m+m'|\leq q^c_{ki}}} \!\!\! \lambda_\text{1d}^{jki} \, \Bigl[{\ts \frac{ (2q^b_{jk}+1) (2q^c_{ki}+1) }{ 2|q^a_{ij}|-1 }} \Bigr]^\frac12 {\scriptstyle (-1)^{-q^a_{ij}-1-m} } \; C\bigl(\begin{smallmatrix}
q^b_{jk} & q^c_{ki} & |q^a_{ij}|-1 \\
m' & -m-m' & -m
\end{smallmatrix}\bigr) \, \phi^{jk}_{b,m'} \,\, \phi^{ki}_{c,-m-m'} \;, \nn
\end{align}
where $C\bigl(\begin{smallmatrix}
l & l' & l'' \\
m & m' & m''
\end{smallmatrix}\!\big)$ are the Clebsch-Gordan coefficients given in \eqref{CB coeff def} and we defined
\be
e_\text{1d}^{ij} = \frac{1}{R\sqrt{k \, m_\sigma^{ij}}} \;,\qquad\qquad \lambda_\text{1d}^{ijk} = \frac{\lambda_\text{3d}}{R \sqrt{4\pi \, m_\sigma^{ij} \, m_\sigma^{jk}} } \;.
\ee
The sign $(-1)^{-q^a_{ij}-1-m}$ in the J-term is necessary for $\rSU(2)$ invariance.
The term $E^{ij}_a$ in (\ref{E-term}) exists for $q_{ij}^a \leq -1$, then the condition $q_{kj}^a \geq 0$ in the first line guarantees that $A^{ij}_{\bar1}$ and $\phi^{kj}_a$ both exist, and the condition $q_{ik}^a \geq 0$ in the second line guarantees that $\phi^{ik}_a$ and $A^{kj}_{\bar1}$ both exist.
Also the term $J^{ji}_a$ in (\ref{J-term}) exists for $q_{ij}^a \leq -1$, which is guaranteed by the two conditions $q_{jk}^b \geq 0$, $q_{ki}^c \geq 0$ on the RHS.
The E-term comes from the reduction of \eqref{3d E term} whereas the J-term from the reduction of the 3d superpotential \eqref{3d superpot}. One can check, by substituting \eqref{binom saturate triangle ineq} and relabeling the indices, that
\be
\label{EJ susy condition}
\sum_{ij,\, a} \, \Theta\bigl( -q^a_{ij}-1 \bigr) \sum_{|m|\leq-q^a_{ij}-1} E^{ij}_{a,m} \, J^{ji}_{a,-m} = 0 \;,
\ee
which is required for supersymmetry. The couplings $e_\text{1d}$ and $\lambda_\text{1d}$ are obtained by tree-level matching.

The complete Lagrangian in terms of the $E$ and $J$ given above is:
\begin{align}
\label{full 1d L}
&\cL_\text{QM} = k \sum_i \fm_i \bigl( A_t^i+\sigma^i \bigr) + \sum_{ij} \biggl\{ \Theta(q_{ij}-1) \! \sum_{|m|\leq q_{ij}-1} \! \Bigl( \, \overline{A_{\bar1,m}^{ji}} \, iD_t^+ \, A^{ji}_{\bar1,m} + \overline{\Lambda^{ji}_{\bar1,m}} \, \Lambda^{ji}_{\bar1,m} \Bigr) \\
&\quad + \Theta(q_{ij} {-} \tfrac12) \! \sum_{|m|\leq q_{ij}} \! \Bigl( \, \overline{ \wt c^{\;ij}_m} \, iD_t^+ \, \wt c^{\;ij}_m + \bigl\lvert g^{ij}_m \bigr\rvert^2 \Bigr) \biggr\} + \sum_{ij,\,a} \biggl\{ \Theta(q^a_{ij}) \! \sum_{|m|\leq q^a_{ij}} \! \Bigl( \, \overline{\phi^{ij}_{a,m}} \, iD^+_t \, \phi^{ij}_{a,m} + \overline{\psi^{ij}_{a,m}} \, \psi^{ij}_{a,m} \Bigr) \nn \\
&\quad + \Theta(-q^a_{ij}-1) \! \sum_{|m|\leq -q^a_{ij}-1} \! \biggl( \overline{\eta^{ij}_{a,m}} \, iD_t^+ \, \eta^{ij}_{a,m} + \bigl\lvert f^{ij}_{a,m} \bigr\rvert^2 - \bigl\lvert E^{ij}_{a,m} \bigr\rvert^2 - \overline{\eta^{ij}_{a,m}} \, QE^{ij}_{a,m} - \overline{Q} \,\overline{E^{ij}_{a,m}} \, \eta^{ij}_{a,m} \nn \\
&\hspace{5cm} - f^{ij}_{a,m} J^{ji}_{a,-m} - \overline{J^{ji}_{a,-m}} \,\,\overline{f^{ij}_{a,m}} - \eta^{ij}_{a,m} \, QJ^{ji}_{a,-m} - \overline{Q} \, \overline{J^{ji}_{a,-m}} \,\, \overline{\eta^{ij}_{a,m}} \biggr) \biggr\} \,, \nn
\end{align}
where $i,j = 1, \dots, N$ whereas $a=1,2,3$. Note that both bosons and fermions have 1-derivative kinetic terms.
The Lagrangian can be more compactly written in superspace:
\begin{align}
\cL_\text{QM} &= \!\int\! d\theta d\bar\theta \; \biggl\{ k \sum_i \fm_i V^i + \sum_{ij} \biggl[ \Theta(q_{ij}-1) \! \sum_{|m|\leq q_{ij}-1} \overline{\Xi^{ji}_{\bar1,m}} \, \Xi^{ji}_{\bar1,m} + \Theta(q_{ij} {-} \tfrac12) \! \sum_{|m|\leq q_{ij}} \overline{C^{ij}_m} \, C^{ij}_m \biggr] \nn \\
&\hspace{2.2cm} + \sum_{ij,\, a} \biggl[ \Theta(q^a_{ij}) \! \sum_{|m|\leq q^a_{ij}} \overline{\Phi^{ij}_{a,m}} \, \Phi^{ij}_{a,m} + \Theta(-q^a_{ij}-1) \! \sum_{|m|\leq -q^a_{ij}-1} \overline{\cY^{ij}_{a,m}} \, \cY^{ij}_{a,m} \biggr] \biggr\} \nn \\
&\quad + \sum_{ij,\, a} \Theta(-q^a_{ij}-1) \! \sum_{|m|\leq q^a_{ij}} \biggl\{ \int\! d\theta \; \cY^{ij}_{a,m} \, J^{ji}_{a,-m}(\Phi) + \int d\bar\theta \; \overline{\cY^{ij}_{a,m}} \;\, \overline{J^{ji}_{a,-m}}(\overline{\Phi}) \biggr\} \,.
\label{full 1d L superspace}
\end{align}
Here we promoted the scalar fields in $J$ to be chiral superfields.

The observables of the 3d theory include the gauge-invariant operators. After gauge fixing by $s\Psi_\text{gf}$, they are the BRST-closed operators, invariant under the residual gauge symmetry, and with ghost number $n_g=0$. The further addition of $\cQ\Psi_\text{gf}$ to the Lagrangian does not modify their correlators, see (\ref{d eq s on res O}). When we go to the effective 1d description (\ref{full 1d L}), the ghost field $c$ is completely integrated out. Any operator containing $\wt c^{\;ij}_m$ should not be regarded as a physical observable, because it will have $n_g < 0$. For instance, one might have noticed that the Lagrangian (\ref{full 1d L}) has a large number of additional global $\rU(1)$ symmetries that rotate each $\wt c^{\;ij}_m$ independently. However, their currents are not physical observables (because they are constructed with $\wt c^{\;ij}_m$), and indeed the symmetries act trivially on the sector of physical observables.%
\footnote{In view of holographic applications of the low-energy quantum mechanics, one should not expect the extra symmetries to appear as gauge fields in AdS$_2$.}
They should not be regarded as emergent symmetries of the physical theory. On the other hand, all $\rU(1)^N$-invariant operators constructed from fields of the low-energy 1d description other than $\wt c^{\;ij}_m$ are physical observables. This is because the BRST transformations of the physical fields $X$ are $sX=\delta_\text{gauge}(c)X$, but $c$ is massive and set to zero in the low-energy description.

\subsection{1-loop determinants and the Witten index}

A simple check that we can perform of the proposed 1d quantum mechanics \eqref{full 1d L superspace} is that its Witten index matches the TT index of the 3d theory, at leading order at large $N$. Indeed, since the Witten index is invariant under RG flow, it must be the same in the UV 3d theory and in the IR 1d effective description. Matching of the indices also ensures that the ground-state degeneracy of the quantum mechanics reproduces the entropy of BPS black holes.

The Witten index of an $\cN=2$ supersymmetric quantum mechanics is defined in exactly the same way as the TT index in \eqref{TTI}. In the Lagrangian formulation, the chemical potentials $\Delta_a$ are introduced as twisted boundary conditions on the fields. For a class of these models, the Witten index has been computed in \cite{Hori:2014tda} (see also \cite{Hwang:2014uwa, Cordova:2014oxa}), and it takes a Jeffrey-Kirwan contour integral form as in (\ref{index full expression from papers}). We want to make sure that the quantum mechanics (\ref{full 1d L superspace}) reproduces the integrand in (\ref{index full expression from papers}) for the value of $\fm_i$ singled out by the saddle-point approximation.

After fixing the 1d gauge $\partial_t \bigl( A_t^i+\sigma^i \bigr) = 0$, the Wilson line gives a classical contribution $\exp\bigl( i \sum_i k\fm_iu_i \bigr)$, where $u$ is the constant mode of the Wick-rotated $A_t+\sigma$. The chirals $\Xi_{\bar 1}$ and Fermi's $C$ coming from the 3d vector multiplet contribute to the 1-loop determinant as
\be
\cZ_{\Xi_{\bar1}} = \prod_{i\neq j} \biggl( \frac{e^{i \, u_{ij}/2} }{ 1-e^{iu_{ij} }} \biggr)^{\Theta(-q_{ij}-1) \, (-2q_{ij}-1) } \,,\qquad
\cZ_{C} =\prod_{i\neq j} \biggl( \frac{e^{iu_{ij}}-1}{ e^{i \, u_{ij}/2} } \biggr)^{\Theta(q_{ij}) \, (2q_{ij}+1)} \;,
\ee
where $u_{ij}=u_i-u_j$. The exponents come from the $2l+1$ degeneracy in each $\rSU(2)$ representation of highest weight $l$, and the $\Theta$ functions ensure that nontrivial contributions only enter when the multiplets exist. Recalling that $q_{ij} \neq 0$ for $i\neq j$, their product simplifies:
\be
\label{3d vect 1-loop}
\cZ_{\Xi_{\bar1}} \, \cZ_{C} = (-1)^{\frac{N(N-1)}{2}} \, \prod_{i\neq j} \biggl(1-\frac{z_i}{z_j} \biggr) \;,
\ee
where $z_i=e^{iu_i}$. The result above matches (up to an inconsequential sign) the 1-loop determinant of a 3d vector multiplet given in \cite{Benini:2015noa} and appearing in (\ref{index full expression from papers}).%
\footnote{The 1-loop determinant of a Fermi multiplet has a sign ambiguity coming from the assignment of fermion number to states in the fermionic Fock space. We have fixed this ambiguity in a specific way to get \eqref{3d vect 1-loop}, but different conventions are possible. Notice, for example, the different choice made in \eqref{chiral and fermi one loop}.}
As opposed to the indirect Higgsing argument which was used in \cite{Benini:2015noa}, the result here provides an explicit derivation based on a careful gauge-fixing procedure. This computation shows that the ghost multiplet $C^{ij}$ appearing in the quantum mechanics is needed to reproduce the correct degeneracy of BPS states.
Lastly, the chirals $\Phi_{a}$ and Fermis $\cY_a$ coming from the 3d chiral multiplets contribute to the 1-loop determinant as
\be
\label{chiral and fermi one loop}
\cZ_{\Phi_{a}} = \prod_{i,j} \biggl( \frac{e^{i (u_{ij}+\Delta_a)/2} } {1-e^{i(u_{ij}+\Delta_a)}} \biggr)^{\Theta(q_{ij}^a) \, (2q_{ij}^a+1)} ,\qquad
\cZ_{\cY_a} = \prod_{i,j} \biggl( \frac{1-e^{i(u_{ij}+\Delta_a)} }{ e^{i(u_{ij}+\Delta_a)/2}} \biggr)^{\Theta(-q^a_{ij}-1) \, (-2q^a_{ij}-1)} .
\ee
Their product is
\bea
\cZ_{\Phi_{a}} \cZ_{\cY_a}
=\prod_{i,j} \biggl( \frac{e^{i (u_{ij}+\Delta_a)/2} }{ 1- e^{i(u_{ij}+\Delta_a)}} \biggr)^{2q_{ij}^a+1} = \frac{ y_a^{N^2(\fn_a+1)/2} }{ (1-y_a)^{N(\fn_a+1)}}\prod_{i\neq j} \biggl( \frac{z_i-y_az_j}{z_j-y_az_i} \biggr)^{\fm_i} \biggl( 1-y_a\frac{z_i}{z_j} \biggr)^{-\fn_a-1} .
\eea
The  complete integrand is thus
\be
\label{final 1d integrand}
\cZ_\text{tot} = e^{ik \sum_i \fm_i u_i} \; \cZ_{\Xi} \, \cZ_C \prod_a \cZ_{\Phi_a} \, \cZ_{\cY_a} \;,
\ee
matching the integrand in (\ref{index full expression from papers}).

Assuming that the JK contour integral formula for the 1d index gets contribution from the same saddle point as in 3d, equality of (\ref{final 1d integrand}) with the integrand in (\ref{index full expression from papers}) guarantees that a large $N$ saddle-point computation of the 3d TT index matches a saddle-point computation of the 1d Witten index, at leading order in $N$ (see Section~\ref{sec: basic idea}).

\section{Stability under quantum corrections}
\label{sec: stability}

The gauge-fixing action $\delta\Psi_\text{gf}$ preserves the real supercharge $\delta$, $\rU(1)_F^2$, and $\rSU(2)$. We first use the $\delta$ invariance of the full action to show that the fermion $\wt c^{\;ij}_m$ only has gauge interactions. This allows us to focus on fields other than $\wt c^{\;ij}_m$. Although the gauge fixing breaks $Q$, $\overline{Q}$, and $\rU(1)_R$, we will then give arguments for why they should be preserved in the effective action. The key observation will be \eqref{d eq s on res O}. Finally, we will use all the symmetries $Q$, $\overline{Q}$, $\rU(1)_F^2$, $\rU(1)_R$ and $\rSU(2)$ to discuss which classical and quantum corrections to the quantum mechanics computed in Section~\ref{sec: QM} one could expect.

\subsection[Interactions involving \texorpdfstring{$\wt c$}{c}]{Interactions involving \matht{\wt c}}
\label{subsec: interactions C}

Using the fermionic symmetry $\delta$, we can argue that the part of the Lagrangian involving the fermions $\wt c^{\;ij}_m$ cannot be anything other than \eqref{full 1d L} at low energies. Let $\langle\cdot\rangle_{\delta}$ denote the gauge-fixed path integral, as in \eqref{d eq s on res O}. For $i,j$ such that $q_{ij}>0$, we consider the quantity
\be
\bigl\langle \, \overline{\wt c^{\;ij}_m}(t) \, D_t^+ \, \wt c^{\;ij}_m(t') \bigr\rangle_\delta = \bigl\langle \, \overline{\wt c^{\;ij}_m}(t) \, \delta b^{ij}_m(t')\bigr\rangle_\delta - \bigl\langle \, \overline{\wt c^{\;ij}_m}(t) \, \delta_\text{gauge}(R) \, \wt c^{\;ij}_{m}(t') \bigr\rangle_\delta \approx \bigl\langle \, \overline{\wt c^{\;ij}_m}(t) \, \delta b^{ij}_m(t') \bigr\rangle_\delta \;.
\ee
Here $b^{ij}_m$ is the $l=q_{ij}$ mode of the auxiliary field $b$ in the gauge-fixing complex. In the first equality we used \eqref{brst restricted cartan} and \eqref{cQ on gf complex}. The approximate equality $\approx$ only holds in the IR limit because the term that was discarded is a correlation function involving massive ghosts $c$ in $R=-\frac{1}{2}\{c,c\}_\fr$, which is exponentially suppressed at large $t-t'$. We continue using the Leibniz rule on $\delta$ and the fact that $\delta$-exact correlators vanish, to write
\be
\bigl\langle \, \overline{\wt c^{\;ij}_m}(t) \, \delta b^{ij}_m(t') \bigr\rangle_\delta = - \bigl\langle \delta \overline{\wt c^{\;ij}_m}(t) \, b^{ij}_m(t') \bigr\rangle_\delta = i \bigl\langle \overline{b^{ij}_m}(t) \, b^{ij}_m(t') \bigr\rangle_\delta \;.
\ee
The path integral over $b^{ij}_m$ is quadratic and can be done exactly, yielding
\be
\bigl\langle \, \overline{\wt c^{\;ij}_m}(t) \, D_t^+ \, \wt c^{ij}_m(t') \bigr\rangle_\delta \approx i \bigl\langle \overline{b^{ij}_m}(t) \, b^{ij}_m(t') \bigr\rangle_\delta = -\delta(t-t') + i \bigl\langle \overline{\cO_H}(t) \, \cO_H(t') \bigr\rangle_\delta \approx -\delta(t-t')\,,
\ee
where
\be
\cO_H = \sqrt{\frac{q_{ij}}{ \xi R^2}} \; A^{ij}_{1,q_{ij},m} - \frac{e_\text{3d}}{R} \; \{\wt c,c\}^{ij}_{l=q_{ij},m} \;.
\ee
The expression $\{\wt c,c\}^{ij}_{l=q_{ij},m}$ stands for the $\bigl( l=q_{ij},m \bigr)$ mode of $\{\wt c,c\}^{ij}$. Both terms inside $\cO_H$ contain massive fields only, therefore $\bigl\langle \overline{\cO_H}(t) \, \cO_H(t') \bigr\rangle_\delta$ is exponentially suppressed at large distances and the approximation holds to increasing accuracy in the IR. Using only symmetry arguments for $\delta$, we have shown that $\wt c^{\;ij}_m$ must satisfy the Schwinger-Dyson equation derived from \eqref{full 1d L} in the IR limit. Any modification of \eqref{full 1d L} containing $\wt c^{\;ij}_m$ would change the Schwinger-Dyson equation, and can thus be excluded.

\subsection[Presence of \texorpdfstring{$\cN=2$}{N=2} supersymmetry and R-symmetry]{Presence of \matht{\cN=2} supersymmetry and R-symmetry}
\label{subsec: 1d susy}

Having taken care of $\wt c^{\;ij}_m$, we want to constrain the effective Lagrangian for the remaining fields. Here we show that in the IR it must preserve 1d $\cN=2$ supersymmetry and $\rU(1)_R$, even though these symmetries are broken by the gauge-fixing term $\delta\Psi_\text{gf}$.

First, we show that the Ward identities for the supercharges $Q$ and $\overline{Q}$ are satisfied on correlators $\cO$ constructed from 1d fields excluding $\wt c^{\;ij}_m$, which are modes of physical fields in 3d. More precisely, we show that $\langle Q\cO\rangle_\delta\approx 0$ (and analogously for $\overline{Q}$). As before, approximate equalities $\approx$ hold in the IR limit. Firstly, since $\cO$ is constructed from modes of physical fields, it has $n_g=0$, and the same goes for $Q\cO$. Then \eqref{d eq s on res O} tells us that $\langle Q\cO\rangle_\delta=\langle Q\cO\rangle_s$. It remains to show that $\langle Q\cO\rangle_s\approx 0$.  

We then follow the standard procedure to derive a Ward identity. In the path integral $\langle\cO\rangle_s$ we perform a field redefinition $X' = X + \epsilon \, QX$ on physical fields $X$ in the form of a supersymmetry transformation, while keeping the fields $Y$ in the gauge-fixing complex unchanged. Let $S_\text{ph}$ be the original action before gauge fixing. At first order in $\epsilon$ we get
\bea
\langle \cO \rangle_s &= \int\! \cD\phi \; \cO \, e^{i \left( S_\text{ph} + s\Psi_\text{gf} \right) } = \int\! \cD\phi \;  \bigl( \cO + \epsilon \, Q\cO \bigr) \, e^{i \left( S_\text{ph} + s\Psi_\text{gf} \right) - i \epsilon \, s \, Q\Psi_\text{gf}} \\
&=\langle\cO\rangle_s + \epsilon \, \Bigl( \langle Q\cO\rangle_s - i \langle\cO\, s \, Q\Psi_\text{gf}\rangle_s \Bigr) + \ldots
\eea
Suppose that $\cO$ is fermionic so that $\langle Q\cO\rangle_s\approx 0$ is a non-trivial statement. At order $\epsilon$, that equality implies
\be
\langle Q\cO\rangle_s = i \langle \cO \, s \, Q\Psi_\text{gf} \rangle_s =
i \bigl\langle (s\cO) \, (Q\Psi_\text{gf}) \bigr\rangle_s = i \Bigl\langle \bigl( \delta_\text{gauge}(c) \cO \bigr) \, (Q\Psi_\text{gf}) \Bigr\rangle_s \approx 0 \;.
\ee
We used that $\bigl\langle s (\cO\,Q\Psi_\text{gf}) \bigr\rangle_s=0$ because the action $S_\text{ph} + s\Psi_\text{gf}$ is $s$-closed. In the last step, $c$ is massive and therefore its correlators vanish in the IR. We can now use \eqref{d eq s on res O} to conclude that $\langle Q\cO\rangle_\delta=\langle Q\cO\rangle_s\approx 0$. 

The Ward identity for $\rU(1)_R$ can be derived with much less work. Any $\cO$ built out of 1d fields excluding $\wt c^{\;ij}_m$ has $n_g=0$, and $\langle \cO\rangle_\delta=\langle\cO\rangle_s$ by \eqref{d eq s on res O}. Since $s\Psi_\text{gf}$ is $\rU(1)_R$ invariant, $\langle\cO\rangle_s=0$ if $\cO$ has nonzero R-charge. Therefore $\langle \cO\rangle_\delta=0$ if $\cO$ has nonzero R-charge.  

Given the above Ward identities, any effective action in the IR should have 1d $\cN=2$ supersymmetry and $\rU(1)_R$ symmetry. For $\rU(1)_R$, we can see this in the following way (the argument for supersymmetry is analogous). Formally, the exact effective action for the fields in the quantum mechanics is given by 
\be
e^{i \left( S_0 + \sum_{r\neq 0} S_r \right)} = \int\! \cD\phi_H \; e^{i \left( S_\text{ph} + \delta \Psi_\text{gf} \right) } \;,
\ee
where $S_r$, $r\in\bZ$ are pieces of the effective action with R-charge $r$, and $\phi_H$ are the massive fields which are integrated out. Note that the $\rU(1)_R$ violating pieces $S_{r\neq 0}$ can in principle be generated%
\footnote{What happens instead, as indicated by the argument below, is that all the generated symmetry-violating pieces involve fields at the scale of the massive ghosts $c$ or higher.}
because $\delta\Psi_\text{gf}$ breaks $\rU(1)_R$. However, the presence of any $S_{r\neq 0}$ would generically violate the $\rU(1)_R$ Ward identity. Indeed, consider an operator $\cO$ with R-charge $-r^*$ which is constructed out of the fields $\phi_L$ in the quantum mechanics excluding $\wt c^{\;ij}_m$. The Ward identity tells us that $\langle\cO\rangle_\delta=0$. However, computing $\langle\cO\rangle_\delta$ directly gives:
\bea
\langle\cO\rangle_\delta &= \int\! \cD\phi_L \; \cO \, e^{i(S_0+\sum_{r\neq0}S_r)} =
\sum_{n=0}^\infty \frac{i^n}{n!} \int\! \cD\phi_L \; \cO \, \biggl( \sum\nolimits_{r\neq0}S_r \biggr)^{\!\!n} \, e^{iS_0}
\\
&= \sum_{n=0}^\infty \frac{i^n}{n!} \int\! \cD\phi_L \; \cO \, \biggl[ \Bigl( \sum\nolimits_{r\neq0}S_r \Bigr)^n \biggr]_{r=r^*} \, e^{iS_0} \neq 0 \;.
\eea
Here $\bigl[ \dots \bigr]_{r=r^*}$ means the sum of the terms with R-charge $r^*$, which, at least for $n=1$, is non-empty if $S_{r^*}$ is present in the effective action. It follows that, in the latter case, the expectation value of $\cO$ would generically be non-zero, violating the Ward identity.

\subsection{Symmetry constraints}
\label{subsec: corrections}

We can use $\rU(1)_R$, $Q$, and $\overline{Q}$, together with the other symmetries, to constrain the interactions that could appear in the effective action. We work within the framework of \cite{Hori:2014tda} (see also \cite{Witten:1993yc}), where the interactions in an $\cN=2$ supersymmetric quantum mechanics are specified by $E$ and $J$ functions, \ie, holomorphic functions of chiral superfields satisfying \eqref{EJ susy condition}. The argument in Section \ref{subsec: interactions C} tells us that the $E$ and $J$ functions corresponding to $C$ must vanish in the IR:
\be
\label{E and J for C}
E^{ij}_{C,m}=0\;,\qquad J^{ji}_{C,-m}=0\;.
\ee
Besides, $C$ cannot appear in the E- and J-terms of the other Fermi multiplets $\cY_a$. Since it is already true classically that $\overline{D}\cY_a\neq 0$ for every $\cY_a$, one expects that $\cY_a$'s cannot appear in $E$ or $J$ functions, because quantum corrections would need to be finely tuned to make them chiral. Therefore, $E$ and $J$ functions can only be holomorphic functions of $\Phi_a$ and $\Xi_{\bar1}$.

Let us neglect gauge charges and $\rSU(2)$ invariance momentarily, and suppress the corresponding indices. To have the same $\rU(1)_F^2$ charges as $\cY_a$ and R-charge $R(\cY_a)+1$, the $E$ function corresponding to $\cY_a$ must have the simple form
\be
E_a \,\sim\, \Phi_a\, h_E(\Xi_{\bar1}) \;,
\ee
where $h_E$ is a holomorphic function. Fleshing out the gauge and $\rSU(2)$ indices, we enforce that $E^{ij}_{a,m}$ have the same gauge charges and be in the same $\rSU(2)$ representation as $\cY^{ij}_{a,m}$. Imposing those conditions on the constant term in $h_E$, we get $E^{ij}_{a,m}\sim\Phi_{a,m}^{ij}$. However, such a term is impossible because $\cY^{ij}_{a,m}$ (and therefore $E^{ij}_{a,m}$) exists when $q^a_{ij}\leq -1$, while $\Phi_{a,m}^{ij}$ exists when $q^a_{ij}\geq 0$. The two conditions are mutually exclusive.%
\footnote{Because of this, the chirals and Fermi's in the quantum mechanics cannot gap each other out through a dynamically generated E-term.}
We remain with terms in $h_E$ which are at least linear in $\Xi_{\bar1}$. Writing the first term explicitly, we find:
\bea
\label{gen E LO}
E_{a,m}^{ij} &= \;\;\, \sum_k e^{ij}_{a,k} \, \Theta(q^a_{kj}) \sum_{|m'|\leq q^a_{kj}} C \bigl( \begin{smallmatrix}
|q_{ik}|-1 & q^a_{kj} & |q_{ij}^a|-1 \\
m-m' & m' & m
\end{smallmatrix}\bigr) \; \Xi^{ik}_{\bar1,m-m'} \, \Phi^{kj}_{a,m'} \\
&\quad + \sum_k \wt e^{\;ij}_{a,k} \, \Theta(q^a_{ik}) \sum_{|m'|\leq q^a_{ik}} C\bigl(\begin{smallmatrix}
|q_{kj}|-1 & q^a_{ik} & |q_{ij}^a|-1 \\
m-m' & m' & m
\end{smallmatrix}\bigr) \; \Phi^{ik}_{a,m'} \, \Xi^{kj}_{\bar1,m-m'} + \ldots
\eea
The $\Theta$ functions are necessary to ensure that the fields $\Phi_a$ and $\Xi_{\bar1}$ exist with their corresponding gauge charges. The Clebsch-Gordan coefficients project the product of $\Xi_{\bar1}$ and $\Phi_a$ to the same $\rSU(2)$ representation carried by $E_{a,m}^{ij}$, \ie, $l=|q_{ij}^a|-1$. Finally, $e^{ij}_{a,k}$ and $\wt e^{\;ij}_{a,k}$ are free coefficients. Analogously, terms of the form $\Phi_a(\Xi_{\overline{1}})^{n\geq 2}$ should contain a product of $n$ Clebsch-Gordan coefficients and balanced gauge indices.   

When constraining the functions $J_a$ corresponding to $\cY_a$, we again start with $\rU(1)_F^2$ and $\rU(1)_R$. Now, $J_a$ must have the opposite $\rU(1)_F^2$ charges to $\cY_a$, and R-charge $-R(\cY_a)+1$. Thus $J_a$ must have the form 
\be
J_a \,\sim\, \Phi_b \, \Phi_c \, h_J(\Xi_{\bar1}) \;,
\ee
where $b$ and $c$ are different flavor indices complementary to $a$. Again, $h_{J}$ is a holomorphic function. We should impose gauge and $\rSU(2)$ invariance. Expanding $h_J$ as a polynomial in $\Xi_{\bar1}$ and writing the first (constant) term explicitly, we have   
\begin{align}
\label{gen J LO}
& J^{ji}_{a,-m} = \sum_k \Biggl[ \frac{ \lambda^{ji}_{a,k} }{ \scriptstyle\sqrt{2|q^a_{ij}|-1}} \, \Theta(q^b_{jk}) \, \Theta(q^c_{ki}) \!\!\! \sum_{\substack{|m'|\leq q^b_{jk}\\ |m+m'|\leq q^c_{ki}}} \!\!\! {\scriptstyle (-1)^{-q^a_{ij}-1-m}} \, C\bigl( \begin{smallmatrix}
q^b_{jk} & q^c_{ki} & |q^a_{ij}|-1 \\
m' & -m-m' & -m
\end{smallmatrix}\bigr) \, \Phi^{jk}_{b,m'} \, \Phi^{ki}_{c,-m-m'}
\nn \\
& + \frac{\wt\lambda^{ji}_{a,k} }{ \scriptstyle\sqrt{2|q^a_{ij}|-1}} \, \Theta(q^c_{jk}) \, \Theta(q^b_{ki}) \!\!\! \sum_{\substack{|m'|\leq q^c_{jk}\\ |m+m'|\leq q^b_{ki}}} \!\!\!\! {\scriptstyle (-1)^{-q^a_{ij}-1-m}} \, C\bigl(\begin{smallmatrix}
q^c_{jk} & q^b_{ki} & |q^a_{ij}|-1 \\
m' & -m-m' & -m
\end{smallmatrix}\bigr) \, \Phi^{jk}_{c,m'} \, \Phi^{ki}_{b,-m-m'} \Biggr] + ... 
\end{align}
The indices $b$ and $c$ above are chosen such that $\epsilon^{abc}=1$, and the factor $1/\sqrt{2|q_{a}^{ij}|-1}$ was added for later convenience. Similarly to the $E$ function, there are two unfixed coefficients $\lambda^{ji}_{a,k}$ and $\wt\lambda^{ji}_{a,k}$. Terms of the form $\Phi_b\Phi_c(\Xi_{\bar1})^{n\geq 1}$ should contain a product of $n+1$ Clebsch-Gordan coefficients and gauge indices should be balanced.

Lastly, supersymmetry requires \eqref{EJ susy condition}. If we restrict $E^{ij}_{a,m}$ and $J^{ji}_{a,-m}$ to the terms written explicitly in \eqref{gen E LO} and \eqref{gen J LO}, this condition implies
\bea
& e^{ij}_{a,k} \, \lambda^{ji}_{a,l} + \wt e^{\;lk}_{c,i} \, \lambda^{kl}_{c,j} = 0 \qquad \text{if} \quad \epsilon^{abc} = 1 \quad \text{and} \quad \Theta(q^a_{kj}) \, \Theta(q^b_{jl}) \, \Theta(q^c_{li}) = 1 \\
& e^{ij}_{a,k} \, \wt\lambda^{ji}_{a,l} + \wt e^{\;lk}_{b,i} \, \wt\lambda^{kl}_{b,j} = 0 \qquad \text{if} \quad \epsilon^{abc} = 1 \quad \text{and}\quad \Theta(q^a_{kj}) \, \Theta(q^c_{jl}) \, \Theta(q^b_{li}) = 1 \;.
\eea
Note that none of the indices above are summed over. The coefficients in \eqref{E-term} and \eqref{J-term} that we found from the reduction satisfy these equations, but they might not be the unique choice. The constraint \eqref{EJ susy condition} would also have to be enforced on terms with higher powers of $\Xi_{\bar1}$, strongly constraining their coefficients.

From classical scaling arguments, we are not able to rule out the presence in \eqref{gen E LO} and \eqref{gen J LO} of terms which have higher powers of $\Xi_{\bar1}$. They could be generated both at tree and at loop level. It would be consistent to neglect those terms if $\Xi_{\bar1}$, which is classically dimensionless, gained a positive anomalous dimension. This is indeed the case for classically dimensionless fermions in SYK models such as \cite{Fu:2016vas, Heydeman:2022lse}, but it remains to be checked in the theory discussed here.

\section*{Acknowledgements}

We are grateful to Dionysios Anninos, Matthew Heydeman, Luca Iliesiu, Juan Maldacena, Gustavo Turiaci, and Alberto Zaffaroni for very useful discussions, and in particular to Itamar Shamir for collaboration in the early stages of this work. We gratefully acknowledge support from the Simons Center for Geometry and Physics, Stony Brook University and the organizers of the SCGP workshop ``Supersymmetric black holes, holography and microstate counting'', as well as the Galileo Galilei Institute (Firenze), the INFN and the organizers of the workshop ``Topological properties of gauge theories and their applications to high-energy and condensed-matter physics'', where part of this work was done.
S.S. especially thanks KU Leuven for hospitality, and Z.Z. thanks the hospitality of the Department of Mathematics at Durham University, during part of this work.
F.B., S.S and Z.Z. are partially supported
by the ERC-COG grant NP-QFT No.~864583 ``Non-perturbative dynamics of quantum fields: from new deconfined phases of matter to quantum black holes'',
by the MIUR-SIR grant RBSI1471GJ,
by the MIUR-FARE grant EmGrav No.~R20E8NR3HX ``The Emergence of Quantum Gravity from Strong Coupling Dynamics'',
by the MIUR-PRIN contract 2015 MP2CX4,
as well as by the INFN ``Iniziativa Specifica ST\&FI''.

\appendix

\section{Large \matht{N} limit computations}
\label{app: large N}

Let us start by studying the first line of \eqref{V prime with m and Omega}, and in particular the terms involving the $\Li_1$ function, whose definition and properties can be found in Appendix~\ref{subsec: polylogs}. We first perform the sum over $j$ (that becomes an integral over $t'$), leaving the sum over $i$ (that becomes an integral over $t$) untouched.

The integral in $t'$ has to be broken in two parts, above and below $t_{\pm\Delta} \equiv t\pm N^{-\alpha}\im\Delta$. When $\im (u_{ji} \mp \Delta)>0$ (for one of the two signs), we can use the series expansion \eqref{polylog series}. This allows us to treat the integral above $t_{\pm\Delta}$:
\begin{align}
\sum_j \Theta\bigl( \im( u_{ji} \mp \Delta) \bigr) \Li_1 \Bigl( e^{i(u_{ji} \mp \Delta)} \Bigr) &\,\to\, N \int_{t_{\pm\Delta}} \!\! dt'\rho(t') \sum_{\ell=1}^\infty\frac{1}{\ell} \, e^{i \ell \left(\str u(t') \,-\, u(t) \,\mp\, \Delta \right)} \nn \\
&\;\,\equiv\; N \sum_{\ell=1}^\infty \frac{e^{\mp i \ell\Delta}}{\ell} \, I_{\text{L},\ell}[\rho](t,\Delta) \;.
\label{first integral complex}
\end{align}
In Appendix~\ref{subsec: large N integrals} we define and manipulate these integrals. Using \eqref{IL tilde final}, we write \eqref{first integral complex} as
\begin{align}
\label{first Li contribution}
& \eqref{first integral complex} = N^{1-\alpha} \Li_2 \Bigl( e^{\mp i (\re \Delta - \dot{v}\im \Delta )} \Bigr) \, \frac{\rho}{1-i\dot{v}} + {} \\
{} + &N^{1-2\alpha} \biggl[ \Li_3 \Bigl( e^{\mp i (\re \Delta - \dot{v}\im \Delta )} \Bigr) \pm (\im\Delta)(1 - i \dot{v}) \Li_2 \Bigl( e^{\mp i (\re \Delta - \dot{v}\im \Delta )} \Bigr) \biggr] \biggl[ \frac{\dot{\rho} }{ (1 - i\dot{v})^2 } + \frac{ i \rho \, \ddot{v} }{ (1-i\dot{v})^3 } \biggr] \nn \\
{} + &\frac{i}{2} N^{1-2\alpha} (\im\Delta)^2 (1 - i\dot{v})^2 \Li_1\Bigl( e^{\mp i (\re \Delta - \dot{v}\im \Delta )} \Bigr) \frac{\rho \, \ddot{v}}{(1-i\dot{v})^3} +\cO(N^{1-3\alpha}) \;. \nn
\end{align}
When $\im (u_{ji} \mp \Delta)<0$, the steps above are not applicable because the series expansion for $\Li_1$ does not converge, but we can use \eqref{Li inversion} so that
\be\label{applied inversion}
\Li_1 \Bigl( e^{i(u_{ji}\mp\Delta)} \Bigr) = \Li_1 \Bigl( e^{i(u_{ij}\pm\Delta)} \Bigr) - i \bigl( u_{ji} \mp \Delta - \pi \bigr) \;.
\ee
Now the $\Li_1$ terms on the RHS can be analyzed in the same way as before using \eqref{IU tilde final}:
\begin{align}
& \sum_j \Theta\bigl( \im( u_{ij} \pm \Delta) \bigr) \Li_1 \Bigl( e^{i(u_{ij} \pm \Delta)} \Bigr) \to
N \!\!\int^{t_{\pm\Delta}} \!\!\! dt' \, \rho(t') \sum_{\ell=1}^\infty \frac{e^{i\ell \left(\str u(t) - u(t') \pm \Delta \right) }}\ell = N \sum_{\ell=1}^\infty \frac{e^{\pm i \ell \Delta}}{\ell} \, I_{\text{U},\ell}[\rho] \nn \\
& = N^{1-\alpha} \Li_2 \Bigl( e^{\pm i (\re \Delta - \dot{v}\im \Delta )} \Bigr)\, \frac{\rho}{1-i\dot{v}} \nn \\
& -N^{1-2\alpha} \biggl[ \Li_3 \Bigl( e^{\pm i (\re \Delta - \dot{v}\im \Delta )} \Bigr) \mp (\im\Delta) (1 - i \dot{v}) \Li_2 \Bigl( e^{\pm i (\re \Delta - \dot{v}\im \Delta )} \Bigr) \biggr] \biggl[ \frac{\dot{\rho}}{(1-i\dot{v})^2} + \frac{i\rho \, \ddot{v}}{(1-i\dot{v})^3} \biggr] \nn \\
&- \frac{i}{2} N^{1-2\alpha} (\im\Delta)^2 (1-i\dot{v})^2 \Li_1 \Bigl( e^{\pm i (\re \Delta - \dot{v}\im \Delta )} \Bigr) \frac{\rho \, \ddot{v}}{(1-i\dot{v})^3} +\cO(N^{-3\alpha}) \;.
\label{second Li contribution}
\end{align}
To obtain the full integral over $t'$, the contributions (\ref{first Li contribution}) and (\ref{second Li contribution}) with upper sign must be summed with minus the ones with lower sign, and the result can be simplified using (\ref{Li inversion}). As in (\ref{V prime with m and Omega}), we then integrate over $t$ together with $\fm(t)$, and sum over $a=1,2,3$. We obtain:
\begin{align}
\label{contribution series}
& i N^{2-2\alpha} \int\! dt \; \frac{i \, \fm \, \rho^2 \, \ddot{v}}{(1-i\dot{v})^3} \, \sum_{a=1}^3 \, (\im\Delta_a)^2 (1-i \dot{v})^2 \, g''_+ \bigl( \re\Delta_a -\dot{v}\im\Delta_a \bigr) \\
& -i N^{2-2\alpha} \int\! dt \; \fm \, \frac{d}{dt} \biggl[ \frac{\rho^2}{(1-i\dot{v})^2} \biggr] \, \sum_{a=1}^3\bigg[ g_+ \bigl( \re\Delta_a - \dot{v}\im\Delta_a \bigr) \nn \\
&\hspace{7cm} + i \, (\im\Delta_a) \, (1 - i \dot{v}) \, g'_+ \bigl( \re\Delta_a-\dot{v}\im\Delta_a \bigr) \bigg]\;. \nn
\end{align}
The function $g_+(u)$ is defined in (\ref{def g+ app}). It remains to add the contribution from the second term on the RHS of \eqref{applied inversion}. We choose the integer ambiguities $n_i$ in \eqref{V prime with m and Omega} such that
\be
\label{amibiguities}
\pi(N - 2n_i)= - \sum_{a=1}^3 \sum_{j=1}^N \biggl[ 2\pi \Bigl( \Theta\bigl( \im(u_{ij} + \Delta_a) \bigr) - \Theta \bigl( \im u_{ij} \bigr) \Bigr) + 2\Delta_a \Theta(\im u_{ij}) \bigg]+\cO(1) \;.
\ee
The subleading $\cO(1)$ term accounts for the possibility that $N$ might be odd and we would not be able to cancel it completely. The contributions from the second term on the RHS of \eqref{applied inversion} and from (\ref{amibiguities}) sum up to 
\begin{align}
\label{contribution linear}
& i \sum_{a,i,j} \fm_i \biggl[ \Bigl( \Theta\bigl( \im(u_{ij} + \Delta_a) \bigr) - \Theta( \im u_{ij}) \Bigr) \bigl( -u_{ji} + \Delta_a - \pi \bigr) + {} \\
& \hspace{2cm} + \Bigl( \Theta \bigl( \im( u_{ij} - \Delta_a) \bigr) - \Theta(\im u_{ij}) \Bigr) \bigl( u_{ji}+\Delta_a-\pi \bigr) \biggr] \nn \\
&\quad = iN^2 \sum_{a=1}^3 \sum_{+,-} \int\! dt \, \fm(t) \, \rho(t) \int_t^{t_{\pm\Delta_a}} \! dt' \, \rho(t') \Bigl[ \pm N^\alpha \bigl( it-it'+v(t)-v(t') \bigr) + \Delta_a - \pi \Bigr] \;. \nn
\end{align}
In each integral we perform the change of variables $t'=t\pm N^{-\alpha}(\im\Delta_a) \varepsilon$, obtaining:
\begin{align}
& (\text{\ref{contribution linear}}) = i N^{2-\alpha} \sum_{a=1}^3 \sum_{+,-} \im\Delta_a \int\! dt\, \fm(t) \, \rho(t) \int_0^1 \! d\varepsilon \times {} \\
&\times \biggl\{ \pm \rho \Bigl( t \pm N^{-\alpha}(\im\Delta_a) \varepsilon \Bigr) \biggl[ -i (\im\Delta_a) \varepsilon \mp N^{\alpha} \, v\Bigl( t \pm N^{-\alpha}(\im\Delta_a) \varepsilon \Bigr) \pm N^\alpha v(t)+\Delta_a-\pi \biggr] \biggr\} \nn \;.
\end{align}
We expand $\rho$ and $v$ in Taylor series and keep only the terms at leading order. Then we integrate in $\varepsilon$ and use that $g_+''(\Delta) = \Delta - \pi$. We obtain the expression:
\begin{align}
\label{contribution linear final}
& (\text{\ref{contribution linear}}) = i N^{2-2\alpha} \sum_{a=1}^3 (\im\Delta_a)^2 \!\int\! dt \, \fm\,\bigg\{ \rho \, \dot\rho  \, g_+'' \bigl( \re\Delta_a - \dot{v}\im\Delta_a \bigr) + {} \\
&\hspace{6.5cm} + i \, \frac{\im\Delta_a}6 \, \frac{d}{dt} \biggl[ \frac{\rho^2}{(1-i\dot{v})^2} \biggr] (1 - i\dot{v})^3 \bigg\}+\mathcal{O}(\fm N^{2-3\alpha}) \;. \nn
\end{align}
We sum (\ref{contribution series}) and (\ref{contribution linear final}). We notice that the various terms can be organized into the Taylor series of $g_+(\Delta_a)$ around the point $\re(\Delta_a) - \dot v \im(\Delta_a)$, which has four terms because $g_+$ is a cubic polynomial. We obtain the compact expression
\be
(\text{\ref{contribution series}}) + (\text{\ref{contribution linear final}}) = - i N^{2-2\alpha} \, G(\Delta) \int\! dt \, \fm \, \frac{d}{dt} \biggl[ \frac{\rho^2}{(1-i\dot{v})^2} \biggr] + \cO\bigl( \fm N^{2-3\alpha},1 \bigr) \;,
\ee
where $G(\Delta)$ is the function defined in (\ref{def G and f+}). It remains to add the first term on the RHS of the first line of \eqref{V prime with m and Omega}. We obtain the final expression:
\be
\label{final mV' app}
\int\! dt\, \fm \, V' = ikN \!\int\! dt\, \rho \, \fm \, u + iN^{2-2\alpha} \, G(\Delta) \!\int\! dt\, \frac{\dot\fm \, \rho^2}{(1-i\dot{v})^2} + \cO\bigl( \fm N^{2-3\alpha} \bigr)\;.
\ee

We apply the same steps to obtain the large $N$ limit of $\Omega$ in \eqref{V prime with m and Omega}. To avoid repetition, we only present the result. We set the integer ambiguity $M$ to $N/2+\cO(1)$. We obtain:
\be
\label{final Omega app}
\Omega = - N^{2-\alpha} \, f_+(\fn,\Delta) \int\! dt \, \frac{\rho^2}{ 1-i\dot{v} } + \cO\bigl( N^{2-2\alpha} \bigr) \;,
\ee
where the function $f_+(\fn,\Delta)$ is defined in (\ref{def G and f+}).

\subsection{Solutions to the saddle-point equations}
\label{sec: sol saddle point}

In this appendix we solve the saddle-point equations (\ref{extremization 1})--(\ref{extremization 3}), in the original parametrization in which $v(t)$ is a real function. Let us first solve \eqref{extremization 3}. After integrating to
\be
k \, (it+v)^2 + \frac{4G\rho}{i+\dot{v}} = A\in\bC \;,
\ee
its real and imaginary parts give
\be
\label{diff_eq_rho_v}
4\rho = - \bigl( 1+\dot{v}^2 \bigr) \im\Bigl[ G^{-1} \, \bigl( A - k \, (it+ v)^2 \bigr) \Bigr] \;,\quad \dot{v} = - \frac{ \re \bigl[ G^{-1} \bigl( A - k (it+ v)^2 \bigr) \bigr] }{ \im \bigl[ G^{-1} \bigl( A - k (it+ v)^2 \bigr) \bigr] } \;.
\ee
We impose that $\rho$ is integrable. This necessarily implies that $\rho\rightarrow 0$ as $t\rightarrow\pm\infty$, or that $\rho$ is defined on compact intervals where $\rho$ is zero at the endpoints. At infinity, or at an endpoint, $\rho=0$ implies $A - k \, (it+v)^2 = 0$. By considering real and imaginary parts, we see that this equation cannot be satisfied as $t\rightarrow\pm\infty$, and $\rho$ must have compact support.  In order for $\rho$ to have two endpoints $t_\pm$ and be defined on the interval $[t_-, t_+]$, $A$ cannot be on the positive real axis. Let $A^\frac12$ be the square root whose imaginary part is positive. The boundary conditions are
\be
\label{boundary_conditions_v}
t_{\pm} = \pm \, k^{-\frac{1}{2}} \, \im (A^\frac{1}{2}) \;,\qquad\qquad v(t_\pm) = \pm \, k^{-\frac{1}{2}} \, \re (A^\frac{1}{2}) \;.
\ee
We then solve the equation for $\dot{v}$ in \eqref{diff_eq_rho_v} using \eqref{boundary_conditions_v} as boundary conditions. The equation can be rewritten and integrated to
\be
\label{im part eq for v}
\im \biggl[ G^{-1} \, (it+v) \, \biggl( A - \frac{k}{3} \, (it+v)^2 \biggr) \biggr] = D \;,
\ee
where $D\in\mathbb{R}$ is an integration constant. The boundary conditions \eqref{boundary_conditions_v} imply $D=0$ and $\im \bigl( G^{-1}A^\frac{3}{2} \bigr) = 0$. Using a real constant $B$ to parametrize the real part of $G^{-1}A^\frac{3}{2}$, we write 
\be
A = k \, (BG)^\frac{2}{3} \;,\qquad B\in\bR \;,
\ee
where $k$ is included for convenience. It is important to keep in mind that there are $3$ branches for $G^\frac{1}{3}$ and the same branch is to be used in every expression. There is a triplet of solutions at this point. The equation \eqref{im part eq for v} can be written as
\be
0 = \im\bigl( G^{-\frac{1}{3}} (it+v) \bigr) \, \biggl[ 3B^\frac{2}{3} + \Bigl( \im\bigl( G^{-\frac{1}{3}} (it+v) \bigr) \Bigr)^2 - 3 \Bigl( \re\bigl( G^{-\frac{1}{3}} (it+v)\bigr) \Bigr)^2 \biggr] \;.
\ee
The solutions obtained by setting to zero the square bracket lead to profiles for $\rho$ with a single zero, and so they have to be discarded. We remain with
\be
\im \Bigl( G^{-\frac{1}{3}} \, (it+v) \Bigr) = 0 \qquad \Rightarrow \qquad v(t) = \frac{\re G^{\frac{1}{3}}}{\im G^{\frac{1}{3}}} \; t \;,
\ee
which through (\ref{diff_eq_rho_v}) gives the following profile for $\rho$:
\be
\rho(t) = \frac{k}{4 \bigl( \im G^{\frac{1}{3}} \bigr)^3} \, \Bigl[ B^\frac{2}{3} \bigl( \im G^{\frac{1}{3}} \bigr)^2 - t^2 \Bigr] \;.
\ee
Requiring that $\rho>0$ within $(t_-,t_+)$ imposes
\be
\label{imposing rho > 0}
\im G^{\frac{1}{3}}>0 \;,
\ee
which restricts the branches we can take for $G^{\frac{1}{3}}$.
Requiring that $\int\! dt\, \rho=1$ fixes $B = 3/k$ and the final result for $u$ and $\rho$ is:
\be
\label{u and rho}
u(t) = N^{\frac{1}{3}} \, \frac{ G^{\frac13} }{ \im G^\frac13 } \, t \;,\quad
\rho(t) = \frac{ (9k)^\frac13 }{ 4\im G^\frac13 } - \frac{k}{ 4 \bigl( \im G^\frac13 \bigr)^3 } \, t^2 \;,\quad
t_{\pm} = \pm \biggl( \frac3k \biggr)^\frac13 \im G^\frac13 \;.
\ee
Notice that if $\Delta_a$ are real and $G>0$, \eqref{imposing rho > 0} fixes the branch of the cube root such that $G^{\frac{1}{3}}$ has phase $e^{\frac{2\pi i}{3}}$, and the solutions for $u$, $\rho$ reduce to those found in \cite{Benini:2017oxt}. We can now solve for $\fm$ using \eqref{extremization 2}. Inserting \eqref{u and rho} for $u$ and $\rho$, the former reduces to:
\be
\bigl( t^2-t_+^2 \bigr) \,\ddot{\wh\fm} + 4t\,\dot{\wh\fm} + 2\wh\fm = \frac{d^2}{dt^2} \Bigl[ \bigl( t^2-t_+^2 \bigr) \, \wh\fm \Bigr] = - 2\,\frac{f_+}{G} \, (it+v) \;,
\ee
whose general solution is
\be
\wh\fm(t) = - \frac{1}{\bigl( t^2-t_+^2 \bigr)} \, \frac{f_+}{3G} \, \frac{G^\frac13}{\im G^\frac13} \, \bigl( t^3 + Ct + D \bigr) \;,
\ee
where $C$ and $D$ are integration constants. The requirement that $\fm$ has compact image, namely that it does not diverge at $t=t_\pm$, fixes $C = - t_+^2$ and $D=0$. This leads to the simple solution
\be
\fm(t) = - \frac{f_+}{3G} \, u(t) \;.
\ee
One can then verify that \eqref{extremization 1} is automatically solved, with the following value for the Lagrange multiplier:
\be
\mu = i f_+ \, \biggl( \frac{k}{3G} \biggr)^\frac13 \;.
\ee
The solution can be expressed more neatly by making use of the reparameterization symmetry \eqref{reparametrization}, performing the transformation $t = (3/k)^{1/3} (\im G^{1/3}) \, t'$. This brings the solution to the form (\ref{u m and rho final}), in which primes have been omitted.

\subsection{Polylogarithms}
\label{subsec: polylogs}

The polylogarithms are defined through their Taylor series around $z=0$:
\be
\label{polylog series}
\text{Li}_k(z)=\sum\nolimits_{\ell = 1}^\infty \frac{z^\ell}{\ell^k} \;,
\ee
which is absolutely convergent for $|z|<1$. This definition can be analytically continued to the whole complex plane, with a branch cut on the real axis from $z=1$ to $z=\infty$. In particular $\Li_1(z)=-\log(1-z)$, where the principal sheet defined by (\ref{polylog series}) is such that $\im \log \in (-\pi, \pi)$.
The functions $\Li_{k\geq 2}$ have an absolutely convergent series (\ref{polylog series}) on the unit circle and are thus continuous at $z=1$, while the functions $\Li_{k\leq 0}$ have a pole at $z=1$ but no branch cut (in particular $\Li_0(z) = \frac z{1-z}$). One can define the single-valued analytic functions
\be
F_k(u) = \Li_k\bigl( 1 - e^{-iu} \bigr)
\ee
defined by (\ref{polylog series}) in the domain $\bigl\lvert 1 - e^{-iu} \bigr\rvert<1$ with $\re u \in \bigl( - \frac\pi2, \frac\pi2 \bigr)$ (implying that $F_k(0) = 0$) and by analytic continuation elsewhere. For instance $F_0(u) = e^{iu}-1$ whereas $F_1(u) = iu$.

Whenever the function is differentiable, we have
\be
z\,\partial_z \text{Li}_k(z) = \text{Li}_{k-1}(z)
\ee
or alternatively
\be
-i\,\partial_u \text{Li}_k(e^{iu}) = \text{Li}_{k-1}(e^{iu}) \qquad\text{or}\qquad
\partial_u F_k(u) = \frac{i}{e^{iu}-1} \, F_{k-1}(u) \;.
\ee
The last relation allows one to define $F_k(u) = \int_0^u \frac{i}{e^{iw}-1} \, F_{k-1}(w)$ which is single-valued because the integrand is analytic with no poles. The polylogarithms satisfy the following identities:
\bea
\label{Li inversion}
\Li_0(e^{iu})+\Li_0(e^{-iu}) &= -g'''_+(u) =-1 \\
\Li_1(e^{iu})-\Li_1(e^{-iu}) &= -ig''_+(u) \\
\Li_2(e^{iu})+\Li_2(e^{-iu}) &= g'_+(u) \\
\Li_3(e^{iu})-\Li_3(e^{-iu}) &= ig_+(u) \;,
\eea
where
\be
\label{def g+ app}
g_+(u) = \frac{1}{6}u^3 - \frac{\pi}{2}u^2 + \frac{\pi^2}{3} u
\ee
is the same function defined in (\ref{def g+}).
These relations are valid for $\re u \in (0,2\pi)$ and the polylogarithms in their principal determination, and can then be extended to the whole complex plane by analytic continuation (notice that the functions on the RHS are polynomials with no branch cuts).

\subsection[Large \texorpdfstring{$N$}{N} integrals]{Large \matht{N} integrals}
\label{subsec: large N integrals}

Let us evaluate, at large $N$, the following integrals:
\bea
I_{\text{L},\ell}[\rho](t,\Delta) &\,\equiv\, \int_{t_{\pm\Delta}} \! dt' \, \rho(t') \, e^{i\ell \left(\str u(t')-u(t) \right)} \;, \\
I_{\text{U},\ell}[\rho](t,\Delta) &\,\equiv\, \int^{t_{\pm\Delta}} \! dt' \, \rho(t') \, e^{i\ell \left(\str u(t)-u(t') \right)} \;,
\eea
where $u(t) = N^\alpha \bigl( it+v(t) \bigr)$ and $t_{\pm\Delta} \equiv t\pm N^{-\alpha}\im\Delta$ (the subscripts L and U stand for lower and upper, respectively). We Taylor expand part of the integrand around $t_{\pm\Delta}$:
\be
\label{integral tilde at power l}
I_{\text{L},\ell}[\rho](t,\Delta) = e^{-i\ell u(t)} \sum_{m=0}^\infty\frac{1}{m!} \, \partial_x^m \Bigl[ \rho(x) \, e^{i \ell N^\alpha v(x)} \Bigr]_{x=t_{\pm\Delta}}\int_{t_{\pm\Delta}} \!\! dt' \, e^{- \ell N^\alpha t'} \, \bigl( t'-t_{\pm\Delta} \bigr)^m \;.
\ee
The integral on the RHS can be evaluated integrating by parts:
\be
\int_{t_{\pm\Delta}} \!\! dt' \, e^{- \ell N^\alpha t'} \, \bigl( t'-t_{\pm\Delta} \bigr)^m = -\sum_{k=0}^m \frac{m! \, (t_+-t_{\pm\Delta})^k }{ k! \, (N^\alpha \ell)^{m-k+1}} \, e^{- \ell N^\alpha t_+} + \frac{m!}{(N^\alpha \ell)^{m+1}} \, e^{- \ell N^\alpha t_{\pm\Delta}} \;,
\ee
where $t_+$ is the upper limit of integration. The boundary terms at $t_+$ can be neglected because of an overall factor $e^{- \ell N^\alpha (t_+-t_{\pm\Delta})}$, which is exponentially suppressed, with respect to the last term. This gives
\be
\int_{t_{\pm\Delta}} \!\! dt' \, e^{- \ell N^\alpha t'} \, \bigl( t'-t_{\pm\Delta} \bigr)^m \,\simeq\, \frac{m!}{(N^\alpha l)^{m+1}} \, e^{- \ell N^\alpha t_{\pm\Delta}} \;.
\ee
For the derivatives in \eqref{integral tilde at power l}, the terms up to NLO in the large $N$ expansion are
\begin{align}
& \partial^m \Bigl[ \rho \, e^{i \ell N^\alpha v} \Bigr]_{x=t_{\pm\Delta}} = {} \\
&\quad = e^{i \ell N^\alpha v} \, (i\ell N^\alpha)^{m-1} \, \biggl( i\ell N^\alpha \, \rho \, \dot v^m + m \, \dot\rho \, \dot v^{m-1} + \tfrac{m(m-1)}2 \, \rho \, \dot v^{m-2}\,  \ddot v + \ldots \biggr) \bigg|_{x = t_{\pm\Delta}} \nn \\
&\quad = e^{i \ell \left(\str N^\alpha v \,\pm\, \im(\Delta) \dot{v} \right) } (i \ell N^\alpha)^{m-1} \biggl[ i \ell N^\alpha \, \rho \, \dot v^m + m \, \dot\rho \, \dot v^{m-1} + \tfrac{m(m-1)}2 \, \rho \, \dot v^{m-2}\,  \ddot v + {} \nn \\
&\hspace{3.5cm} \pm i \ell \im(\Delta) \Bigl( \dot{\rho} \, \dot{v}^m + m \, \rho \, \dot{v}^{m-1} \, \ddot{v} \pm \tfrac12 \, i\ell \im(\Delta) \, \rho \, \dot{v}^m \, \ddot{v} \Bigr) + \ldots \biggr] \;. \nn
\end{align}
In the last expression $\rho$ and $v$ are functions of $t$. Other contributions are subleading by powers of $N^{-\alpha}$. Plugging this back in \eqref{integral tilde at power l}, we get
\begin{align}
\label{IL tilde final}
& I_{\text{L}, \ell}[\rho](t,\Delta) = e^{\mp \ell \, \im(\Delta) \, (1-i \dot{v})} \biggl[ \frac{1}{\ell N^\alpha} \, \frac{\rho}{1-i\dot{v}} + {} \\
&\qquad\quad + \frac{1}{\ell^2 N^{2\alpha}} \Bigl( 1 \pm \ell \, \im(\Delta) \, (1-i\dot{v}) \Bigr) \biggl( \frac{\dot{\rho}}{(1-i\dot{v})^2} + \frac{i \, \rho \, \ddot{v}}{(1-i\dot{v})^3} \biggr) + \frac{1}{2N^{2\alpha}} (\im\Delta)^2 \frac{i \, \rho \, \ddot{v}}{1-i\dot{v}} \biggr] \;. \nn
\end{align}
Repeating the same steps for the other integral we find
\begin{align}
\label{IU tilde final}
& I_{\text{U}, \ell}[\rho](t,\Delta) = e^{\pm \ell \, \im(\Delta) \, (1-i \dot{v})} \biggl[ \frac{1}{\ell N^\alpha} \, \frac{\rho}{1-i\dot{v}}  \\
&\qquad\quad - \frac{1}{\ell^2 N^{2\alpha}} \Bigl( 1 \mp \ell \, \im(\Delta) \, (1-i\dot{v}) \Bigr) \biggl( \frac{\dot{\rho}}{(1-i\dot{v})^2} + \frac{i \, \rho \, \ddot{v}}{(1-i\dot{v})^3} \biggr) - \frac{1}{2N^{2\alpha}} (\im\Delta)^2 \frac{i \, \rho \, \ddot{v}}{1-i\dot{v}} \biggr] \;. \nn
\end{align}

\section{3d SUSY variations}
\label{app: 3d SUSY variations}

In terms of a single Dirac spinor $\epsilon$, the 3d supersymmetry transformations under which the Lagrangians in (\ref{initial 3d Lagrangians}) are invariant, for chiral and vector multiplets, respectively, are:
\bea
\label{3d susy chiral m}
Q \Phi &= 0 \hspace{1.5cm}& Q\Psi &= \bigl( i\gamma^\mu D_\mu\Phi - i \sigma\Phi \bigr) \, \epsilon \hspace{2.5cm}& \wt Q \Psi &= \epsilon^c F \\
\wt Q \Phi &= -\overline\epsilon \, \Psi & \wt Q\overline\Psi &= -\overline\epsilon \, \bigl( i\gamma^\mu D_\mu\Phi^\dagger + i\Phi^\dagger \sigma \bigr) & Q \overline \Psi &=- \overline{\epsilon^c} F^\dagger \\
Q \Phi^\dagger &= \overline \Psi \, \epsilon & Q F &= -\overline{\epsilon^c} \bigl( i\gamma^\mu D_\mu \Psi +i\sigma\Psi-i\lambda\Phi \bigr) & \wt Q F & = 0 \\
\wt Q \Phi^\dagger &= 0 & \wt Q F^\dagger &= \bigl( iD_\mu \overline\Psi \gamma^\mu-i\overline\Psi\sigma+i\Phi^\dagger\overline\lambda \bigr) \, \epsilon^c & Q F^\dagger & = 0
\eea
and
\bea
\label{3d susy vector m}
Q A_\mu &= -\frac{i}{2} \, \overline\lambda \gamma_\mu \epsilon \hspace{1cm}& Q\lambda &= \biggl( \frac{1}{2}\gamma^{\mu\nu}F_{\mu\nu}+iD- i\gamma^\mu D_\mu\sigma \biggr) \epsilon \hspace{1cm}& \wt Q \lambda &= 0\\
\wt Q A_\mu &= \frac{i}{2} \, \overline\epsilon\gamma_\mu\lambda & \wt Q\overline\lambda &= \overline\epsilon \, \biggl( \frac{1}{2}\gamma^{\mu\nu}F_{\mu\nu}+iD+ i\gamma^\mu D_\mu\sigma \biggr) & Q \overline\lambda &= 0\\
Q \sigma &= -\frac{1}{2} \, \overline\lambda \epsilon & Q D &= -\frac{1}{2} \bigl( D_\mu\overline\lambda\gamma^\mu - \sigma\overline\lambda \bigr) \epsilon \\
\wt Q \sigma &= \frac{1}{2} \, \overline\epsilon \lambda & \wt Q D &= -\frac{1}{2}\overline\epsilon \bigl( \gamma^\mu D_\mu\lambda - \sigma\lambda \bigr) \;.
\eea

\section{Monopole spherical harmonics on \texorpdfstring{\matht{S^2}}{S\^2}}
\label{app: harmonics}

We use complex coordinates on $S^2$ to perform the reduction. We define stereographic coordinates
\be
z = e^{i\varphi} \tan\frac{\theta}{2} \qquad \text{for } \theta< \pi \;,\qquad
v = e^{-i\varphi} \cot\frac{\theta}{2} \qquad \text{for } \theta> 0 \;,
\ee
related by $v=1/z$, which exhibit $S^2$ as $\bC\bP^1$.
The round metric with radius $R$ is proportional to the Fubini-Study metric, and the Lorentzian metric on $S^2\times \bR$ is
\be
ds^2 = \frac{4R^2}{(1+ z \bar z)^2} \, dz \, d\bar z - dt^2 \,\equiv\, g^\frac{1}{2}dz\,d\bar z - dt^2 \,= e^1e^{\bar 1}-(e^3)^2 \;,
\ee
where we defined the vielbein
\be
e^3 = dt \;,\qquad e^1 = g^\frac{1}{4} dz \;,\qquad e^{\bar 1} = g^\frac{1}{4} d\bar z \;.
\ee
Here $e^1$ and $e^{\bar 1}$ are complex conjugates of each other and therefore any real $p$-form expressed in this basis has components satisfying the reality property $X_{1\cdots}^*=X_{\bar 1\cdots}$. Flat indices are lowered and raised by the flat metric $\eta_{ab}$ with $\eta_{1\bar 1} = \eta_{\bar 11} = \frac12$.
The volume form has flat components $\epsilon_{01\bar 1}=i/2$.

Let us now move to spinors. We choose the set of gamma matrices
\be
\gamma_t = \begin{pmatrix} i & 0 \\ 0 & -i \end{pmatrix} \;,\qquad \gamma_1 = \begin{pmatrix} 0 & 0 \\ 1 & 0 \end{pmatrix} \;,\qquad \gamma_{\bar 1}= \begin{pmatrix} 0 & 1 \\ 0 & 0 \end{pmatrix} \;,
\ee
satisfying $\{\gamma_a,\gamma_b\}=2\eta_{ab} \unit$. The generators of the Dirac representation are $\gamma_{ab}=\frac{1}{2}[\gamma_a,\gamma_b]$.
On $S^2\times \bR$ the 3d Lorentz group $SO(2,1)$ is broken to the $\rU(1)$ generated by $\gamma^{1\bar 1}$, and fields are characterized by a spin that is the charge under this $\rU(1)$.
The spin connection, defined by $(\omega^a{}_b)_\mu = e^a{}_\nu \bigl( \partial_\mu e^\nu{}_b + \Gamma^\nu_{\mu\rho} e^\rho{}_b \bigr)$, has non-zero components
\be
(\omega^1{}_1)_z = -(\omega^{\bar 1}{}_{\bar 1})_z = -\frac{\bar z}{1+z\bar z} \;,\qquad\qquad (\omega^1{}_1)_{\bar z} = -(\omega^{\bar 1}{}_{\bar 1})_{\bar z} = \frac{z}{1+z\bar z} \;.
\ee
The spinor covariant derivative (without gauge connections) $D_\mu \smat{ \psi_+ \\ \psi_-} \equiv (D_\mu \psi_+, D_\mu \psi_-)^\sT$ can be written as
\be
\label{spin conn}
D = d - is\omega \qquad\text{with}\qquad \omega = i \, \frac{\bar z \, dz - z \, d\bar z}{1+z\bar z}=(\cos\theta-1) \,d\varphi
\ee
and $s=\pm\frac{1}{2}$ is the spin.
Note that $\frac1{2\pi} \int_{S^2} d\omega = -2$.
The components $\psi_\pm$ are sections of the $\rU(1)$ bundles associated to the line bundles $\cK^{\pm\frac{1}{2}}\cong \cO(\mp 1)$, where $\cK$ is the canonical bundle. A generic $\rU(1)$ bundle is labeled by a half-integer monopole charge $q$, and has covariant derivative $D = d - i q a$. To conform with the conventions of \cite{Wu:1976ge} for the monopole harmonics, we write the connection as a half-integer multiple of $a=-\omega$.

Similarly, the Levi-Civita connection on 1-forms is a $\rU(1)$ connection when projected onto the frame fields:
\be
e^z_1 \, \nabla_\mu A_z=(\partial_\mu-i\omega_\mu)e^z_1A_z\equiv D_\mu A_1 \;,\qquad e^{\bar z}_{\bar 1} \, \nabla_\mu A_{\bar z} = ( \partial_\mu + i \omega_\mu) e^{\bar z}_{\bar 1}A_{\bar z}\equiv D_\mu A_{\bar 1} \;.
\ee
Thus $A_1=e^z_1 A_z$ and $A_{\bar 1}=e^{\bar z}_{\bar 1}A_{\bar z}$ are sections with $q=-1$ and $q=+1$, respectively. On the other hand, $D_\mu A_3 = \partial_\mu A_3$ and thus $A_3$ is a section of the trivial bundle, like a scalar. Defining $D_a=e^\mu_aD_\mu$, one finds $(dA)_{ab} = e^\mu_a e^\nu_b(\nabla_\mu A_\nu -\nabla_\nu A_\mu) = D_aA_b-D_bA_a$.
If, in addition, the fields are in the adjoint representation of the gauge group and there is a background gauge field with fluxes,
\be
A = \frac{1}{2} \fm_i H^i \,a \qquad\Rightarrow\qquad \frac{1}{2\pi} \int_{S^2} dA = \fm_i H^i \;,
\ee
then including this background in the covariant derivatives $D_\mu$ shifts the spin $s\rightarrow s-\frac{\alpha(\fm)}{2}$, or equivalently $q\rightarrow q+\frac{\alpha(\fm)}{2}$, where $\alpha$ are the roots.

The derivatives $D_1$ and $D_{\bar 1}$ raise and lower the spin by $1$, respectively. This is opposite in terms of the charge $q$.
Their explicit expressions are
\be
\label{rasing lowering charge}
D^{(q)}_1 = \frac{1}{2R} \Bigl( (1+z\bar z) \, \partial_z-q\bar z \Bigr) \;,\qquad\qquad D^{(q)}_{\bar 1} = \frac{1}{2R} \Bigl( (1+z\bar z) \, \partial_{\bar z} + qz \Bigr) \;,
\ee
where the superscript indicates the charge of the section they act on, whereas under complex conjugation $D_1^{(q)\,*} = D^{(-q)}_{\bar 1}$ and $D_{\bar 1}^{(q)\,*} = D^{(-q)}_1$.
We define the operators 
\be
\label{su(2) rot op}
L_+ = z^2\partial_z+\partial_{\bar z}-qz \;,\qquad L_- = - \bar z^2\partial_{\bar z}-\partial_z-q\bar z \;,\qquad L_z = z\partial_z-\bar z\partial_{\bar z}-q \;,
\ee
satisfying the $\fsu{(2)}$ algebra $[L_z,L_\pm]=\pm L_\pm$ and $[L_+,L_-]=2L_z$. The covariant Laplacian is
\bea
-D^2 &\,\equiv\, L^2-q^2 = \frac{1}{2}\{L_+,L_-\}+L_z^2-q^2 = - \bigl( 1+z\bar z \bigr)^2 \partial_z\partial_{\bar z} - q ( 1+z\bar z) L_z - q^2 \\
	&\,=\, -\frac1{\sin\theta} \partial_\theta \, \bigl( \sin\theta \, \partial_\theta \bigr) + \frac1{\sin^2\theta} \bigl( -i\partial_\varphi - q + q \cos\theta \bigr)^2 \;,
\eea
which can be diagonalized simultaneously with $L^2$ and $L_z$. Its eigenfunctions are the monopole spherical harmonics $Y_{q,l,m}$ with $|m|\leq l$, that we choose to be orthonormal on an $S^2$ of radius $1$:
\be
\int_{S^2}\! \sqrt{g} \;\; \wb{Y_{q,l,m}} \; Y_{q,l',m'} = \delta_{l,l'} \, \delta_{m,m'} \;.
\ee
The highest harmonic with $m=l$, annihilated by $L_+$, is
\be
Y_{q,l,l}(z,\bar{z}) \,\propto\, \frac{z^{l+q}}{(1+z\bar z)^l} \;.
\ee
Regularity at the poles implies $l+q\in \bZ_{\geq 0}$ and $l \geq |q|$.

The Laplacian can be written in terms of the derivatives as
\be
\label{comm D and laplacian}
{-D}^2 = - 4 R^2 D_1D_{\bar 1} + q = - 4 R^2 D_{\bar 1} D_1 - q = - 2 R^2 \{D_1,D_{\bar 1} \} \;.
\ee
Besides, one can verify that
\be
[D_1,L_z]=[D_1,L_\pm]=[D_{\bar 1},L_z]=[D_{\bar 1},L_\pm]=0 \;.
\ee
Therefore the derivatives act as bundle-changing operators mapping $Y_{q,m,l}$ to $Y_{q\pm1, m, l}$. The exact relations can be derived integrating by parts the orthonormality conditions. For a suitable choice of phases one finds \cite{Wu:1976ge,Wu:1977qk}:
\bea
\label{raise lower coeffs}
D^{(q)}_1 \, Y_{q,l,m} &= -\frac{s_-(q,l)}{2R} \, Y_{q-1,l,m} \qquad& \text{with } \quad s_-(q,l) &= \bigl[ l(l+1)-q(q-1) \bigr]^\frac{1}{2} \;, \\
D^{(q)}_{\bar 1} \, Y_{q,l,m} &= \;\; \frac{s_+(q,l)}{2R} \, Y_{q+1,l,m} \qquad& \text{with } \quad s_+(q,l) &= \bigl[ l(l+1)-q(q+1) \bigr]^\frac{1}{2} \;.
\eea
Following the same conventions as in \cite{Wu:1977qk}, the monopole harmonics satisfy
\be
\label{harmonics reality property}
\wb{ Y_{q,l,m}}=(-1)^{q+m} \; Y_{-q,l,-m}
\ee
under complex conjugation.

Finally, the triple overlap of harmonics is given in terms of Wigner 3$j$-symbols:
\begin{multline}
\label{triple overlap}
\int\! d\Omega\; Y_{q,l,m}Y_{q',l',m'}Y_{q'',l'',m''} = {} \\
{} = (-1)^{l+l'+l''} \biggl[ \frac{ (2l+1) (2l'+1) (2l''+1) }{4\pi} \biggr]^\frac12 \begin{pmatrix}
			l & l' & l'' \\ q & q' & q''
		\end{pmatrix} \begin{pmatrix}
			l & l' & l'' \\ m & m' & m''
		\end{pmatrix} \;,
\end{multline}
or equivalently
\begin{align}
\label{clebsch gordan}
& Y_{q,l,m} \, Y_{q',l',m'} = {} \\
& \sum_{l''} (-1)^{l+l'+l''+q''+m''} \biggl[ \frac{ (2l+1)(2l'+1)(2l''+1) }{4\pi} \biggr]^\frac12 \begin{pmatrix}
		l & l' & l'' \\ q & q' & q''
	\end{pmatrix} \begin{pmatrix}
		l & l' & l'' \\ m & m' & m''
	\end{pmatrix}Y_{-q'',l'',-m''} \nn
\end{align}
The $3j$-symbols are directly related to Clebsch-Gordan coefficients that decompose the angular momentum state $\lvert l''\,m''\rangle$ in terms of $\lvert l\,m\,l'\,m'\rangle=\lvert l\,m\rangle\otimes\lvert l'\,m'\rangle$:
\be
\label{CB coeff def}
C\bigl(\begin{smallmatrix}
l & l' & l''\\
m & m' & m''
\end{smallmatrix}\bigr) \,\equiv\, \langle l\,m\,l'\,m'\vert\,l''\,m''\rangle = (-1)^{l-l'+m''} \sqrt{2l''+1} \begin{pmatrix} l & l' & l'' \\ m & m' & -m'' \end{pmatrix} \,.
\ee
In particular, the Clebsch-Gordan coefficients are zero unless $m + m' = m''$, $\bigl\lvert m^{(i)} \bigr\rvert \leq l^{(i)}$ with $m^{(i)} = l^{(i)} \text{ mod }1$, and $l^{(i)} \leq l^{(j)} + l^{(k)}$. The $3j$-symbol is symmetric under even permutations of its columns, and gains a sign $(-1)^{l+l'+l''}$ under odd permutations. It also gains a sign $(-1)^{l+l'+l''}$ when one changes sign to $m$, $m'$ and $m''$ simultaneously.
This implies the following relations among Clebsch-Gordan coefficients:
\bea
C\bigl(\begin{smallmatrix} l' & l'' & l \\ m' & -m'' & -m \end{smallmatrix}\bigr) &= (-1)^{l-l''+m'} \left[ \resizebox{0.8\width}{!}{$ \dfrac{2l+1}{2l''+1} $} \right]^{1/2} \, C\bigl( \begin{smallmatrix} l & l' & l'' \\ m & m' & m'' \end{smallmatrix}\bigr) \;, \\
C\bigl(\begin{smallmatrix} l'' & l & l' \\ -m'' & m & -m' \end{smallmatrix}\bigr) &= (-1)^{l''-l'+m} \left[ \resizebox{0.8\width}{!}{$ \dfrac{2l'+1}{2l''+1} $} \right]^{1/2} \, C\bigl(\begin{smallmatrix} l & l' & l'' \\ m & m' & m'' \end{smallmatrix}\bigr) \;, \\[.4em]
C\bigl(\begin{smallmatrix} l' & l & l'' \\ m' & m & m'' \end{smallmatrix}\bigr) &= (-1)^{l + l' - l''} C\bigl(\begin{smallmatrix} l & l' & l'' \\ m & m' & m'' \end{smallmatrix}\bigr) \;.
\eea
In the special case that $l'' = l + l'\equiv L$ (and $m + m' =- m'' \equiv M$ as in the general case):
\bea
\label{binom saturate triangle ineq}
\begin{pmatrix}
l & l' & L \\ m & m' & -M
\end{pmatrix} &= (-1)^{l-l'+M} \biggl[ \frac{1}{2L+1} \binom{2L}{L+M}^{-1} \binom{2l}{l+m} \binom{2l'}{l'+m'} \biggr]^\frac12 \;, \\[.3em]
C\bigl(\begin{smallmatrix} l & l' & L \\ m & m' & M \end{smallmatrix}\bigr) &= \biggl[ \binom{2L}{L+M}^{-1} \binom{2l}{l+m} \binom{2l'}{l'+m'} \biggr]^\frac12 \;.
\eea

\section{1d \matht{\cN=2} superspace}
\label{app: 1d susy}
	
We review here the 1d $\cN=2$ superspace formalism, drawing from Appendix~A of \cite{Hori:2014tda}. The $\cN=2$ superspace in quantum mechanics, which we denote as $\bR^{1|2}$, has coordinates $(t,\theta,\bar\theta)$, where $\theta$ is a complex fermionic coordinate. A supersymmetry transformation is
$\delta =-\epsilon \, Q + \overline{\epsilon} \, \overline{Q}$,
where $\epsilon$, $\overline{\epsilon}$ are anticommuting parameters, and $Q$, $\overline{Q}$ are anticommuting generators so that $\delta$ is commuting. Here $Q$ and $\overline{Q}$ are defined as differential operators acting on superfields:
\be
\label{1d scharges diff ops}
Q \equiv \partial_\theta + \frac{i}{2} \, \bar\theta \, \partial_t \;,\qquad \overline{Q} \equiv -\partial_{\bar\theta} - \frac{i}{2} \, \theta \, \partial_t \;.
\ee
They satisfy the algebra $Q^2 = \overline{Q}^2 = 0$ and $\{Q, \overline{Q} \} = - i \partial_t$. Moreover, $Q$ and $\overline{Q}$ anticommute with another set of differential operators
\be
D \equiv \partial_\theta - \frac{i}{2} \, \bar\theta \, \partial_t \;,\qquad \overline{D} \equiv -\partial_{\bar\theta} + \frac{i}{2} \, \theta \, \partial_t \;,
\ee
which satisfy the algebra $D^2 = \overline{D}^2 = 0$ and $\{D, \overline{D} \} = i\partial_t$.
One has $\overline{(DX)}=(-1)^F\overline{D}\,\overline{X}$ and $\overline{(\overline{D}X)}=(-1)^F D\overline{X}$. 

\subsection{Matter multiplets}

A chiral superfield $\Phi_h$ is defined by $\overline{D}\Phi_h=0$. Gauge transformations act as
\be
\Phi_h \rightarrow h \, \Phi_h \;,\qquad h=e^{\chi} \;,\qquad \chi:\bR^{1|2}\rightarrow\bC\otimes r \;,\qquad \overline{D} \chi = 0 \;,
\ee
where $r$ is some representation of the gauge group. $\overline{D}\Phi_h=0$ implies that $\Phi_h$ and its complex conjugate anti-chiral superfield $\overline{\Phi}_h$ have expansion:
\be
\label{chiral theta exp}
\Phi_h = \phi + \theta \psi - \frac{i}{2} \theta \bar\theta \, \partial_t\phi \;,\qquad\qquad \overline{\Phi}_h = \overline{\phi} - \bar\theta \,\overline{\psi} + \frac{i}{2} \theta \bar\theta \, \partial_t \overline{\phi} \;.
\ee
Acting with \eqref{1d scharges diff ops} on $\Phi_h$ and $\overline{\Phi}_h$, we find the following supersymmetry variations:
\bea
\label{1d chiral susy vars}
Q\phi &=\psi \;,\qquad& Q\psi &= 0 \;,\qquad& \overline{Q}\phi &= 0 \;,\qquad& \overline{Q}\psi &= i\partial_t\phi\;.
\eea
Suppose that $\Phi_{a,h}$ are a collection of bosonic chiral superfields. We can also have fermionic Fermi superfields $\cY_h$, satisfying $\overline{D}\cY_h=E(\Phi_h)$ for some holomorphic function $E(\Phi_h)$, and transforming as $\cY_h\rightarrow h\cY_h$ under some representation of the gauge group. $\overline{D}\cY_h=E(\Phi_h)$ implies that $\cY_h$ and its conjugate $\overline{\cY}_h$ have expansion:
\bea
\cY_h &= \eta - \theta f - \bar\theta E(\phi) + \theta\bar\theta \Bigl( \partial_aE(\phi)\psi_a - \tfrac{i}{2} \partial_t\eta \Bigr) = \eta-\theta f -\bar\theta E(\Phi) - \tfrac{i}{2} \theta\bar\theta \partial_t\eta \;, \\
\overline{\cY}_h &= \overline{\eta} - \bar\theta \,\overline{f} - \theta\overline{E}(\overline{\phi}) + \theta\bar\theta \Bigl( \overline{\psi}_a \overline{\partial}_a \overline{E}(\overline{\phi}) + \tfrac{i}{2} \partial_t \overline{\eta} \Bigr) = \overline{\eta} - \bar\theta \, \overline{f} - \theta \overline{E}(\overline{\Phi}) + \tfrac{i}{2} \theta\bar\theta \partial_t\overline{\eta} \;.
\eea
Acting with \eqref{1d scharges diff ops} gives the supersymmetry variations:
\bea
\label{1d fermi susy vars}
Q\eta &=-f \;,\qquad& Qf &= 0 \;,\qquad& \overline{Q} \eta &= E(\phi) \;,\qquad& \overline{Q}f &= - i \partial_t\eta + \partial_aE(\phi) \, \psi_a \;.
\eea

\subsection{Vector multiplet}

We assume that the gauge group $G$ is semi-simple (inclusion of $\rU(1)$ factors is trivial) with Lie algebra $\fg$. Denote the complexified algebra as $\fg_\bC = \fg\otimes\bC=\fg\oplus_\bR i\fg$, with Killing form given by the trace operation $\Tr$. It admits a root space decomposition $\fg_\bC = \fh_\bC \oplus_{\alpha\in\Phi} L_\alpha$, where $\fh_\bC$ is a Cartan subalgebra and $\Phi$ is the set of all roots. We can use the Chevalley basis $\fg_\bC = \text{span}_\bC \{ H^{i=1,\ldots,\rk G},E^\alpha \;|\;\alpha\in\Phi\}$, where $i$ indexes a set of simple roots $\alpha^i$ and $H^i$ is defined in the following way:
\be
\exists! \; H^i \in \fh_\bC \; \big| \; \alpha^i(h)=\Tr (H^ih) \,,\;\forall \;h\in\fh_\bC \;.
\ee
The element $E^\alpha$ is also normalized so that $\Tr E^\alpha E^{-\alpha}=1$. The compact real form is
\be
\fg=\text{span}_\bR \bigl\{ iH^i,E^\alpha-E^{-\alpha},i(E^\alpha+E^{-\alpha}) \,\bigm|\, \alpha\in\Phi^+ \bigr\} \;,
\ee
where $\Phi^+$ is the set of positive roots. Using the fact that $\Tr$ splits between each summand in $\fh_\bC\oplus_{\alpha\in\Phi^+}(L_\alpha\oplus L_{-\alpha})$, and that $\Tr$ is positive definite on $H^i$, it quickly follows that $\Tr$ is negative (positive) definite on $\fg$ ($i\fg$). Any $\Lambda\in i\fg$ can be expressed with $\Lambda^i$, $\Lambda_1^\alpha$, $\Lambda_2^\alpha\in\bR$ as 
\bea
\Lambda &= \sum\nolimits_i \Lambda^iH^i+\sum\nolimits_{\alpha\in\Phi^+} \Bigl[ \Lambda_1^\alpha(E^\alpha+E^{-\alpha})+\Lambda_2^\alpha i(E^\alpha-E^{-\alpha}) \Bigr] \\
&= \sum\nolimits_i \Lambda^iH^i + \sum\nolimits_{\alpha\in\Phi^+} \bigl( \Lambda^\alpha E^\alpha+\overline{\Lambda^\alpha}E^{-\alpha} \bigr) \,,\qquad \Lambda^\alpha\equiv\Lambda_1^\alpha+i\Lambda_2^\alpha \;.
\eea
Therefore, defining a formal Hermitian conjugation on elements of $\fg_\bC$ as $\overline{H^i}\equiv H^i$, $\overline{E^\alpha}\equiv E^{-\alpha}$, we can alternatively define $i\fg$ as $i\fg = \bigl\{ \Lambda\in\fg_\bC\,\big|\,\overline{\Lambda}=\Lambda \bigr\}$. A generic group element $k=e^{i\Lambda}$ then satisfies $\overline{k}=e^{-i\Lambda}=k^{-1}$. If $G=\rU(N)$, this formal Hermitian conjugation becomes the actual conjugate transpose on $N\times N$ matrices. 
	
To build gauge interactions, we introduce the independent superfields $\Omega$ and $V^-$. $\Omega$ is valued in $\fg_{\bC}$, while $V^-$ is valued in $i\fg$, \ie, $\overline{V^-}=V^-$. One can either use $\Omega$ alone, or include \emph{both} $\Omega$ and $V^-$ in the theory. The crucial role played by $\Omega$ is to allow for gauge-covariant chiral and Fermi conditions. Under gauge transformations, they transform as:
\bea
\label{1d vect supergauge transf}
& e^\Omega \rightarrow k \, e^\Omega \, h^{-1} \;,\quad& & V^-\rightarrow k V^-k^{-1}+ik(\partial_tk^{-1}) \;, \\
& h = e^\chi \;,&& \chi:\bR^{1|2}\rightarrow\fg_\bC \;,\qquad \overline{D}\chi=0 \;,\\
& k = e^{i\Lambda} \;,&& \Lambda:\bR^{1|2}\rightarrow i\fg \;,\qquad\quad \overline{\Lambda}=\Lambda \;.
\eea
Without loss of generality, $V^-$ can be expanded as
\be
V^- = A_t - \sigma - i\theta\overline{\lambda} - i \bar\theta \lambda + \theta\bar\theta D \;,
\ee
where $(A_t-\sigma,\, D)$ are valued in $i\fg$ and $\lambda$ is valued in $\fg_\bC$. We now define the various ingredients used to construct supersymmetric actions. The gauge-covariant superspace derivatives are defined as
\be
\cD \,\equiv\, e^{-\overline{\Omega}} \, D \, e^{\overline{\Omega}} \;,\qquad \overline{\cD} \,\equiv\, e^\Omega \, \overline{D} \, e^{-\Omega} \;,\qquad \cD_t^- \,\equiv\, \partial_t - i V^- \;,
\ee
which, according to \eqref{1d vect supergauge transf} and using $\overline{D}h = D\overline{h} = 0$, transform as
\be
\label{gauge transf cov superspace deriv}
\cD \,\rightarrow\, k\cD k^{-1} \;,\qquad \overline{\cD} \,\rightarrow\, k\overline{\cD}k^{-1} \;,\qquad \cD_t^- \,\rightarrow\, k \cD_t^- k^{-1} \;.
\ee
They satisfy the algebra
\be
\label{cD_squares}
\cD^2 = \overline{\cD}^2 = 0 \;,\qquad \{\cD,\overline{\cD}\} = i (\partial_t-iV^+) \,\equiv\, i \cD_t^+ \;,
\ee
where $V^+$ is an $i\fg$-valued superfield constructed out of $\Omega$ only:
\be
V^+ \equiv D \bigl[ e^\Omega \bigl( \overline{D} e^{-\Omega} \bigr) \bigr] + \overline{D} \bigl[ e^{-\overline{\Omega}} \bigl( D e^{\overline{\Omega}} \bigr) \bigr] + \bigl\{ e^\Omega \bigl( \overline{D} e^{-\Omega} \bigr) , e^{-\overline{\Omega}} \bigl( De^{\overline{\Omega}} \bigr) \bigr\} \;.
\ee
If the gauge group is Abelian this simplifies to $V^+ = - [D,\overline{D}] \, \Omega$. As it was for $D$ and $\overline{D}$, one has $\overline{(\cD X)}=(-1)^F \, \overline{\cD} \,\overline{X}$ and $\overline{(\overline{\cD}X)} = (-1)^F \, \cD \overline{X}$. One can check that the gauge transformation of $V^+$ is identical to that of $V^-$:
\be
V^+ \,\rightarrow\, kV^+k^{-1}+ik(\partial_tk^{-1}) \;,
\ee
which is consistent with \eqref{gauge transf cov superspace deriv} and \eqref{cD_squares}. We will also have occasion to use the field strength superfield
\be
\label{field strength superfield def}
\Upsilon \equiv [\overline{\cD},\cD_t^-] = - i \overline{D}V^- - \partial_t \bigl[ e^\Omega \bigr( \overline{D}e^{-\Omega} \bigr) \bigr] -i \bigl[ e^\Omega \bigl( \overline{D} e^{-\Omega} \bigr) , V^- \bigr] \;,
\ee
which also transforms covariantly as $\Upsilon\rightarrow k\Upsilon k^{-1}$. From the definition, it follows directly that $\overline{\cD}\Upsilon=0$. 

Instead of $\Omega$ and $V^-$, we can equivalently use two other superfields $V$ and $V^-_h$ defined as
\be
e^V \equiv e^{\overline{\Omega}} e^\Omega \;,\qquad
V^-_{h} \,\equiv\, e^{\overline{\Omega}} \, V^- e^{\Omega} + \frac{i}{2}e^{\overline{\Omega}}\partial_te^{\Omega} - \frac{i}{2} \bigl( \partial_t e^{\overline{\Omega}} \,\bigr) e^{\Omega} \;,\qquad \overline{V^-_h}=V^-_{h} \;,
\ee
which only transform under the complexified gauge transformations as:
\be
e^V \,\rightarrow\, \overline{h}^{-1}e^Vh^{-1} \;,\qquad
V^-_h \,\rightarrow\, \overline{h}^{-1}V^-_hh^{-1} + \frac{i}{2}\overline{h}^{-1} e^V\partial_th^{-1} - \frac{i}{2} \bigl( \partial_t\overline{h}^{-1} \bigr) e^Vh^{-1} \;. 
\ee
Note that $V$ is constructed solely out of $\Omega$, while $V^-_h$ is built out of both $V^-$ and $\Omega$. In this formulation, the theory might contain $V$ only, or both $V^-_h$ and $V$. Analogously to the above, out of $V$ and $V^-_h$ we can construct
\bea
V^+_h &\,\equiv\, \frac{1}{2} e^V \overline{D} \bigl( e^{-V}De^V \bigr) + \frac{1}{2} D \bigl( e^V\overline{D} e^{-V} \bigr) e^V = e^{\overline{\Omega}}V^+ e^{\Omega} + \frac{i}{2} e^{\overline{\Omega}}\partial_te^{\Omega} - \frac{i}{2} \bigl( \partial_te^{\overline{\Omega}} \,\bigr) e^{\Omega} \;, \\
\Upsilon_h &\,\equiv\, -i \, e^V\overline{D} \Bigl[ e^{-V} \Bigl( V^-_h + \frac{i}{2} \partial_te^V \Bigr) \Bigr] = e^{\overline{\Omega}} \Upsilon e^{\Omega} \;.
\eea
One can check that $V^+_h$ transforms in the same way as $V^-_h$, and $\Upsilon_h$ transforms in the same way as $e^V$. In an Abelian theory,
\be
V^+_h = \frac{1}{2} e^V \bigl( \overline{D}D-D\overline{D} \bigr) V \;.
\ee
When writing matter Lagrangians in terms of $\Phi_h$ and $\cY_h$ which transform with chiral gauge transformations $h$, it will be convenient to use $V$ and $V^-_h$.

Given any chiral or Fermi superfield, one can define covariantly-chiral counterparts
\be
\label{covariantly chiral superfields}
\Phi_k \,\equiv\, e^\Omega \, \Phi_h \;,\qquad \cY_k \,\equiv\, e^\Omega\cY_h \;,\qquad \overline{\cD}\Phi_k = 0 \;,\qquad \overline{\cD}\cY_k = E(\Phi_k) \;,
\ee
which transform under the gauge group as $\Phi_k \rightarrow k \, \Phi_k$ and $\cY_k \rightarrow k\, \cY_k$. These fields are useful when one is using $\Omega$ and $V^-$ to describe the vector multiplet.

\subsection{Wess-Zumino gauge}
\label{app: WZ gauge}

We can expand $\Omega$ and the gauge transformation parameters $\chi$, $\Lambda$ as:
\be
\Omega = \Omega_0 + \theta\Omega_{\theta} + \bar\theta \Omega_{\bar\theta} + \theta \bar\theta \Omega_{\theta\bar\theta} \,,\quad\;\;
\chi = \chi_0 + \theta\chi_\theta - \frac{i}{2} \theta\bar\theta \partial_t\chi_0 \,,\quad\;\;
\Lambda = \Lambda_0 + \theta\Lambda_\theta - \bar\theta \,\overline{\Lambda}_\theta + \theta\bar\theta \Lambda_{\theta\bar\theta} \,.
\ee
We show that, using gauge transformations, every component of $\Omega$ can be canceled except for $\Omega_{\theta\bar\theta}$, and we can further set $\overline{\Omega}_{\theta\bar\theta} = \Omega_{\theta\bar\theta}$, \ie, $\Omega_{\theta\bar\theta}$ is valued in $i\fg$. We shall call this component $-\frac{1}{2}(A_t+\sigma)$, where both $A_t$ and $\sigma$ are valued in $i\fg$. Due to the relative sign, this is independent of $(A_t-\sigma)$ in $V^-$. In other words, we can bring $\Omega$ to the form 
\be
\Omega = - \frac{1}{2} \, \theta\bar\theta \, (A_t+\sigma) \;,
\ee
that we dub the Wess-Zumino gauge. First, we use the transformation $\chi = \Omega_0 - \frac{i}{2} \theta\bar\theta \partial_t\Omega_0$, $\Lambda=0$ to set $\Omega_0\rightarrow 0$, after which only transformations with $\chi_0 = i\Lambda_0$ preserve $\Omega_0 = 0$ and are allowed. Next, performing the transformation $\chi = \theta (\Omega_\theta + \overline{\Omega}_{\bar\theta})$, $\Lambda = i \theta \overline{\Omega}_{\bar\theta} + i \bar\theta \Omega_{\bar\theta}$ sets $\Omega_\theta$, $\Omega_{\bar\theta} \rightarrow 0$. Further transformation parameters cannot have $\theta$ or $\bar\theta$ components since otherwise a nonzero $\Omega_{\bar\theta}$ would be generated. Lastly, we perform $\chi=0$, $\Lambda=\frac{i}{2} \theta \bar\theta (\Omega_{\theta\bar\theta} - \overline{\Omega}_{\theta\bar\theta})$, after which $\Omega_{\theta\bar\theta} \rightarrow \frac12 (\Omega_{\theta\bar\theta} + \overline{\Omega}_{\theta\bar\theta})$ is valued in $i\fg$. The residual gauge transformations are $\chi = i\Lambda_0 + \frac{1}{2}\theta\bar\theta \partial_t\Lambda_0$, $\Lambda = \Lambda_0$, under which 
\be
A_t+\sigma \;\rightarrow\; e^{i\Lambda_0}(A_t+\sigma)e^{-i\Lambda_0} + i \, e^{i\Lambda_0}\partial_te^{-i\Lambda_0} \;.
\ee
These are purely time-dependent gauge transformations, as expected. In this gauge, the gauge-covariant superspace derivatives simplify to
\be
\cD_t^+ = D_t^+ = \partial_t - i(A_t+\sigma) \;,\qquad \cD=\partial_\theta - \frac{i}{2} \bar\theta D_t^+ \;,\qquad \overline{\cD}= -\partial_{\bar\theta} + \frac{i}{2} \theta D_t^+ \;,
\ee
and 
\be
V^+ = A_t + \sigma \;,\qquad\qquad \Upsilon = \lambda-\theta \bigl( D_t\sigma+iD \bigr) - \frac{i}{2} \theta \bar\theta D_t^+ \lambda \;.
\ee
	
The action of supersymmetry on $\Omega$, using \eqref{1d scharges diff ops}, is $\delta\Omega = \frac12 \epsilon \bar\theta (A_t+\sigma) - \frac{1}{2} \bar\epsilon \theta(A_t+\sigma)$ and the Wess-Zumino gauge is not preserved. This can be compensated by an infinitesimal gauge transformation with parameters
\be
\label{compensating gt WZ gauge}
\Lambda = \frac{i}{2} \, \epsilon \bar\theta (A_t+\sigma) + \frac{i}{2} \bar\epsilon \theta (A_t+\sigma) + \cO(\epsilon^2) \;,\qquad \chi = - \bar\epsilon \theta (A_t+\sigma) + \cO(\epsilon^2) \;.
\ee
The supersymmetry transformations that preserve Wess-Zumino gauge are computed using $\delta$ with the addition of the compensating gauge transformation above. For $\Omega$, its variation under the combined supersymmetry and gauge transformation is $\delta\Omega+i\Lambda-\chi=0$ by construction. The superfields $\Phi_k$, $\cY_k$ are only sensitive to the gauge transformations generated by $\Lambda$, and not to those generated by $\chi$. The addition of the $\Lambda$-transformation \eqref{compensating gt WZ gauge} to $\delta$ can be directly absorbed into the supercharges:
\be
\label{1d scharges redef w gauge}
Q_\text{WZ} \equiv \partial_\theta + \frac{i}{2} \, \bar\theta \bigl[ \partial_t - \delta_\text{gauge}(A_t+\sigma) \bigr] \;,\qquad
\overline{Q}_\text{WZ} \equiv - \partial_{\bar\theta} - \frac{i}{2} \, \theta \bigl[ \partial_t - \delta_\text{gauge}(A_t+\sigma) \bigr] \;.
\ee
Note that $\delta_\text{gauge}(\Lambda)$ acts according to the gauge representation of each superfield, except for $V^\pm$, on which $\delta_\text{gauge} (\Lambda) \, V^\pm = \partial_t \Lambda - i [V^\pm,\Lambda]$. The modified supercharges satisfy the algebra
\be
\label{1d susy alg in WZ gauge}
Q_\text{WZ}^2 = \overline{Q}_\text{WZ}^2 = 0 \;,\qquad \{ Q_\text{WZ}, \overline{Q}_\text{WZ} \} = - i \bigl[ \partial_t - \delta_\text{gauge}(A_t+\sigma) \bigr] \;.
\ee

\subsection{Transformations in Wess-Zumino gauge}
\label{app: transf in WZ gauge}

Acting with \eqref{1d scharges redef w gauge} on $V^\pm$ and reading off the variations of each component, we find the following supersymmetry variations (and their complex conjugate) for the vector multiplet:
\bea
\label{1d vect susy vars}
Q_\text{WZ} \, A_t &= - Q_\text{WZ} \, \sigma = - \frac{i}{2} \, \overline{\lambda} \;,\qquad& Q_\text{WZ} \, \lambda &= - D_t\sigma - iD \;, \\
Q_\text{WZ} \, D &= - \frac{1}{2} D_t^+ \overline{\lambda} \;,& \overline{Q}_\text{WZ} \, \lambda &= 0 \;.
\eea
Note that $Q_\text{WZ} (A_t+\sigma) = \overline{Q}_\text{WZ}(A_t+\sigma)=0$, consistently with \eqref{1d susy alg in WZ gauge}. In Wess-Zumino gauge, $\Phi_k$ and its conjugate $\overline{\Phi}_k$ have expansion:
\be
\Phi_k = \phi + \theta\psi - \frac{i}{2} \theta\bar\theta D_t^+\phi \;,\qquad\qquad \overline{\Phi}_k = \overline{\phi} - \bar\theta \,\overline{\psi} + \frac{i}{2} \theta \bar\theta D_t^+ \overline{\phi} \;.
\ee
Acting with \eqref{1d scharges redef w gauge} on $\Phi_k$ we find the following supersymmetry variations:
\be
\label{1d chiral susy vars WZ}
Q_\text{WZ} \, \phi =\psi \;,\qquad Q_\text{WZ} \, \psi = 0 \;,\qquad \overline{Q}_\text{WZ} \, \phi = 0 \;,\qquad \overline{Q}_\text{WZ} \, \psi = iD_t^+ \phi \;.
\ee
Alternatively, we can obtain the same variations by acting with $\delta + \chi = - \epsilon \, Q_\text{WZ} + \bar\epsilon \, \overline{Q}_\text{WZ}$ on $\Phi_h$, with $\chi$ given in \eqref{compensating gt WZ gauge}. Analogously, $\cY_k$ and its conjugate $\overline{\cY}_k$ have the expansions
\bea
\cY_k &= \eta - \theta f - \bar\theta E(\phi) + \theta\bar\theta \Bigl( \partial_aE(\phi) \psi_a - \tfrac{i}{2}D_t^+\eta \Bigr) = \eta-\theta f - \bar\theta E(\Phi) - \tfrac{i}{2} \theta\bar\theta D_t^+ \eta \\
\overline{\cY}_k &= \overline{\eta} - \bar\theta \,\overline{f} - \theta\overline{E}(\overline{\phi}) + \theta\bar\theta \Bigl( \overline{\psi}_a \overline{\partial}_a \overline{E}(\overline{\phi}) + \tfrac{i}{2} D_t^+ \overline{\eta} \Bigr) = \overline{\eta} - \bar\theta \,\overline{f} - \theta\overline{E}(\overline{\Phi}) + \tfrac{i}{2} \theta\bar\theta D_t^+ \overline{\eta} \;,
\eea
and acting with \eqref{1d scharges redef w gauge} gives the supersymmetry variations:
\bea
\label{1d fermi susy vars WZ}
Q_\text{WZ} \, \eta &= -f \;,\quad& Q_\text{WZ} \, f &= 0 \;,\quad& \overline{Q}_\text{WZ} \, \eta &=E(\phi) \;,\quad& \overline{Q}_\text{WZ} \, f &= - iD_t^+ \eta + \partial_a E(\phi) \, \psi_a \;. 
\eea
Again, we can obtain the same variations by acting with $\delta+\chi$ on $\cY_h$.

\subsection{Supersymmetric Lagrangians}
\label{app: susy lagrangians}

As with the prototypical $4$d $\cN=1$ supersymmetry, there are two broad classes of supersymmetric terms: D-terms and F-terms. Let $X$ be a bosonic, gauge-invariant, real-valued superfield with expansion
\be
\label{generic real superfield theta expansion}
X = X_0 + \theta X_\theta - \bar\theta \,\overline{X_\theta} + \theta\bar\theta X_{\theta\bar\theta} \;.
\ee
Acting with $Q$ and $\overline{Q}$, we find that $QX_{\theta\bar\theta} = -\frac{i}{2}\partial_tX_\theta$ and $\overline{Q} X_{\theta\bar\theta} = \frac{i}{2} \partial_t \overline{X_\theta}$ are total derivatives. Moreover, $Q \overline{Q} X_0 = X_{\theta\bar\theta}$ up to a total derivative. Therefore,
\be
\label{generic D term}
\int\! d\theta d\bar\theta \, X = - X_{\theta\bar\theta} = Q\overline{Q} \,(-X_0)
\ee
is supersymmetric, and we call such terms D-terms. They are always $Q$ and $\overline{Q}$ exact. Conversely, suppose there is a term in the Lagrangian of the form $Q\overline{Q}(-X_0)$ where $X_0$ is real and gauge invariant. If there is a real-valued superfield $X$ with bottom component $X_0$, it must have the same expansion \eqref{generic real superfield theta expansion}. Therefore \eqref{generic D term} holds and this term can be written as a D-term in superspace.  

Let $Y$ be a fermionic, gauge-invariant, complex-valued chiral superfield, $\overline{\cD}Y=\overline{D}Y=0$. Its complex conjugate $\overline{Y}$ is anti-chiral and satisfies $D\overline{Y}=0$. They have expansion:
\be
Y = Y_0 + \theta Y_\theta - \frac{i}{2} \, \theta\bar\theta \, \partial_tY_0\;,\qquad\qquad \overline{Y} = \overline{Y_0} + \bar\theta \, \overline{Y_\theta} + \frac{i}{2} \, \theta\bar\theta \, \partial_t\overline{Y_0} \;.
\ee
Acting with $Q$ and $\overline{Q}$ on $Y$ and $\overline{Y}$, one finds that $Y_\theta$ and $\overline{Y_\theta}$ are separately supersymmetric up to total derivatives. Moreover, $Y_\theta=QY_0$ and $\overline{Y_\theta}=-\overline{Q}\,\overline{Y_0}$. Therefore:
\be
\int\! d\theta\, Y + \int\! d\bar\theta \, \overline{Y} = Y_\theta + \overline{Y_\theta} = Q \, Y_0 - \overline{Q}\,\overline{Y_0} = (Q+\overline{Q}) (Y_0 - \overline{Y_0})
\ee
is supersymmetric, and we call such terms F-terms. They are always $(Q+\overline{Q})$ exact.

We can now write the following supersymmetric Lagrangians, with component expressions in Wess-Zumino gauge. In the gauge sector, if the theory only contains $\Omega$ or equivalently $V$, the only term we can think of is a Wilson line in $A_t+\sigma$. For a $\rU(1)$ gauge group, the supersymmetric Wilson loop of charge $q$ can be written as
\be
\label{susy wilson loop}
\exp \biggl( iq\oint dt \int\! d\theta d\bar\theta \; V \biggr) \,\stackrel{\text{WZ}}{=}\, \exp\biggl( i q \oint dt \, (A_t+\sigma) \biggr) \;.
\ee
If both $V^-$ and $\Omega$ are present, we can write the following terms. The conventional gauge kinetic term is
\bea
\label{1d vec kin term}
\frac{1}{2e^2_\text{1d}} \int\! d\theta d\bar\theta \, \Tr\overline{\Upsilon}\Upsilon = \frac{1}{2e^2_\text{1d}} \int\! d\theta d\bar\theta \, \Tr \overline{\Upsilon_h} e^{-V} \Upsilon_h e^{-V}
\stackrel{\text{WZ}}{=} \frac{1}{2e^2_\text{1d}} \Tr \Bigl[ (D_t\sigma)^2 + D^2 + i\overline{\lambda} D_t^+ \lambda \Bigr] \;.
\eea
Note that the superfield $V^--V^+$ transforms covariantly, $V^--V^+\rightarrow k \, (V^--V^+) \, k^{-1}$, under gauge transformations. For an adjoint-invariant form $\zeta:i\fg\rightarrow \bR$, the Fayet-Iliopoulos term is:
\be
\int\! d\theta d\bar\theta \; \zeta\bigl( V^- - V^+ \bigr) = \int\! d\theta d\bar\theta \; \zeta\Bigl( \bigl( V^-_h - V^+_h \bigr) \, e^{-V} \Bigr) \,\stackrel{\text{WZ}}{=}\, - \zeta(D) \;.
\ee
If the gauge group is Abelian, $V^+_he^{-V}=\frac{1}{2}(\overline{D}D-D\overline{D})V$ becomes a total derivative under the superspace integral. Therefore, FI terms for Abelian gauge groups can be written as
\be
\int\! d\theta d\bar\theta \; \zeta\bigl( V^-_h e^{-V} \bigr) \;.
\ee
We can also write a mass term that gaps $V^-$ (or equivalently the gaugino and $\sigma$):
\be
\label{1d vect mass term}
-\frac12 \int\! d\theta d\bar\theta \, \Tr \bigl( V^- - V^+ \bigr)^2 = - \frac12 \int\! d\theta d\bar\theta \, \Tr\Bigl( \bigl( V^-_h - V^+_h \bigr) \, e^{-V} \Bigr)^2 \,\stackrel{\text{WZ}}{=}\, \Tr \bigl( \overline{\lambda}\lambda - 2\sigma D \bigr) \;.
\ee

Moving on to the matter sector, the conventional kinetic term for a chiral multiplet is:
\bea
\label{1d chiral kin term}
i \int\! d\theta d\bar\theta \; \overline{\Phi}_k \cD_t^- \Phi_k &= \int\! d\theta d\bar\theta \, \biggl( \frac{i}{2} \, \overline{\Phi_h} \, e^V \partial_t \Phi_h - \frac{i}{2} \, \partial_t \overline{\Phi_h} \, e^V \Phi_h + \overline{\Phi_h} \, V^-_h \, \Phi_h \\
& \stackrel{\text{WZ}}{=}
- \overline{\phi} \bigl( D_t^2 + \sigma^2 + D \bigr) \phi + i\overline{\psi}D_t^-\psi + i\overline{\phi}\lambda\psi - i \overline{\psi} \, \overline{\lambda}\phi \;,
\eea
where $D_t^-\equiv\partial_t-i(A_t-\sigma)$. It requires the presence of both $V^-$ and $\Omega$. Alternatively, we can write a kinetic term that couples to $V^+$ in place of $V^-$, in which case only $\Omega$ (or $V$) is required:
\bea
i \!\int\!\! d\theta d\bar\theta \; \overline{\Phi}_k \cD_t^+ \Phi_k &= \!\int\! d\theta d\bar\theta \, \biggl( \frac{i}{2} \, \overline{\Phi_h} \, e^V \partial_t \Phi_h - \frac{i}{2} \, \partial_t \overline{\Phi_h} \, e^V \Phi_h + \overline{\Phi_h} \, V^+_h \, \Phi_h \biggr) \\
& \stackrel{\text{WZ}}{=} D_t^+\overline{\phi} \, D_t^+\phi + i \overline{\psi}D_t^+\psi \;.
\eea
We can also write a term with a first-order action for $\phi$, and it only requires $\Omega$:
\be
\label{1d chiral unconv kin term}
\int\! d\theta d\bar\theta \; \overline{\Phi}_k \Phi_k = \int\! d\theta d\bar\theta \; \overline{\Phi_h} \, e^V \Phi_h \,\stackrel{\text{WZ}}{=}\, i \, \overline{\phi} D_t^+ \phi+\overline{\psi}\psi \;.
\ee
The conventional kinetic term for a Fermi multiplet is
\be
\label{1d Fermi kin term}
\int\!\! d\theta d\bar\theta \; \overline{\cY}_k \cY_k = \int\!\! d\theta d\bar\theta \; \overline{\cY_h} e^V \cY_h \,\stackrel{\text{WZ}}{=} \, i\overline{\eta}D_t^+\eta + \overline{f}f - \bigl\lvert E(\phi) \bigr\rvert^2 - \overline{\eta} \, \partial_aE(\phi) \, \psi_a - \overline{\psi}_a \, \overline{\partial}_a\overline{E}(\overline{\phi}) \,\eta \,,
\ee
and it only requires $\Omega$. If present, terms in $E(\Phi)$ that are linear in the chiral superfields $\Phi_a$ give rise to mass terms which gap out the chiral and Fermi multiplets together. Quadratic or higher-order terms in $E(\Phi)$ produce cubic or higher-order interactions. We shall call them E-interactions. Suppose now that we have a collection of Fermi superfields $\cY_i$ with $\overline{D}\cY_i=E_i(\Phi)$. In addition to $E_i$, we associate another holomorphic function $J_i(\Phi)$ of the chiral superfields to each Fermi such that $E_iJ_i$ (with repeated indices summed) is gauge invariant and $E_iJ_i=0$. Then $\cY_i \, J_i(\Phi)$ is a gauge-invariant fermionic chiral superfield. We can therefore write the F-terms:
\be
\label{1d J term}
\int\!\! d\theta \; \cY_i \, J_i(\Phi) + \!\int\!\! d\bar\theta \; \overline{\cY_i} \,\overline{J}_i(\overline{\Phi}) = -f_iJ_i(\phi) - \eta_i \, \partial_aJ_i(\phi) \, \psi_a - \overline{f}_i \, \overline{J}_i(\overline{\phi}) - \overline{\psi}_a \, \overline{\partial}_a\overline{J}_i(\overline{\phi}) \, \overline{\eta}_i \;.
\ee
Note that because $\cY_iJ_i$ is gauge invariant, $\cY_{i,h}J_i(\Phi_h)=\cY_{i,k}J_i(\Phi_k)$. We will call interactions that are constructed in this way J-interactions.

\subsection{Twisted 3d Yang-Mills and Chern-Simons terms}
\label{app: twisted ym and cs}

In this subsection, we show how the parts of the topologically twisted 3d Yang-Mills and Chern-Simons Lagrangians containing $\Xi_{\bar1}$ can be written in 1d superspace. The terms lie slightly beyond the scope of the exposition above, because $\Xi_{\bar1}$ transforms as a connection on $S^2$ under gauge transformations, as reported in \eqref{gauge chiral super gt h}.

\paragraph{Yang-Mills.}
The first line in \eqref{3d YM in 1d susy parts} can be written in superspace as:
\bea
\label{YM chiral part k transf}
& \Tr \Bigl[ 4 \lvert F_{t\bar1} \rvert^2 + 4iDF_{1\bar1} - 4|D_{\bar1}\sigma|^2 + i\overline\Lambda_1 (D_t+i\sigma \Lambda_{\bar1} + 2 \Lambda_t \, D_1\Lambda_{\bar 1} - 2\overline\Lambda_1 \, D_{\bar 1} \overline\Lambda_t \Bigr] \\
&\quad \stackrel{\text{WZ}}{=}\, 4i \int\! d\theta d\bar\theta \; \Tr \Bigl( \Xi_{1,k} \, \partial_t\Xi_{\bar1,k} - \cF_{1\bar1,k} \, V^- \Bigr) \,,
\eea
where we defined the superfield
\be
\cF_{1\bar1,k} \,\equiv\, \partial_1 \Xi_{\bar1,k} - \partial_{\bar1} \Xi_{1,k} - i \bigl[ \Xi_{1,k}, \Xi_{\bar1,k} \bigr] \;. 
\ee
Here $\cF_{1\bar1,k}$ transforms covariantly under super-gauge transformations as $\cF_{1\bar1,k} \mapsto k \cF_{1\bar1,k} k^{-1}$. Note that the superspace expression has the same form as a Chern-Simons term for superfields, with $V^-$ playing the role of the connection along $t$. Therefore, under finite gauge transformations:
\begin{align}
& \delta_\text{gauge} \; 4i \!\int\! d\theta d\bar\theta \,  \Tr \Bigl( \Xi_{1,k} \, \partial_t \Xi_{\bar1,k} - \cF_{1\bar1,k} \, V^- \Bigr) = 2i \!\int\! d\theta d\bar\theta \, \Tr k^{-1}\partial_tk \, \bigl[ k^{-1}\partial_1k ,\, k^{-1}\partial_{\bar1}k] \;, \nn\\
&\qquad = 2 i \Tr \partial_t \partial_\theta \Bigl( k^{-1}\partial_{\bar\theta} k \bigl[ k^{-1}\partial_1k ,\, k^{-1}\partial_{\bar1} k \bigr] \Bigr) + \text{cyclic} \;.
\label{superspace int winding}
\end{align}
The omitted terms contain cyclic permutations of $(t,1,\bar1)$. This gauge variation looks like a winding number for super-gauge transformations. Since we are taking derivatives of the winding number density (albeit with respect to fermionic variables), a total derivative is expected because the winding number is homotopy invariant. 

Alternatively, we can use superfields which are only sensitive to complexified gauge transformations. The superspace expression in \eqref{YM chiral part k transf} can then be written as
\be
\label{YM chiral part h transf}
\eqref{YM chiral part k transf} = 4i \!\int\! d\theta d\bar\theta \, \Tr \Bigl( \Xi_{1,h} \, \partial_t \Xi_{\bar1,h} - \cF_{1\bar1,h} \, e^{-V} V^-_h \Bigr) \;,
\ee
where total derivatives of the kind \eqref{superspace int winding} have been neglected. One can check that \eqref{YM chiral part h transf} is real and gauge invariant up to total derivatives.

\paragraph{Chern-Simons.}
We now want to write the first piece of \eqref{3d CS in 1d susy parts} in superspace. To do this, we follow a similar procedure as in \cite{Zupnik:1988ry}. First, the fields $X$ are extended to be functions $\wh{X}$ of an auxiliary coordinate $y\in(0,1)$ in an arbitrary way, except that they must fulfil boundary conditions
\be
\label{boundary cond extension}
\wh{X}(\theta,\varphi,t,y=0)=0 \;,\qquad \wh{X}(\theta,\varphi,t,y=1)=X(\theta,\varphi,t) \;.
\ee
Extended quantities will be denoted with a hat. Given \eqref{boundary cond extension}, we have:
\be
\cL_{\text{CS},\Xi} \Big|_\text{WZ} = \wh{\cL}_{\text{CS},\Xi}(y=1) \Big|_\text{WZ} = \int_0^1 \! dy \; \partial_y \wh{\cL}_{\text{CS},\Xi} \Big|_\text{WZ} \;.
\ee
Now, $\partial_y \wh{\cL}_{\text{CS},\Xi}$ can be written in superspace as:
\begin{align}
\partial_y \wh{\cL}_{\text{CS},\Xi} \Big|_\text{WZ} &= \Tr \Bigl[ -4i\partial_y(\hat{A}_t+\hat{\sigma}) \hat{F}_{1\bar1} + 4\partial_y\hat{A}_1 \bigl( i\partial_t \hat{A}_{\bar1} - i\hat{D}_{\bar1} (\hat{A}_t+\hat{\sigma}) \bigr) + \partial_y\hat{\overline{\Lambda}}_1 \hat{\Lambda}_{\bar1} \nn \\
&\qquad\quad {} + 4\partial_y\hat{A}_{\bar1} \bigl( -i\partial_t \hat{A}_1 + i\hat{D}_1(\hat{A}_t+\hat{\sigma}) \bigr) - \partial_y\hat{\Lambda}_{\bar1} \hat{\overline{\Lambda}}_1 \Bigr] \nn \\
&= 4 \partial_y \!\int\! d\theta d\bar\theta \, \Tr\Bigl[ \hat{\Xi}_{1,\overline{h}} \, \hat{\Xi}_{\bar1,h} - i\hat{V} \Bigl( \partial_1 \hat{\Xi}_{\bar1,h} -\partial_{\bar1} \hat{\Xi}_{1,\overline{h}} - i \bigl[ \hat{\Xi}_{1,\overline{h}} ,\, \hat{\Xi}_{\bar1,h} \bigr] \Bigr) \Bigr] \;.
\end{align}
This superspace expression is only valid in Wess-Zumino gauge where $V = - \theta\bar\theta (A_t+\sigma)$, and it is not invariant under super-gauge transformations. Even so, we can take it as a starting point for constructing the gauge-invariant completion. A gauge-invariant expression that reduces to the above in Wess-Zumino gauge is
\be
\label{CS superspace integrand}
\partial_y \wh{\cL}_{\text{CS},\Xi} = 4 \int\! d\theta d\bar\theta \, \Tr \Bigr[ -i \, e^{-\hat{V}}\partial_y\big(e^{\hat{V}}\big) \hat{\cF}_{1\bar1,h} + \hat{\Xi}_{1,h} \, \partial_y \hat{\Xi}_{\bar1,h} + \partial_y\hat{\Xi}_{1,\overline{h}} \, \hat{\Xi}_{\bar1,\overline{h}} \Bigr] \;.
\ee
One can check that the first term is Hermitian, while the second and third terms are Hermitian conjugates of each other. Therefore
\bea
\label{CS superspace}
\cL_{\text{CS},\Xi} &\,=\, \Tr \Bigl[ 4iA_1\partial_tA_{\bar1} - 4i(A_t+\sigma)F_{1\bar1} + \overline{\Lambda}_1 \Lambda_{\bar1} \Bigr] \\
&\stackrel{\text{WZ}}{=} 4\int_0^1 \! dy \, d\theta d\bar\theta \, \Tr\Bigl[ -i \, e^{-\hat{V}} \partial_y \big(e^{\hat{V}}\big) \hat{\cF}_{1\bar1,h} + \hat{\Xi}_{1,h} \, \partial_y \hat{\Xi}_{\bar1,h} + \partial_y\hat{\Xi}_{1,\overline{h}} \, \hat{\Xi}_{\bar1,\overline{h}} \Bigr] \;.
\eea
If the gauge group is Abelian, \eqref{CS superspace integrand} is a total derivative in $y$ and the auxiliary coordinate $y$ can be eliminated to give
\be
\cL_{\text{CS},\Xi} = 4 \int\! d\theta d\bar\theta \, \biggl[ \Xi_{1,\overline{h}} \, \Xi_{\bar1,h} - iV \bigl( \partial_1\Xi_{\bar1,h} - \partial_{\bar1} \Xi_{1,\overline{h}} \bigr) + \frac{1}{2}\partial_1 V \, \partial_{\bar1} V \biggr] \;.
\ee
For non-Abelian gauge groups there is no compact expression for the integral in $y$, but we can expand in powers of $V$. Choosing
\be
\hat{\Xi}_{\bar1,h} = y \, \Xi_{\bar1,h} \;,\qquad\qquad \hat{V} = y \, V \;,
\ee
one obtains the following expression up to quadratic terms in $V$:
\begin{multline}
\cL_{\text{CS},\Xi} = 4 \!\int\! d\theta d\bar\theta \, \Tr\biggl[ \Xi_{1,\overline{h}} \, \Xi_{\bar1,h} - iV \Bigl( \partial_1 \Xi_{\bar1,h} - \partial_{\bar1} \Xi_{1,\overline{h}} - i \bigl[ \Xi_{1,\overline{h}} ,\, \Xi_{\bar1,h} \bigr] \Bigr) \\
+ \frac{1}{2} \Bigl( \partial_1 V- i \bigl[ \Xi_{1,\overline{h}},V \bigr] \Bigr) \Bigl( \partial_{\bar1}V - i \bigl[ \Xi_{\bar1,h}, V \bigr] \Bigr) + \cO(V^3) \biggr] \;.
\end{multline}

\section{Partial gauge fixing}
\label{app: partial gauge fixing}

In this appendix we follow \cite{Ferrari:2013aza} and review the general procedure for partial gauge fixing. Let $\cG$ be the infinite-dimensional group of gauge transformations, and $\{e_A\}$ a Hermitian basis for its algebra $\fg$. Denote the structure constants of $\fg$ as $[e_A,e_B]=if_{ABC}\,e_C$. The basis $\{e_A\}$ is also chosen such that it is orthonormal under the inner product 
\be
\int\Tr \, (e_A \, e_B) = \delta_{AB} \;.
\ee
Let $\mathcal{R}\subset\cG$ be a subgroup, which will be the group of residual gauge transformations after partial gauge fixing. We call its algebra $\fr \subset \fg$ ($\fr$ stands for residual). We split the basis as $\{e_A\} = \{e_i,e_a\}$, where $\{e_i\}$ is a basis for $\fr$ whereas $\{e_a\}$ is a basis for $\ff \cong \fg/\fr$ ($\ff$ stands for gauge fixed). Since $\mathcal{R}$ is a subgroup, $\fr$ is a subalgebra and $[\fr,\fr] \subset \fr$, or $f_{ija}=0$. By anti-symmetry of the structure constants this implies $f_{iaj}=0$, or $[\fr, \ff] \subset \ff$. In summary, the algebra of $\fg$ decomposes as
\be
[e_i,e_j] = i f_{ijk} \, e_k \;,\qquad [e_i,e_a] = i f_{iab} \, e_b \;,\qquad [e_a,e_b] = i f_{abi} \, e_i + i f_{abc} \, e_c \;.
\ee
In particular, this implies that the $e_a$'s transform under the adjoint action in a real orthogonal representation of $\mathcal{R}$, which we call $R_{f}$.

In order to fix $\cG$ to $\mathcal{R}$, we need to choose as many gauge-fixing conditions as there are generators in $\ff$. In other words, we need to choose gauge-fixing functions $G_\text{gf}^a(X)$, where $X$ collectively denotes the physical fields in chiral and vector multiplets. Notice that $G_\text{gf}^a(X)$ should transform in $R_{f}$ under $\mathcal{R}$. This is true for all the gauge-fixing functions we can think of. The first step in the gauge-fixing procedure is to integrate in an adjoint scalar $\Lambda\in \fg$, and add $\int\frac{1}{2}\Tr\Lambda^2$ to the action. Notice that $\Lambda$ will have mass dimension $[\Lambda]=3/2$. Since $\Lambda$ is completely decoupled from everything else, introducing it does not change the path integral. We then insert $1$ in the path integral, written as
\be
1 = \Delta(X,\Lambda) \int_{\cG} \cD g \, \prod_a \, \delta \bigl( G^a_\text{gf}(X^g)-(\Lambda^g)^a \bigr) \;,
\ee
where superscripts $(\cdot)^g$ denote a finite gauge transformation by $g$. Suppose that $g_{X,\Lambda}\in\cG$ satisfies $G^a_\text{gf}(X^{g_{X,\Lambda}})-(\Lambda^{g_{X,\Lambda}})^a=0$, then so does $rg_{X,\Lambda}$ for any $r\in\mathcal{R}$, due to the covariant transformations of $G^a_\text{gf}$ and $\Lambda^a$ under $\mathcal{R}$. Therefore, $\mathcal{R}$ remains as the residual gauge group. Notice that it is necessary for $\Lambda$ to transform under gauge transformations. This is different from the standard Faddeev-Popov procedure, in which $\Lambda$ is only integrated over at the very last step. That would have been sufficient if the gauge were completely fixed ($\mathcal{R}=0$). The slightly different procedure described here will produce extra interaction terms in the ghost action. Now, as usual, the invariance of $\cD g$ ensures that the determinant $\Delta$ is gauge invariant, and
\be
\Delta(X,\Lambda)^{-1} = \Delta (X^{g_{X,\Lambda}}, \Lambda^{g_{X,\Lambda}})^{-1} = \int_{\cG} \cD g \, \prod_a \, \delta \Bigl( G^a_\text{gf}(X^{g \cdot g_{X,\Lambda}})-(\Lambda^{g \cdot g_{X,\Lambda}})^a \Bigr) \;.
\ee
Assuming no Gribov copies and writing $g=1+\epsilon^Ae_A$, $\delta_A\equiv\delta_\text{gauge}(e_A)$, one can expand the argument of the delta function to linear order in $\epsilon^A$ and obtain $\epsilon^b \, \delta_b \bigl[ G_\text{gf}(X^{g_{X,\Lambda}}) - \Lambda^{g_{X,\Lambda}} \bigr]^a$.
The fact that the terms with $\epsilon^i$ disappear ensures that $\text{Vol}(\mathcal{R})$ is factorized as an overall factor in the Faddeev-Popov determinant:
\be
\Delta(X,\Lambda) = \det \delta_b \Bigl[ G^a_\text{gf}(X^{g_{X,\Lambda}}) - (\Lambda^{g_{X,\Lambda}})^a \Bigr]/\,\text{Vol}(\mathcal{R}) \;.
\ee
The determinant can be shown to be well-defined on the coset $\mathcal{R} g_{X,\Lambda}$. Having determined $\Delta(X,\Lambda)$, inserting $1$ in the path integral gives
\be
\int\! \cD X \, \cD\Lambda \, \cD g \; e^{iS(X) - \frac{i}{2} \!\int \Tr\Lambda^2} \; \Delta(X,\Lambda) \, \prod_a \, \delta \bigl( G^a_\text{gf}(X^g) - (\Lambda^g)^a \bigr) \;.
\ee
Undoing the gauge transformation in the delta function, the integral over the gauge group factorizes and one gets
\be
\int\! \cD X \, \cD\Lambda \; e^{iS(X) - \frac{i}{2} \!\int \Tr \Lambda^2} \det\bigl( \delta_bG^a_\text{gf}(X) - \delta_b\Lambda^a \bigr)\, \prod_a \, \delta\bigl( G^a_\text{gf}(X) - \Lambda^a \bigr) \;.
\ee
By means of $\delta_b\Lambda^a=i\Lambda^A[e_b,e_A]^a=-\Lambda^Af_{bAa}=-f_{abi}\Lambda^i-f_{abc}\Lambda^c$ we can explicitly write:
\bea
\det\Bigl( \delta_bG^a_\text{gf}(X) - \delta_b\Lambda^a \Bigr) = \int \biggl( \prod_a \cD \wt c^{\,a} \, \cD c^a \biggr) \, \exp \!\Bigl[ -\wt c^{\,a} \Bigl( \delta_bG^a_\text{gf}(X)+f_{abi}\Lambda^i+f_{abc}\Lambda^c \Bigr) \, c^b \Bigr] \;,
\eea
where we have introduced the Grassmann scalars $c^a$, $\wt c^{\,a}$. Note that they are valued in $\ff$ and not in $\fg$: modes corresponding to residual gauge transformations are not present. Also note that by dimensional analysis, $[\,\wt c\,]+[c]=[G_\text{gf}] =3/2$. Without loss of generality, we can take $[c]=0$, $[\,\wt c\,]=3/2$. Integrating out $\Lambda^i$ and imposing the delta functions for $\Lambda^a$, one gets the action:
\be
S(X) + \int \Tr \biggl[ -\frac{G^2_\text{gf}}{2} + G_\text{gf} \bigl\{ \wt c, c \bigr\} + i \, \wt c \, \delta_{\text{gauge}}(c) \, G_\text{gf} + \frac{1}{2} \{\wt c, c\}_\fr \{\wt c, c \}_\fr \biggr] \;.
\ee
This is equivalent to the following action with extra scalars $b^a$ integrated in:
\be
\label{action after partial gf}
S(X) + \int \Tr \biggl[ \frac{b^2}{2} + b \Bigl( G_\text{gf}-\{\wt c,c\} \Bigr) + i \, \wt c \, \delta_\text{gauge}(c) \, G_\text{gf} + \frac{1}{2}\{\wt c,c\}^2 \biggr] \;.
\ee
Notice that $b^a$ have dimension $[b]=3/2$. One should keep in mind that $c,\wt c,b$ only contain modes in $\ff$. We will now rescale
\be
G_\text{gf} \,\rightarrow\, e^{-1}_\text{3d} \, G_\text{gf} \;\qquad b \,\rightarrow\, e^{-1}_\text{3d} \, b \;,\qquad  c \,\rightarrow\, e^{-1}_\text{3d} \, c \;,
\ee
after which $[G_\text{gf}]=2$ , $[c]=\frac{1}{2}$, and $[b]=2$. The gauge-fixing action gains an overall factor of $1/e^{2}_\text{3d}$. This is useful because the background Coulomb gauge $G_\text{gf}=D_i^BA^i/\sqrt{\xi}$ (with $\xi$ a positive dimensionless parameter) that we choose in the main text has dimension $[G_\text{gf}]=2$. This is true for many other standard gauge-fixing functions, such as the Lorenz gauge $\partial_\mu A^\mu/\sqrt{\xi}$ and the background Lorenz gauge $D_\mu^BA^\mu/\sqrt{\xi}$.


\bibliographystyle{ytphys}
\baselineskip=0.99\baselineskip
\bibliography{BHEntropy}


\end{document}